%% file: wbest-170202.tex
\documentclass[preprint]{aastex}
\usepackage{graphics}
\usepackage{natbib}
\usepackage{amsmath,amsthm,amssymb}
\usepackage{color}
\usepackage{ulem}
\usepackage[T1]{fontenc}
\usepackage{epstopdf}
\newcommand{\ips}{\ensuremath{i_{\rm P1}}}
\newcommand{\zps}{\ensuremath{z_{\rm P1}}}
\newcommand{\yps}{\ensuremath{y_{\rm P1}}}

\newcommand{\jtwo}{\ensuremath{J_{\rm 2MASS}}}
\newcommand{\htwo}{\ensuremath{H_{\rm 2MASS}}}
\newcommand{\ktwo}{\ensuremath{K_{S,{\rm 2MASS}}}}
\newcommand{\PS}{\protect \hbox {Pan-STARRS1}}
\newcommand{\WISE}{{\it WISE}}
\newcommand{\um}{\ensuremath{\mu}m}                        % microns
\newcommand{\mjup}{\ensuremath{M_{\mathrm{Jup}}}}           % Jupiter masses
\newcommand{\rjup}{\ensuremath{R_{\mathrm{Jup}}}}           % Jupiter radii
\newcommand{\msun}{\ensuremath{M_{\odot}}}                   % solar masses
\newcommand{\lsun}{\ensuremath{L_{\odot}}}                   % solar luminosities
\newcommand{\teff}{\ensuremath{T_{\rm eff}}}                 % effective temperature
\newcommand{\my}{\protect \hbox {mas yr$^{-1}$}}   % mas/yr
\newcommand{\kms}{\protect \hbox {km s$^{-1}$}}   % km/s
\newcommand{\hto}{H$_2$O}
\newcommand{\htoa}{H$_2$O--1}
\newcommand{\htob}{H$_2$O--2}
\newcommand{\htod}{H$_2$OD}
\newcommand{\fehz}{FeH$_{\rm z}$}
\newcommand{\voz}{VO$_{\rm z}$}
\newcommand{\nai}{Na~\textsc{i}}
\newcommand{\ki}{K\,\textsc{i}}
\newcommand{\kij}{K\,\textsc{i}$_{\rm J}$}

\newcommand{\jht}{\ensuremath{(J-H)_{\rm 2MASS}}}
\newcommand{\jkt}{\ensuremath{(J-K_S)_{\rm 2MASS}}}
\newcommand{\hkt}{\ensuremath{(H-K_S)_{\rm 2MASS}}}
\newcommand{\ywa}{\ensuremath{y_{\rm P1}-W1}}
\newcommand{\wawb}{\ensuremath{W1-W2}}

\newcommand{\fldg}{\mbox{\textsc{fld-g}}}
\newcommand{\intg}{\mbox{\textsc{int-g}}}
\newcommand{\vlg}{\mbox{\textsc{vl-g}}}
\newcommand{\rchi}{\ensuremath{\chi^2_\nu}}
\newcommand{\mua}{\ensuremath{\mu_\alpha{\rm cos}\,\delta}}     % pm_ra * cos(dec)
\newcommand{\mud}{\ensuremath{\mu_\delta}}                                % pm_dec

\shorttitle{Young L dwarfs in Taurus and Sco-Cen}
\shortauthors{Best, W. M. J. et al}

\slugcomment{\large Accepted by {\it The Astrophysical Journal}, 2017 January 31}

\begin{document}

\title{A Search for L/T Transition Dwarfs With \PS\ and \WISE. \\
  III. Young L Dwarf Discoveries and Proper Motion Catalogs in Taurus and Scorpius-Centaurus}
\author{William M. J. Best\altaffilmark{1,7}, 
  Michael C. Liu\altaffilmark{1,7},
  Eugene A. Magnier\altaffilmark{1}, 
  Brendan P. Bowler\altaffilmark{2,3}, 
  Kimberly M. Aller\altaffilmark{1,7}, 
  Zhoujian Zhang\altaffilmark{1},
  Michael C. Kotson\altaffilmark{4}, 
W. S. Burgett\altaffilmark{5}, 
K. C. Chambers\altaffilmark{1}, 
P. W. Draper\altaffilmark{6}, 
H. Flewelling\altaffilmark{1}, 
K. W. Hodapp\altaffilmark{1}, 
N. Kaiser\altaffilmark{1}, 
N. Metcalfe\altaffilmark{6}, 
R. J. Wainscoat\altaffilmark{1},
C. Waters\altaffilmark{1}
}

% The ordering here should be sequential, matching the sequence in the list of authors:
\altaffiltext{1}{Institute for Astronomy, University of Hawaii, 2680 Woodlawn Drive, Honolulu, HI 96822, USA; wbest@ifa.hawaii.edu}
\altaffiltext{2}{McDonald Observatory and the University of Texas at Austin, Department of Astronomy, 2515 Speedway C1400, Austin, TX 78712, USA}
\altaffiltext{3}{McDonald Fellow}
\altaffiltext{4}{Lincoln Laboratory, Massachusetts Institute of Technology, 244 Wood Street, Lexington, MA 02420, USA}
\altaffiltext{5}{GMTO Corporation, 251 S. Lake Ave., Suite 300, Pasadena, CA 91101, USA}
\altaffiltext{6}{Department of Physics, Durham University, South Road, Durham DH1 3LE, UK}
\altaffiltext{7}{Visiting Astronomer at the Infrared Telescope Facility, which
  is operated by the University of Hawaii under Cooperative Agreement
  no. NNX-08AE38A with the National Aeronautics and Space Administration,
  Science Mission Directorate, Planetary Astronomy Program.}

\begin{abstract}

  We present the discovery of eight young M7--L2 dwarfs in the Taurus
  star-forming region and the Scorpius-Centaurus OB Association, serendipitously
  found during a wide-field search for L/T transition dwarfs using \PS\
  (optical) and \WISE\ (mid-infrared) photometry.  We identify
  PSO~J060.3200+25.9644 (near-infrared spectral type L1) and
  PSO~J077.1033+24.3809 (L2) as new members of Taurus based on their \vlg\
  gravity classifications, the consistency of their photometry and proper
  motions with previously known Taurus objects, and the low probability of
  contamination by field objects.  PSO~J077.1033+24.3809 is the coolest
  substellar member of Taurus found to date.  Both Taurus objects are among the
  lowest mass free-floating objects ever discovered, with estimated masses
  $\approx$6~\mjup, and provide further evidence that isolated planetary-mass
  objects can form as part of normal star-formation processes.
  PSO~J060.3200+25.9644 (a.k.a. DANCe~J040116.80+255752.2) was previously
  identified as a likely member of the Pleiades (age~$\approx125$~Myr) based on
  photometry and astrometry, but its \vlg\ gravity classification and
  near-infrared photometry imply a much younger age and thus point to Taurus
  membership.  We have also discovered six M7--L1 dwarfs in outlying regions of
  Scorpius-Centaurus with photometry, proper motions, and low-gravity spectral
  signatures consistent with membership.  These objects have estimated masses
  $\approx$15--36~\mjup.  The M7 dwarf, PSO~J237.1470$-$23.1489, shows excess
  mid-infrared flux implying the presence of a circumstellar disk.  Finally, we
  present catalogs of \PS\ proper motions for low-mass members of Taurus and
  Upper Scorpius with median precisions of $\approx$3~\my, including 67 objects with no
  previous proper motion and 359 measurements that improve on literature values.

\end{abstract}

\keywords{brown dwarfs --- proper motions --- stars: formation --- stars:
  individual (PSO~J060.3200+25.9644, PSO~J077.1033+24.3809,
  PSO~J237.1470$-$23.1489)}

\section{Introduction}
\label{intro}

Brown dwarfs with ages $\lesssim100$~Myr are valuable laboratories for testing
both the youngest substellar evolutionary models and the lowest-gravity
(therefore lowest-mass) atmospheric models.  For instance, brown dwarfs with
T$_{\rm eff}\lesssim1400$~K and ages $\lesssim20$~Myr will have masses
$\lesssim10$~\mjup\ \citep[e.g.,][]{Chabrier:2000hq}, firmly in the planetary
regime ($\lesssim13$~\mjup).  These young, very low mass brown dwarfs therefore
serve as vital templates for understanding directly imaged planets.

Star-forming regions and young open clusters offer the opportunity to identify
multiple young brown dwarfs in small areas of the sky, at the age when these
objects are brightest.  Planetary-mass objects in star-forming regions have been
discovered both as companions to stars
\citep[e.g.,][]{Luhman:2006ge,Lafreniere:2008jt} and as free-floating objects
\citep[e.g.,][]{Luhman:2009cn,Weights:2009ht}.  Wide-field, red-sensitive
surveys such as the \PS\ $3\pi$~Survey \citep[PS1;][K. Chambers et al., 2017, in
prep]{Kaiser:2010gr}, the Two Micron All Sky Survey
\citep[2MASS;][]{Skrutskie:2006hl}, the UKIRT Infrared Deep Sky Survey
\citep[UKIDSS;][]{Lawrence:2007hu}, and the Wide-Field Infrared Survey Explorer
\citep[\WISE;][]{Wright:2010in} have the ability to detect free-floating very
low mass brown dwarfs in nearby star-forming regions
\citep[e.g.,][]{Lodieu:2013eo,Esplin:2014he}, although interstellar reddening at
optical and near-infrared wavelengths can make brown dwarfs difficult to
distinguish from background giant stars.  More discoveries of these objects
would improve our understanding of the early evolution of low mass brown dwarfs
and giant planets.

At a distance of $\approx$145~pc and an age of $\approx$1--2~Myr
\citep{Kraus:2009dh}, the Taurus-Auriga molecular cloud (hereinafter Taurus) is
one of the closest star-forming regions to the Sun.  A comprehensive review of
Taurus and its observational history can be found in \citet{Kenyon:2008vj}.
Taurus has been searched extensively for substellar members from optical to
mid-infrared wavelengths
\citep[e.g.,][]{Luhman:2006cd,Slesnick:2006ij,Guieu:2006fl,Luhman:2009cn,Quanz:2010hp,Rebull:2010js}.
\citet{Esplin:2014he} cataloged 74 members with spectral types M6--L0, which at
the young age of Taurus span the full brown dwarf mass regime from the
stellar/sub-stellar boundary down to planetary masses ($\lesssim$13~\mjup).  The
coolest known substellar objects in Taurus to date are the free-floating
$\approx$4--7~\mjup\ 2MASS~J04373705+2331080 (hereinafter 2MASS~J0437+2331),
discovered and classified as L0 by \citet{Luhman:2009cn}, and the planetary-mass
companion 2MASS~J04414489+2301513 Bb \citep{Todorov:2010cn}, with a near-IR
spectral type of L$1\pm1$ on the \citet{Allers:2013hk} system and an estimated
mass of $\approx10\pm2$~\mjup\ \citet{Bowler:2015en}.

The Scorpius-Centaurus Association (hereinafter Sco-Cen) is the nearest OB
association to the Sun.  The Sco-Cen complex, reviewed in detail by
\citet{Preibisch:2008wx}, has a distance similar to Taurus but an older age
($\approx$10--20~Myr).  With no significant ongoing star formation, Sco-Cen is
much less affected by interstellar reddening, but any planetary-mass objects
will also have cooled and become fainter than equivalent-mass objects in Taurus.
The Upper Scorpius subgroup in particular has been the target of many searches
for ultracool dwarfs
\citep[e.g.,][]{Martin:2004ki,Lodieu:2006di,Lodieu:2011df,Slesnick:2006ij,Slesnick:2008ci,Dawson:2014hl}.
\citet{Lodieu:2008hm} have probed the deepest into the substellar regime,
spectroscopically confirming over 20 M8--L2 dwarfs in Upper Scorpius with masses
down to $\approx$15~\mjup.

Here we present the discovery of two early-L dwarfs in Taurus and six M7--L1
dwarfs in Sco-Cen, serendipitously identified in a wide-field search for L/T
transition dwarfs in the \PS\ and \WISE\ catalogs.  In Section~\ref{photometry}
we explain how these objects were initially identified, and we describe
follow-up observations in Section~\ref{obser}.  We detail the observed features
of our discoveries in Section~\ref{results}.  We discuss their membership in
Taurus (Section~\ref{taurus}) or Sco-Cen (Section~\ref{scocen}), and provide
estimated masses and comparisons with model spectra.  We summarize our
discoveries in Section~\ref{summary}.

\section{Photometric Selection}
\label{photometry}
We conducted a search over $\approx$28,000~deg$^2$ for L/T transition dwarfs in
the field using a merged catalog of PS1 and \WISE\ photometry.  The search is
described in detail in \citet[hereinafter Paper I]{Best:2013bp}, and the full
spectroscopic follow-up results are presented in \citet[hereinafter Paper
II]{Best:2015em}, including our discovery of 130 ultracool dwarfs.  Among these
discoveries were 23 late-M and L dwarfs with spectroscopic signs of low gravity,
implying youth.  This was a surprisingly large number given that we were
searching for objects with field ages and cooler spectral types
($\approx$L6--T5) and that objects with ages $\lesssim200$~Myr should comprise
at most a few percent of the local population in a galaxy $\gtrsim$10~Gyr old.
In Paper II, we determined that our search could find late-M and L dwarfs with
\wawb\ colors redder than average for their spectral types, bringing younger
objects into our sample.

Our search also specifically avoided the heavily reddened areas of the sky
defined in \citet{Cruz:2003fi}, which include Taurus and Sco-Cen.  The eight
discoveries described in this paper lie just outside these reddened regions,
with one exception: PSO~J060.3200+25.9644 (hereinafter PSO~J060.3+25) lies
$\approx$2$^\circ$ inside the \citet{Cruz:2003fi} Taurus boundaries, but we
pursued follow-up observations because its spectral energy distribution (SED)
from \zps\ through $W2$ (0.9--4.6~\um) strongly suggested an unreddened
ultracool object.

The PS1 \zps\ and \yps, \WISE, 2MASS, and MKO photometry for our discoveries was
originally presented in Paper II.  In Table~\ref{tbl.ps1.wise}, we update the
earlier version of PS1 photometry from Paper II with photometry from the PS1
Data Release 1 (DR1; Chambers et al., 2017, in prep; Magnier et al., 2017, in
prep) and include \ips\ magnitudes.  The photometry used in both Paper II and
this work is the mean PSF photometry from individually calibrated images; the
DR1 photometry in this work includes many more epochs.  We also replace the WISE
All-Sky photometry \citep{Cutri:2012wm} used in Paper II with AllWISE magnitudes
\citep{Cutri:2014wx}.  For reference, we reproduce the 2MASS and MKO photometry
here in Table~\ref{tbl.nir}.  We also include photometry for the
previously-identified Taurus L dwarf 2MASS~J0437+2331.

\section{Near-Infrared Spectroscopy}
\label{obser}
We obtained near-infrared spectra for our candidates between 2013 January and
2015 May using the NASA Infrared Telescope Facility (IRTF).  We used the
facility spectrograph SpeX \citep{Rayner:2003hf} in prism mode with the $0.5''$
($R\approx120$) and $0.8''$ ($R\approx75$) slits.  Details of our observations
are listed in Table~\ref{tbl.obslog}.  For each science target we observed a
nearby A0V star contemporaneously for telluric calibration.  All spectra were
reduced in standard fashion using versions 3.4 and 4.0 of the Spextool software
package \citep{Vacca:2003fw,Cushing:2004bq}.  These observations were part of a
large program (Paper II) in which our desired S/N was $\gtrsim$30, sufficient
for accurate spectral typing based on overall spectral morphology but not
necessarily for robust analysis of specific features.  We therefore observed
PSO~J060.3+25 and PSO~J077.1033+24.3809 (hereinafter PSO~J077.1+24) a second
time to achieve higher S/N, and for each object we combined the first and second
epochs using the Spextool \textit{xcombspec} routine to obtain a single
higher-S/N spectrum, which we present in this paper.

While comparing the spectra of our discoveries to spectral standards, we noticed
a small wavelength offset in the spectrum of the field L0 standard
2MASS~J03454316+2540233 (hereinafter 2MASS~J0345+2540) from
\citet{Burgasser:2006jj}.  The offset is large enough to affect calculations of
the \citet{Allers:2013hk} gravity-sensitive spectral indices that we used to
analyze our discoveries (Section~\ref{results.indices}).  We therefore used
IRTF/SpeX to obtain a new spectrum of 2MASS~J0345+2540 with more accurate
wavelength calibration, which we used for our analysis.
Appendix~\ref{appendixc} presents this new spectrum and provides details of the
observations.

\section{Results}
\label{results}

\subsection{Ultracool Discoveries}
\begin{figure}
\begin{center}
  \includegraphics[width=0.54\columnwidth]{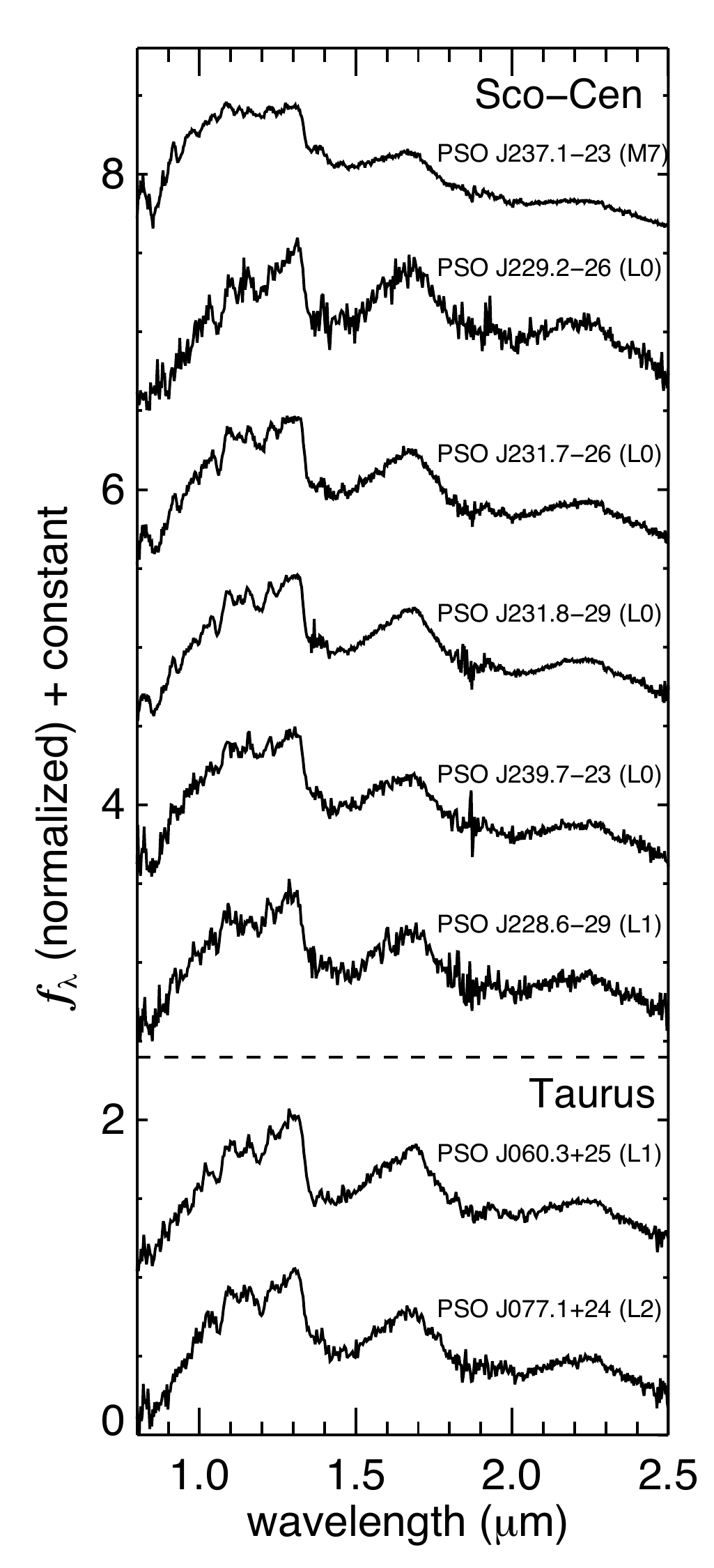}
  \caption{SpeX prism spectra for our eight discoveries, normalized at the
    $J$-band peak ($1.27\,\mu$m), arranged from earliest to latest spectral
    type, and offset by a constant.  Our two Taurus discoveries are at the
    bottom.}
  \label{fig.tausco.stack}
\end{center}
\end{figure}

\begin{figure}
\begin{center}
  \includegraphics[width=0.80\columnwidth, trim = 0 0 0 10mm]{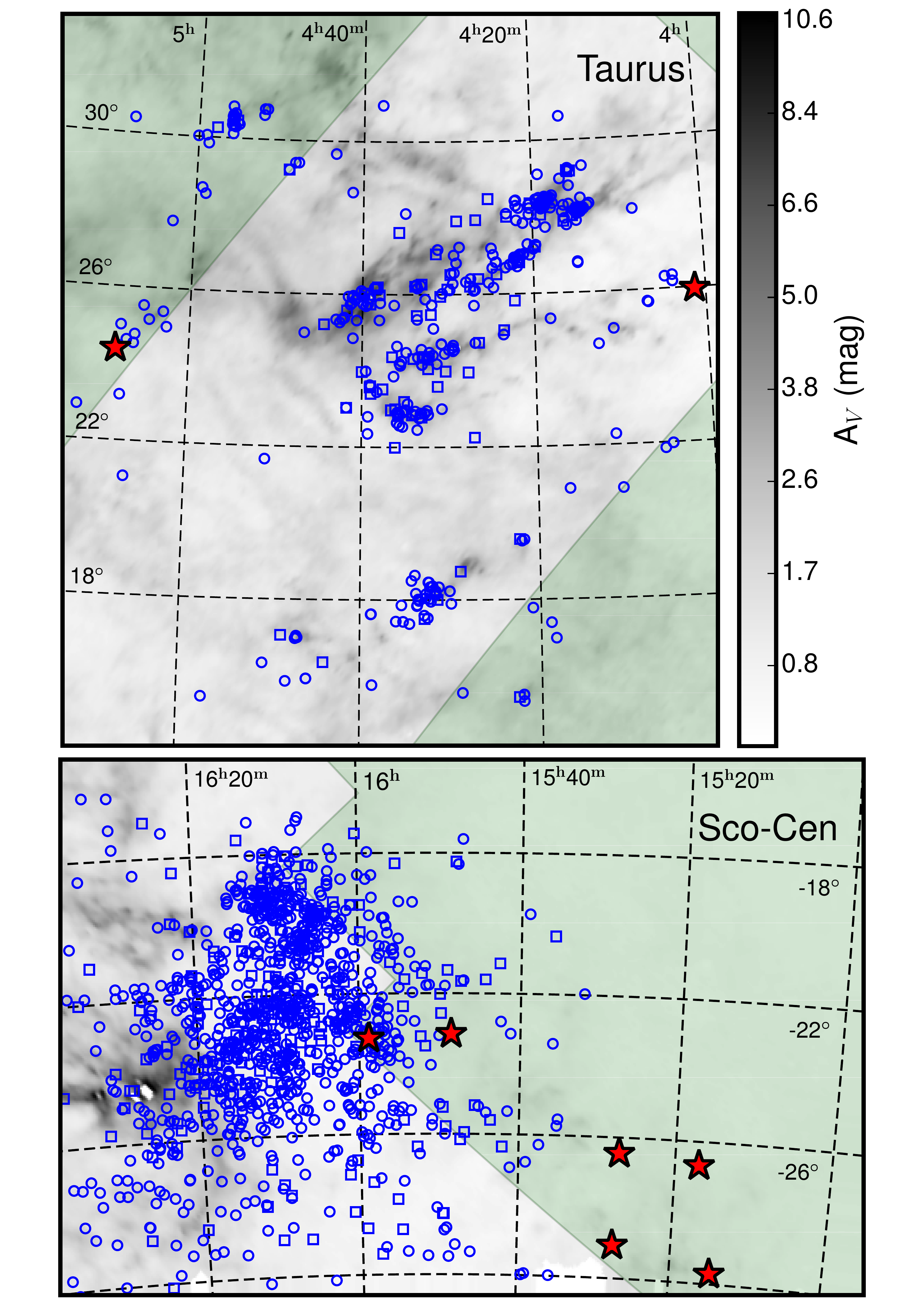}
  \caption{{\it Top}: The locations in Taurus of our discoveries (red stars) and
    known stars (blue circles) and ultracool dwarfs (${\rm SpT}\ge{\rm M6}$,
    blue squares) from \citet{Esplin:2014he}.  The grayscale background shows
    visual extinction (scale at right) from the reddening map of
    \citet{Schlafly:2014hh}.  The green shading marks the regions included in
    our search (Paper II), i.e., outside of reddened regions identified by
    \citet{Cruz:2003fi}.  Our two discoveries lie on the outskirts of Taurus in
    regions of low extinction.  {\it Bottom}: The portion of Sco-Cen surveyed by
    PS1 (north of $-30^\circ$) shown in the same format, with known members of
    Upper Sco from \citet{Luhman:2012hj}, \citet{Dawson:2014hl}, and
    \citet{Rizzuto:2015bs}.  The two leftmost discoveries are in Upper Sco,
    while the other four appear to be members of Upper Centaurus-Lupus.  The
    knot of high extinction near ($16^h30^m$,~$-24^\circ$) is the
    $\rho$~Ophiuchi star-forming region.}
  \label{disc.map}
\end{center}
\end{figure}

The SpeX prism spectra for our eight young ultracool discoveries are presented
in Figure~\ref{fig.tausco.stack}, and their spectral types are listed in
Table~\ref{tbl.indices.allers}.  We show their locations in the sky in
Figure~\ref{disc.map}.  PSO~J060.3+25 was previously identified by
\citet{Sarro:2014ci} and \citet{Bouy:2015gl} as a likely very low-mass member of
the Pleiades cluster, based on photometry and astrometry.
PSO~J237.1471$-$23.1489 (hereinafter PSO~J237.1$-$23) was identified by
\citet{Lodieu:2013eo} as a photometric and astrometric candidate member of Upper
Sco.  We independently discovered these objects and present here spectral
confirmation of their ultracool nature.  We also find that PSO~J060.3+25 is more
likely to be a Taurus member than a Pleiad (Section~\ref{taurus.evidence.pm}).
The other six objects are new discoveries.

\begin{figure}
\begin{center}
  \includegraphics[width=1.00\columnwidth, trim = 0 0 9mm 0]{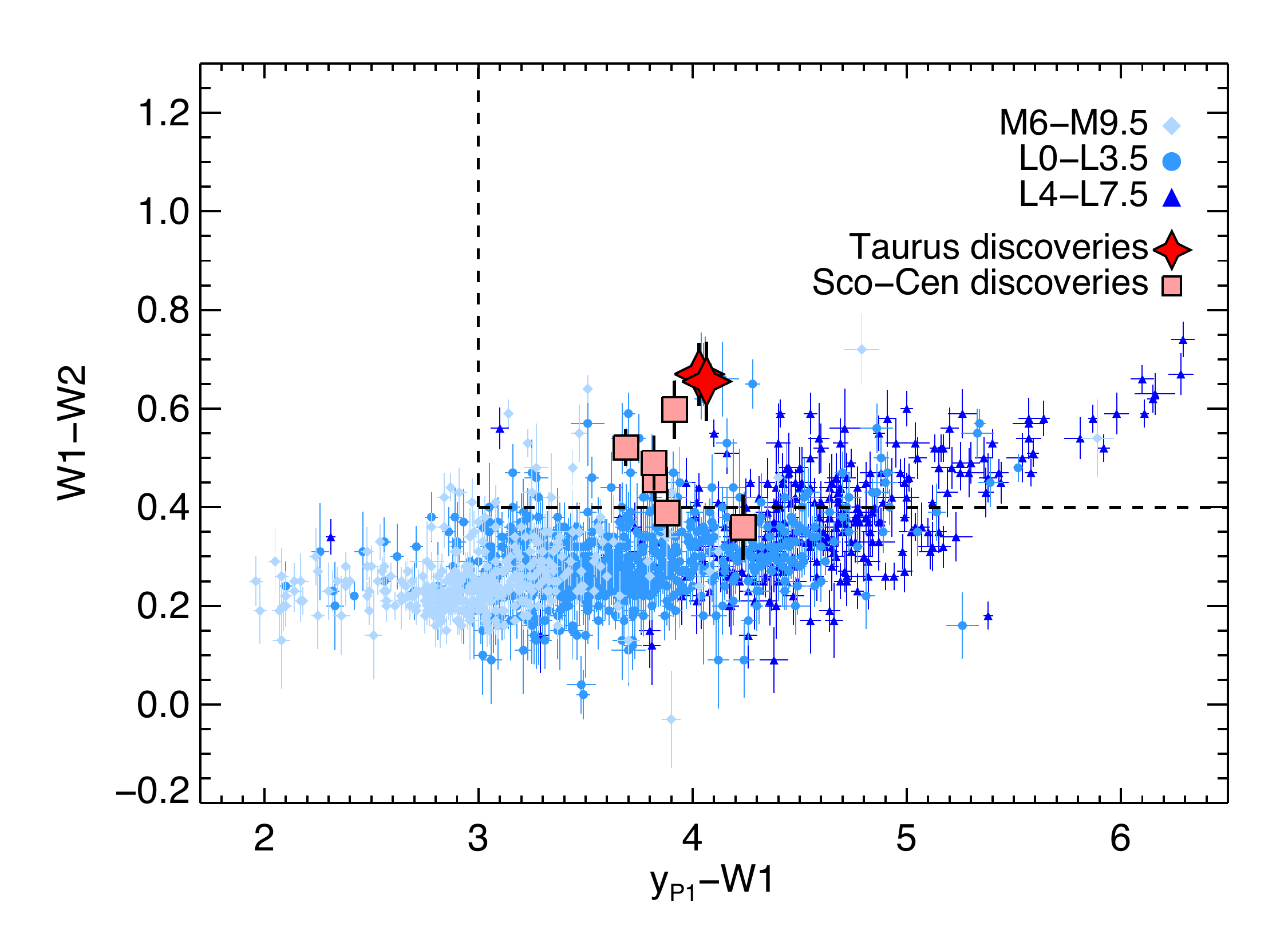}
  \caption{\wawb\ vs. \ywa\ (AllWISE) diagram featuring our discoveries in
    Taurus (red stars) and Sco-Cen (pink triangles), plotted over previously
    known ultracool dwarfs in shades of blue (cooler spectral types have darker
    shades). The black dashed lines represent the color cuts we used in our
    search (Paper II), for which we used WISE All-Sky photometry.  We chose
    objects above and to the right of the dashed lines.  (The two Sco-Cen
    discoveries with AllWISE $W1-W2<0.4$~mag were included in our search because
    they have WISE All-Sky $W1-W2>0.4$~mag.)  Our young M7--L2 discoveries have
    somewhat redder \wawb\ colors than field objects of the same spectral types,
    which explains why our search for L/T transition dwarfs found these
    earlier-type objects.}
\label{fig.w1w2.yw1}
\end{center}
\end{figure}

\begin{figure}
\begin{center}
  \includegraphics[width=1.00\columnwidth, trim = 10mm 0 5mm 0]{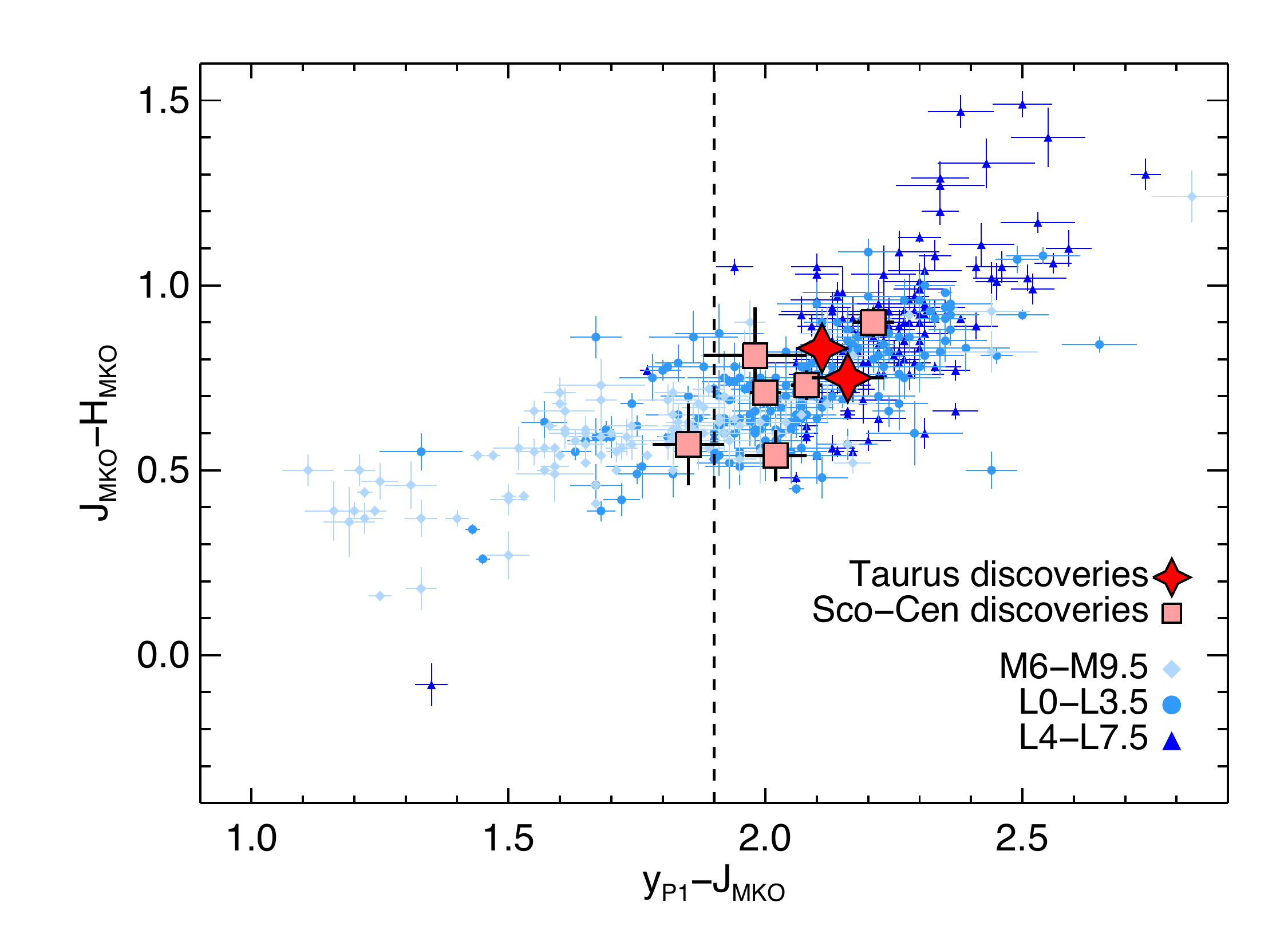}
  \caption{$(J-H)_{\rm MKO}$ vs. $y_{\rm P1}-J_{\rm MKO}$ diagram for our
    discoveries, using the same format as Figure~\ref{fig.w1w2.yw1}.  We chose
    objects in our search to the right of the dashed line (using an earlier,
    preliminary version of \yps\ photometry); $J-H$ was not used to select
    objects.  Our discoveries have normal $y_{\rm P1}-J_{\rm MKO}$ and
    $(J-H)_{\rm MKO}$ colors compared with field objects of the same spectral
    types.}
\label{fig.JH.yJ}
\end{center}
\end{figure}

In Figures \ref{fig.w1w2.yw1} and~\ref{fig.JH.yJ} we compare the colors of our
discoveries with those of previously known late-M and early-L dwarfs.  Our
discoveries in Taurus and Sco-Cen have red \wawb\ colors compared with field
objects of similar spectral types, which led to the fortuitous discovery of
these young M7--L2 dwarfs even though our search was designed to find L/T
transition dwarfs (spectral types $\approx$ L6--T4).  The \ywa,
$y_{\rm P1}-J_{\rm MKO}$ and $(J-H)_{\rm MKO}$ colors of our discoveries are
normal compared with field objects of similar spectral types.

\subsection{Spectral Indices and Spectral Types}
\label{results.indices}
We used three methods to assign spectral types for our discoveries: visual
comparison with low-gravity field standards and two index-based methods.
Table~\ref{tbl.indices.allers} gives the spectral types calculated from the
index-based system of \citet[hereinafter AL13]{Allers:2013hk}.  The AL13 indices
were designed to assign near-infrared spectral types consistent with optical
spectral types, independent of surface gravity.  Since all of our discoveries
show clear spectral signs of low gravity (Section~\ref{results.gravity}), we
adopted the AL13 index-based types as our final spectral types, rounded to the
nearest subtype and assigned an uncertainty of 1~subtype (following AL13).  For
confirmation, we visually compared our spectra to the \vlg\ standards of AL13.
All of our visually-determined types are within 1 subtype of the adopted
index-based types.

In Table~\ref{tbl.indices.burg}, we list the spectral types determined using the
index-based system of \citet[hereinafter B06]{Burgasser:2006cf}, compared with
our adopted AL13 spectral types.  The B06 and AL13 spectral types are consistent
within their $2\sigma$ uncertainties, although the B06 types are mostly 1--3
subtypes later.  This is probably a consequence of the fact that the B06 indices
are not defined for spectral types earlier than L0, so the B06 spectral types
are averages only of L~types.

\subsection{Low-Gravity Signatures}
\label{results.gravity}
Low-gravity signatures in ultracool dwarf spectra are a well-established
indication of ages $\lesssim$200~Myr
\citep[e.g.,][]{Kirkpatrick:2008ec,Allers:2013hk}.  We used the AL13 system
based on gravity-sensitive near-IR spectral indices to assess whether our
spectra display signs of low gravity.  In this system, an object is assigned a
score of 0 for field gravity (\fldg, ages $\gtrsim200$~Myr), 1 for intermediate
gravity (\intg, ages $\approx50-200$~Myr), or 2 for very low gravity (\vlg, ages
$\approx10-30$~Myr).  We calculated indices and gravity scores for our
discoveries following \citet{Aller:2016kg}, who adapt AL13 to incorporate Monte
Carlo assessment of the uncertainties in the indices and gravity classes.  We
also visually examined the gravity-sensitive features in our spectra as a check
on the gravity scores.

We classify six of our discoveries as \vlg: PSO~J060.3+25, PSO~J077.1+24,
PSO~J231.7899$-$26.4494 (hereinafter PSO~J231.7$-$26), PSO~J231.8942$-$29.0600
(hereinafter PSO~J231.8$-$29), PSO~J237.1$-$23, and PSO~J239.7016$-$23.2664
(hereinafter PSO~J239.7$-$23).  Table~\ref{tbl.gravity.allers} lists their
indices and gravity scores.  Figure~\ref{fig.gravplots} compares the spectra of
these six \vlg\ objects with field standards from \citet{Kirkpatrick:2010dc} and
\vlg\ standards from AL13 having the same spectral types.  For the L0 field
standard 2MASS~J0345+2540, we use our new SpeX prism spectrum
(Appendix~\ref{appendixc}).  All six of our spectra display weak 0.99~\um\ FeH
and 1.25~\um\ \ki\ absorption, and a triangular H band shape, all signs of
youth.  PSO~J060.3+25, PSO~J077.1+24, PSO~J231.7$-$26, and PSO~J231.8$-$29 also
show strong 1.06~\um\ VO absorption, which AL13 identify as an additional sign
of youth for L0--L4 dwarfs.

\begin{figure}
\begin{center}
  \epsscale{1.8}
  \plottwo{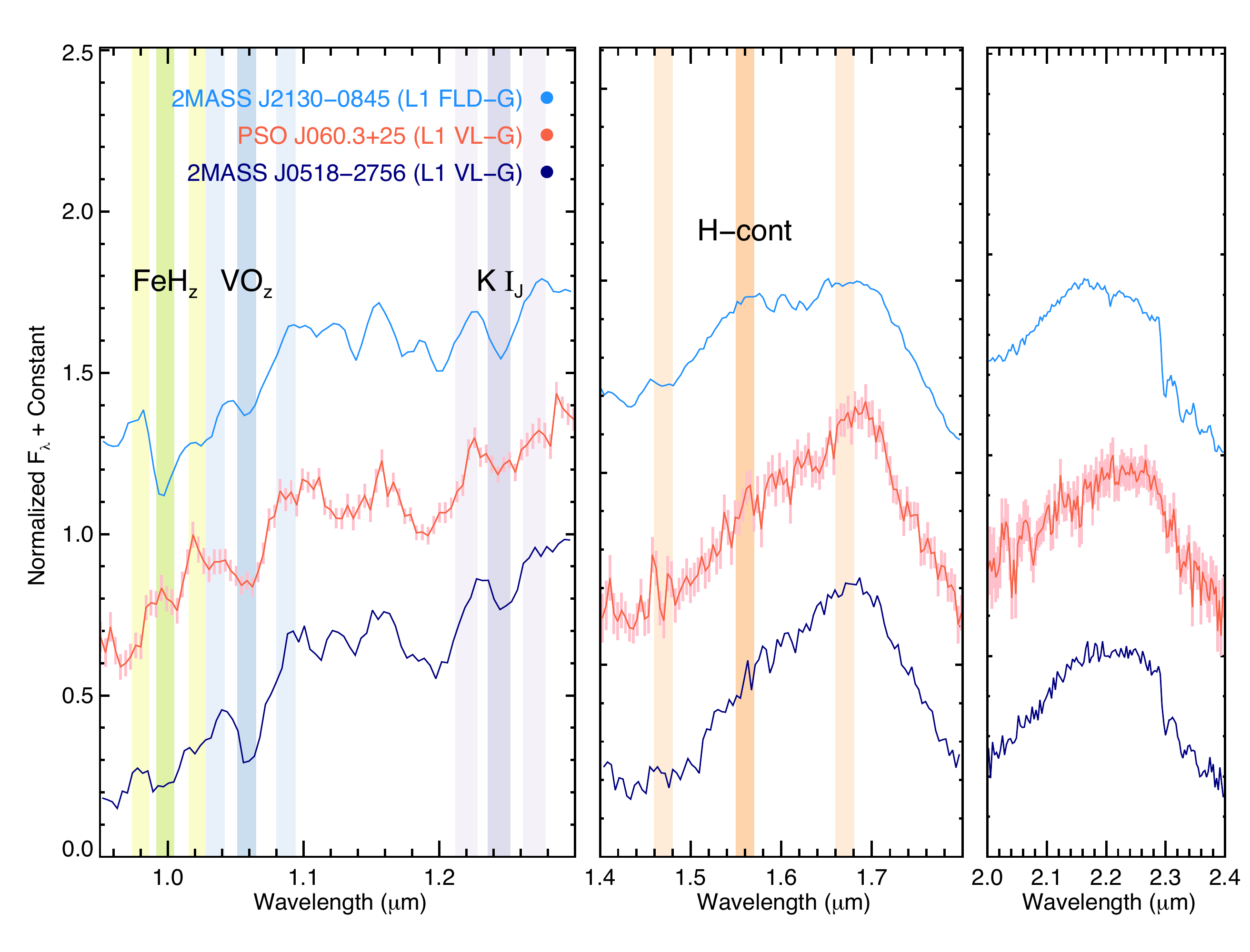}{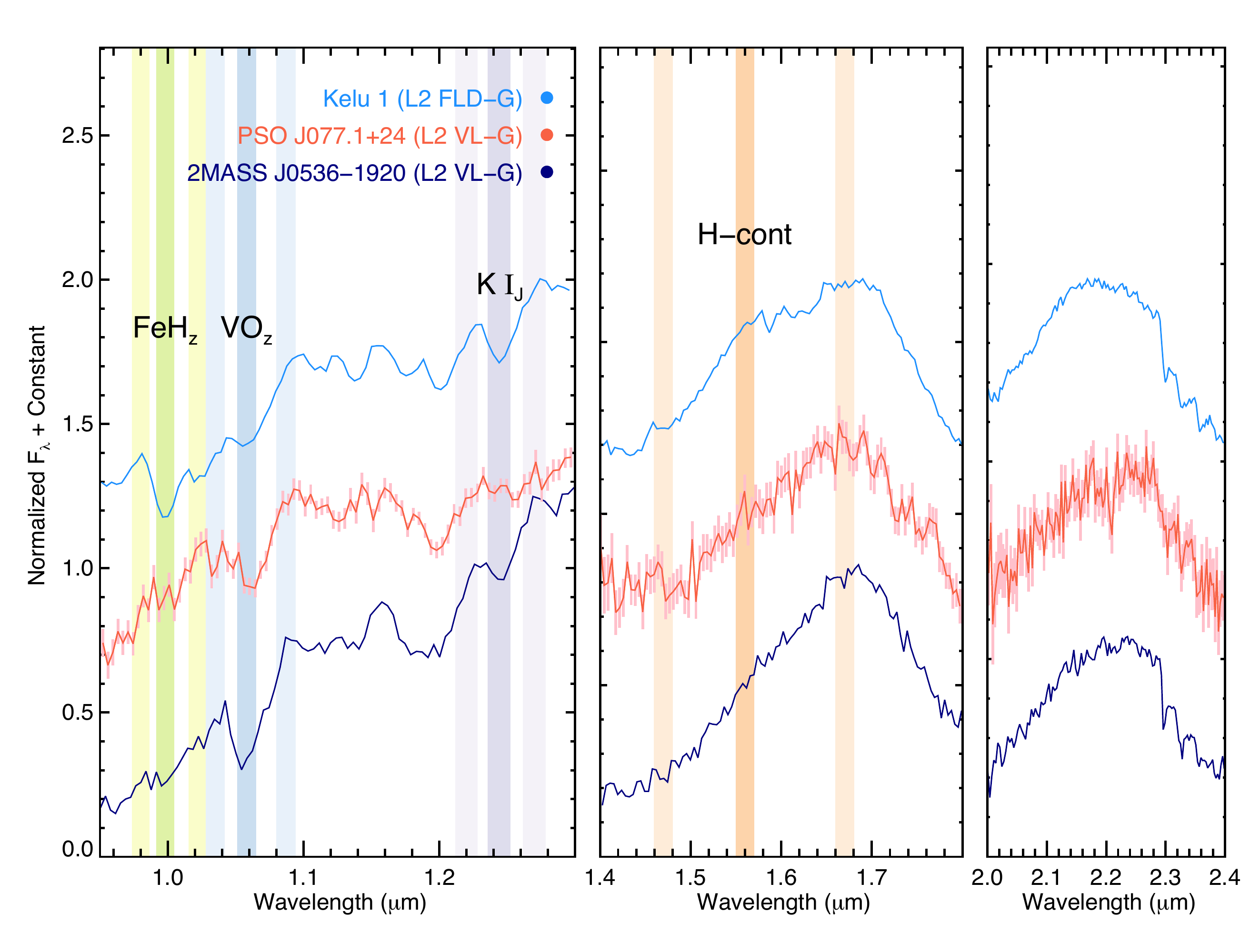}
  \caption{Our six \vlg\ discoveries (middle/red, with error bars), compared
    with field standards (top/light blue) from \citet{Kirkpatrick:2010dc} and
    \vlg\ standards (bottom/dark blue) from AL13 of the same spectral type. The
    vertical colored bands show the wavelength intervals used to calculate the
    labeled spectral indices.  For all six objects, the \fehz, \kij, and H-cont
    features are more similar to the \vlg\ standards, and the \voz\ absorption
    also indicates \vlg\ for the L dwarfs. (\voz\ is not a valid gravity
    indicator for M dwarfs.)}
\figurenum{fig.gravplots.1}
\end{center}
\end{figure}

\begin{figure}
\begin{center}
  \epsscale{1.8}
  \plottwo{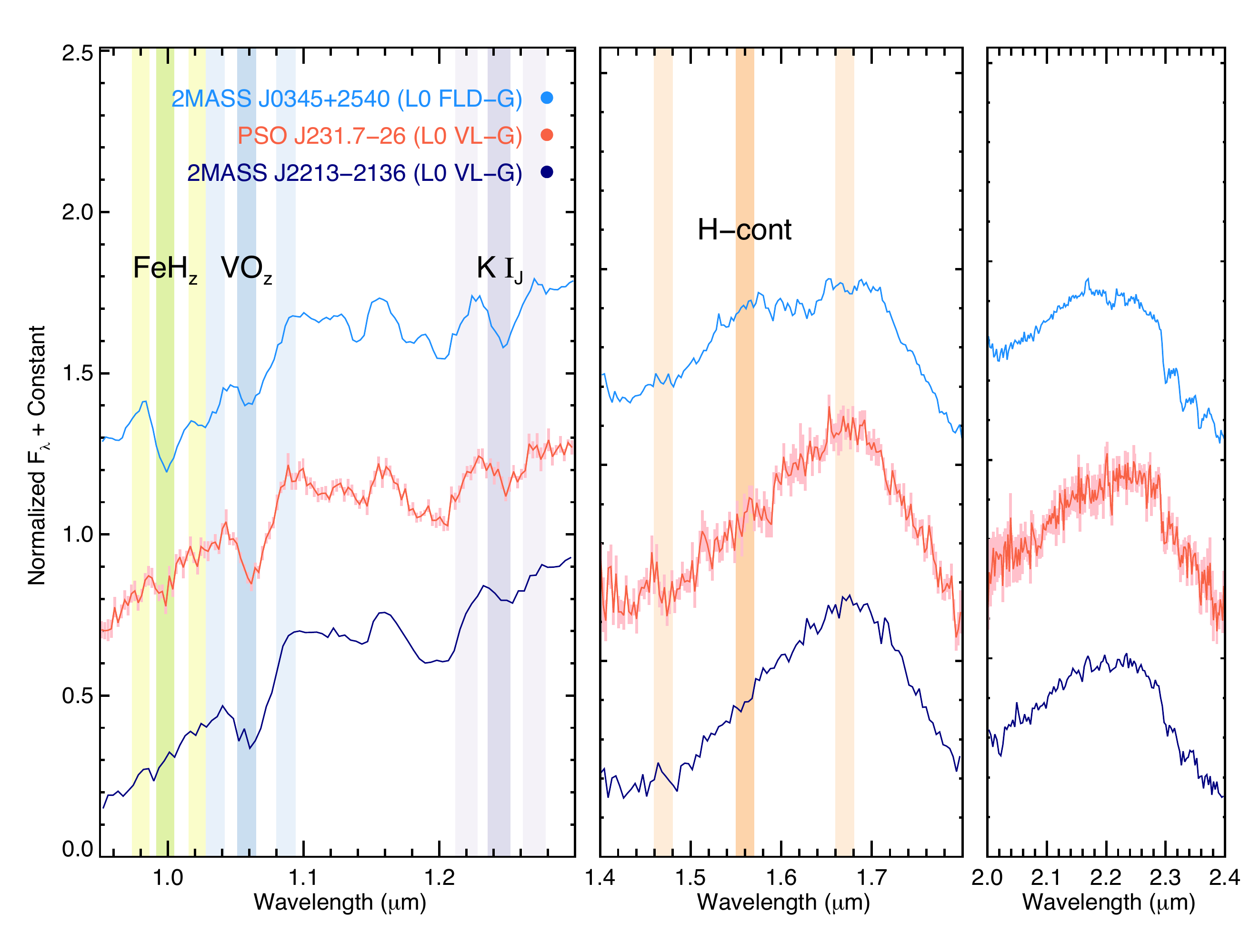}{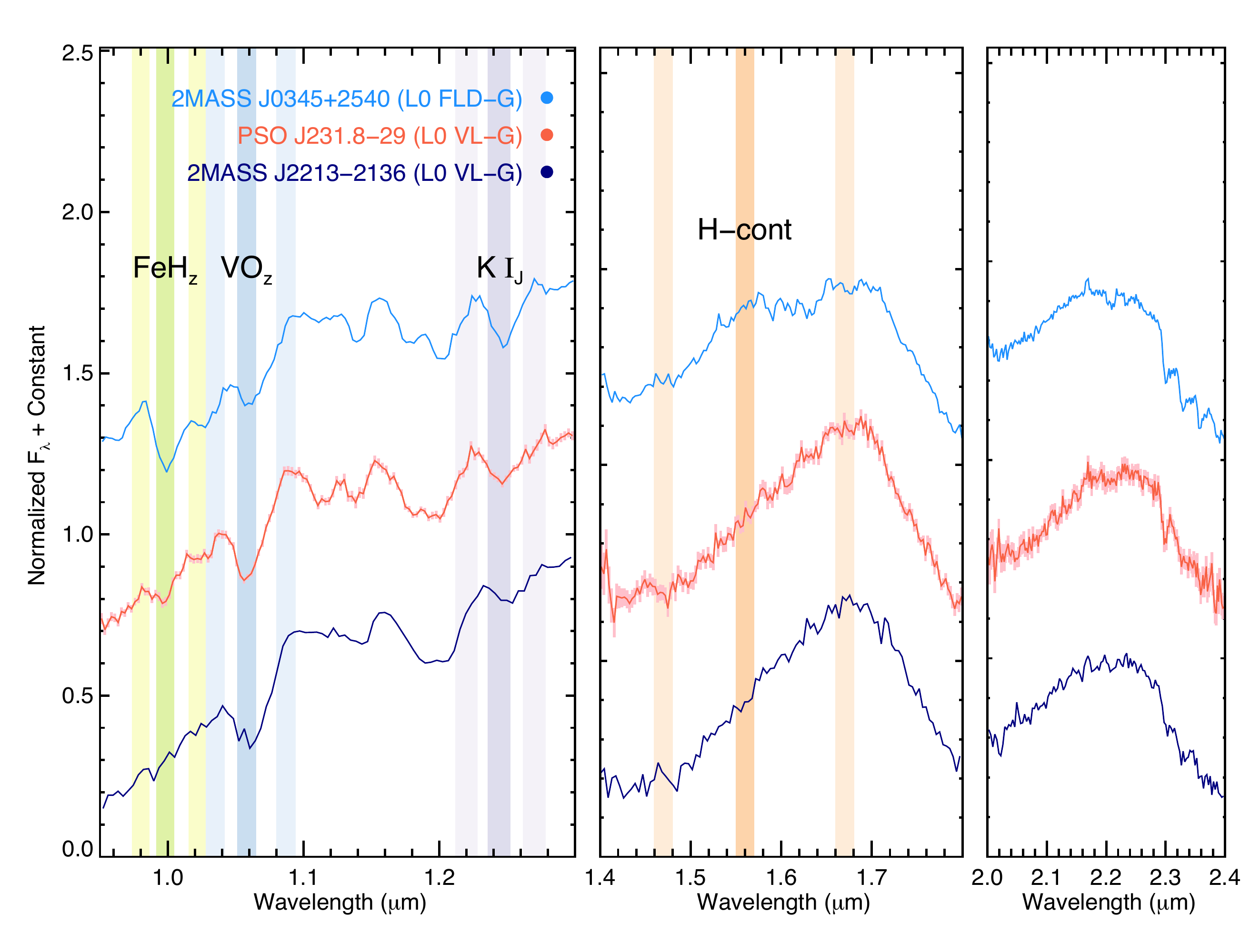}
  \caption{continued.}
  \figurenum{fig.gravplots.2}
\end{center}
\end{figure}

\begin{figure}
\begin{center}
  \epsscale{1.8}
  \plottwo{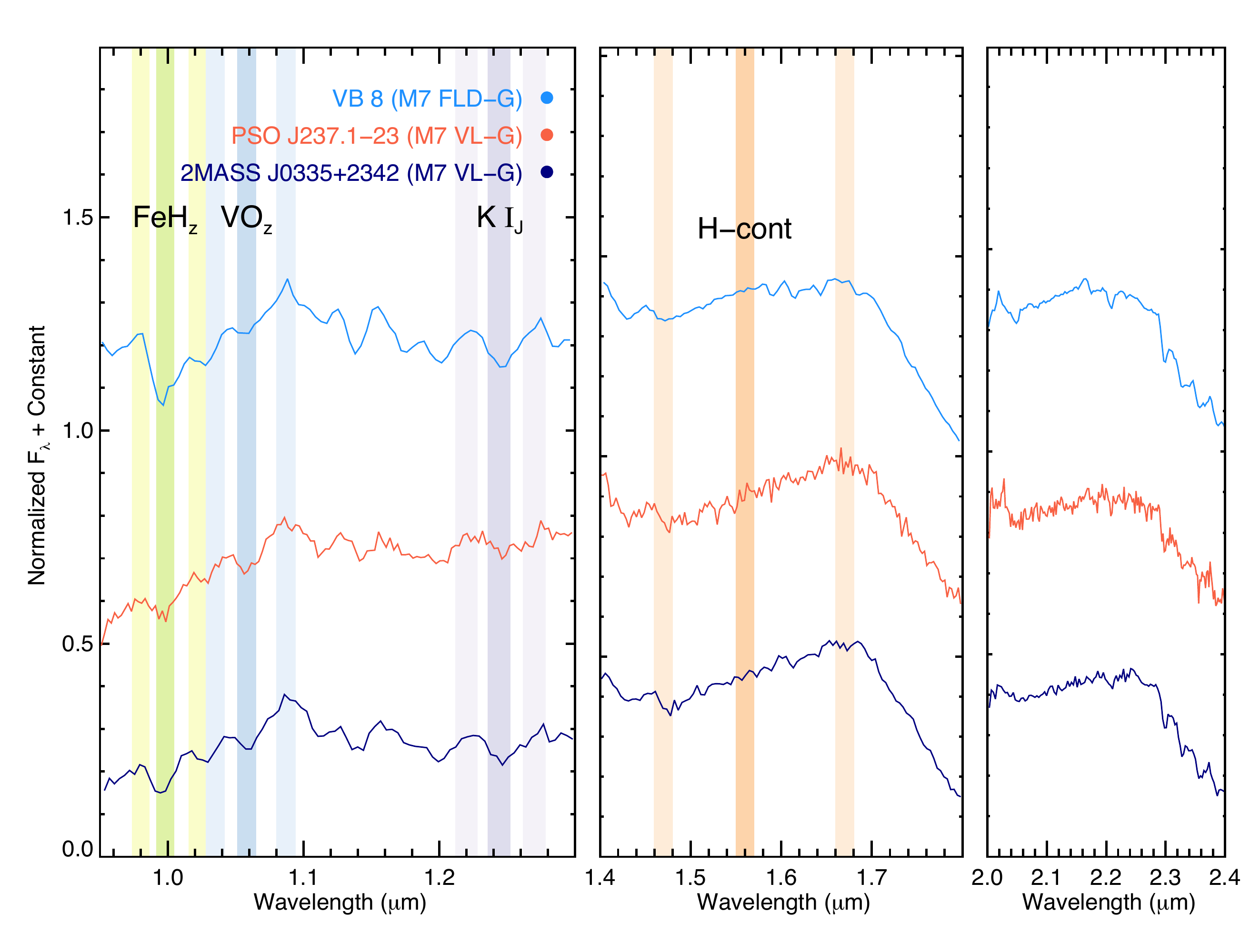}{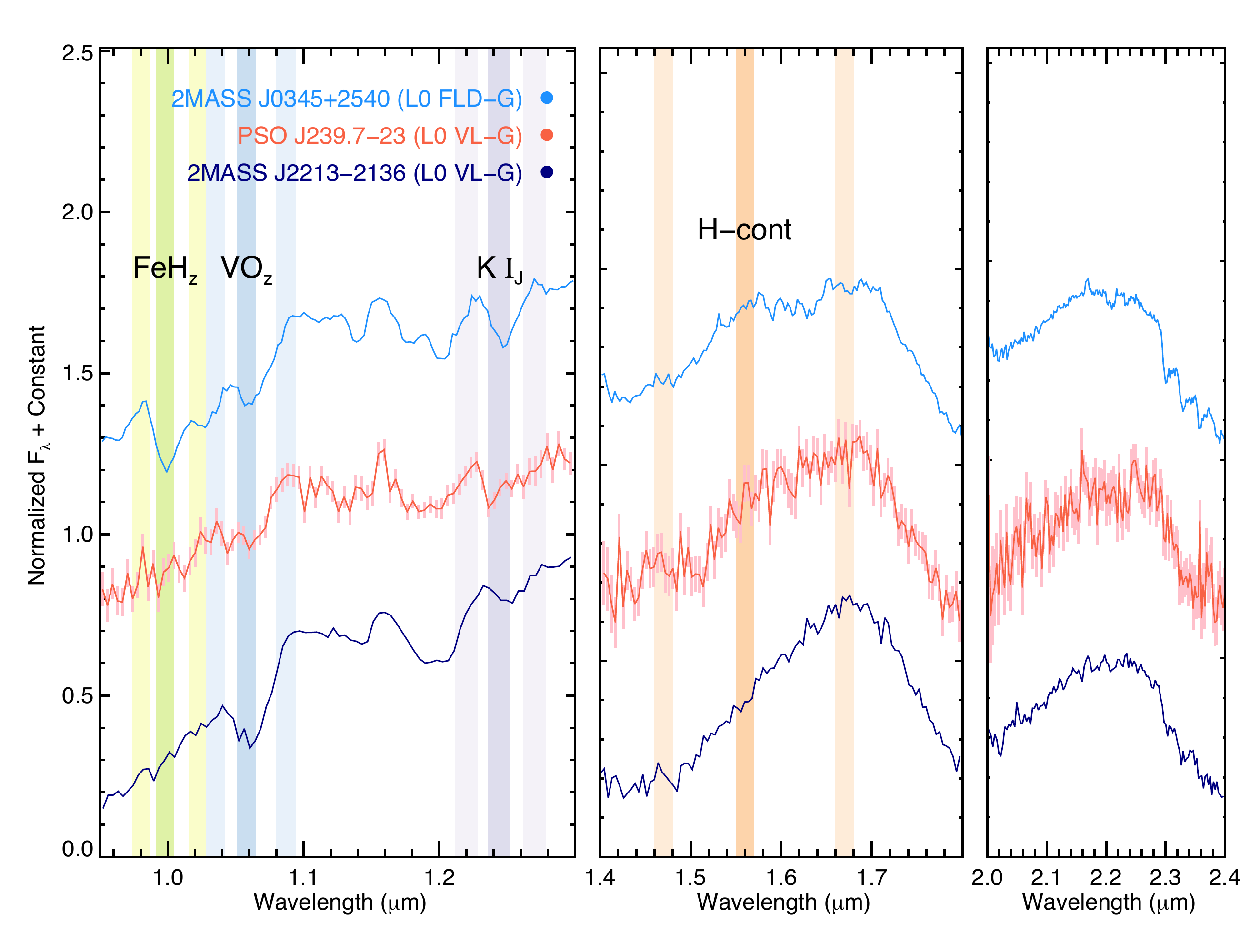}
  \caption{continued.}
  \label{fig.gravplots}
\end{center}
\end{figure}

\begin{figure}
\begin{center}
  \epsscale{1.8}
  \plottwo{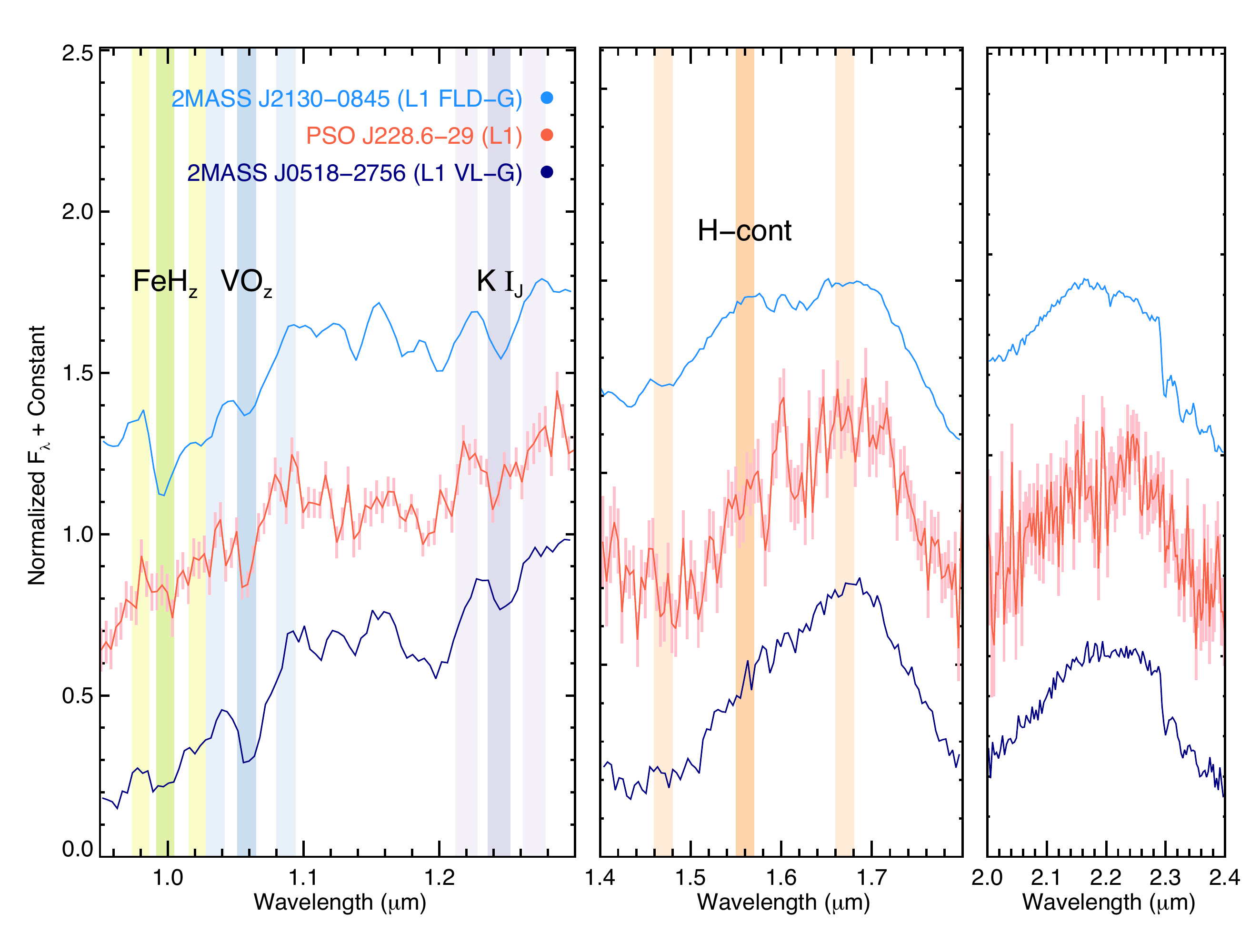}{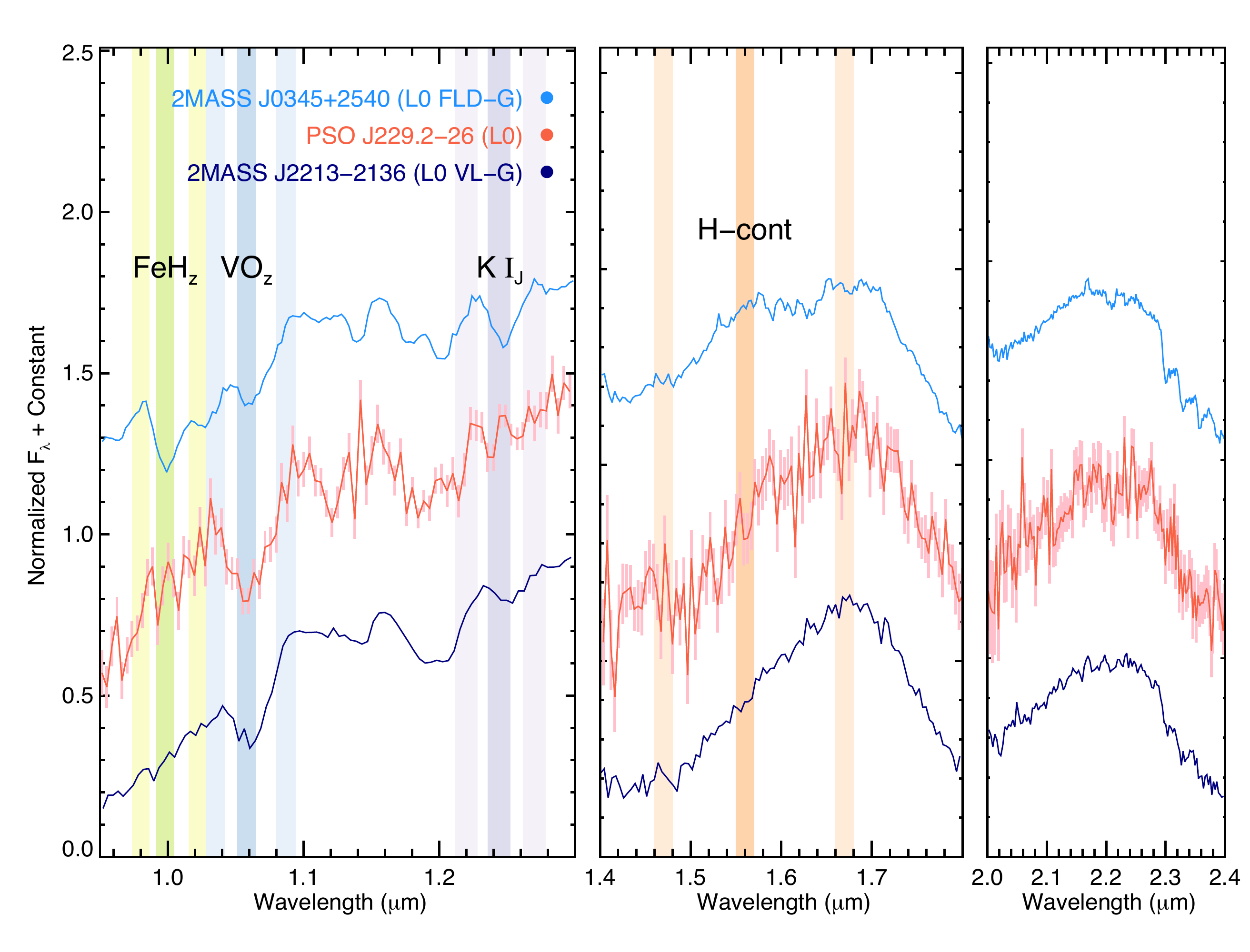}
  \caption{Same as Figure~\ref{fig.gravplots}, but showing the two objects
    (PSO~J228.6$-$29 and PSO~J229.2$-$26) for which we did not calculate gravity
    classes due to low S/N. These two spectra nevertheless show \fehz, \voz,
    \kij, and H-cont features resembling those of the \vlg\ standards.}
  \label{fig.gravplots.noisy}
\end{center}
\end{figure}

For the other two objects, PSO~J228.6773$-$29.7088 (hereinafter PSO~J228.6$-$29)
and PSO~J229.2354$-$26.6738 (hereinafter PSO~J229.2$-$26), the S/N of our
spectra was too low ($\lesssim$30) to yield robust gravity scores from the AL13
indices.  Figure~\ref{fig.gravplots.noisy} shows these spectra, again compared
with the appropriate field and \vlg\ standards.  In spite of the measurement
uncertainties, visual inspection confirms that both objects have weak 0.99~\um\
FeH and 1.25~\um\ \ki\ absorption, strong 1.06~\um\ VO absorption, and
triangular H band shapes, so we regard them as strong candidate \vlg\ objects.

\subsection{Proper Motions}
\label{kinematics}
Proper motions are key to establishing membership in star-forming regions and
clusters whose bulk motion through space is well defined.  We use proper motions
from PS1 Processing Version~3.2 (PV3.2), the version used for the photometry and
mean positions in PS1~DR1.  (PS1 proper motions will be part of a future public
release.)  PV3.2 astrometry is calibrated to the \textit{Gaia}~DR1
\citep{GaiaCollaboration:2016cu} astrometric reference frame.  We present the
proper motions for our discoveries in Table~\ref{tbl.kin.mass}, and we discuss
their consistency with members of Taurus and Sco-Cen in Sections \ref{taurus}
and~\ref{scocen}.  We present catalogs of proper motions for low-mass members of
Taurus and Upper Sco in Appendices \ref{appendixa} and~\ref{appendixb}, along
with a brief summary of the method used to calculate PS1 proper motions.

\subsection{No Candidate Binaries}
\label{results.binaries}
As in Paper II, we used the spectral index criteria of \citet[][hereinafter
BG14]{BardalezGagliuffi:2014fl} and visual inspection to search for spectral
features indicating that our discoveries may be unresolved binaries.
PSO~J228.6$-$29 satisfies only one of the twelve BG14 criteria, and our other
seven discoveries meet none of the BG14 criteria.  Similarly, we found no
evidence of spectral blends via visual inspection.  None of these objects appear
to be candidate unresolved binaries.

In addition, we investigated whether any of our discoveries could be members of
wide binary or multiple systems.  We searched for nearby known members of the
star-forming regions in which our discoveries reside, using the catalogs of
\citet{Esplin:2014he} for Taurus and \citet{Luhman:2012hj} for Upper Scorpius.
As several of our Sco-Cen discoveries lie outside the classical boundaries of
Upper Scorpius (Section~\ref{scocen}), we also searched across all catalogs in
\textit{Vizier}\footnote{http://vizier.u-strasbg.fr/viz-bin/VizieR} for objects
near these discoveries.  We found no known members of Taurus or Sco-Cen within
$70''$ (corresponding to a projected separation $\approx10,000$~AU) of our
discoveries.  At wider separations, it is still possible for a pair of low-mass
stars and/or brown dwarfs to be physically bound
\citep{Dhital:2010ir,Deacon:2014ey}.  However, such a binary is likely to have
formed through capture in the natal cluster rather than as an initially bound
system \citep{Kouwenhoven:2010kj}.  Thus, we conclude that all of our
discoveries are likely to be free-floating brown dwarfs that formed as single
objects.

\section{Taurus Discoveries}
\label{taurus}
Two of our discoveries, PSO~J060.3+25 (L1) and PSO~J077.1+24 (L2), reside at the
projected outer edges of the nearby Taurus star-forming region.
Figure~\ref{disc.map} shows their sky locations in Taurus.  In
Section~\ref{taurus.evidence} we present confirmation that they are bona fide
members of Taurus.  We also estimate their masses (Section~\ref{taurus.masses})
and assess whether they have circumstellar disks (Section~\ref{taurus.disks}).

We first explore why these objects were not identified in previous studies of
Taurus, a region that has been repeatedly searched for ultracool dwarfs.  Our
objects lie $\approx$8$^\circ$ ($\approx$20~pc) from the projected center of
Taurus, on opposite sides from each other.  Many searches surveyed smaller
and/or more central regions of Taurus that did not include our objects
\citep[e.g.,][]{Guieu:2006fl,Luhman:2009cn,Quanz:2010hp,Rebull:2010js}.  Our
objects also lie just outside the footprint of Spitzer images analyzed by
\citet{Esplin:2014he}.  Our objects are very faint, especially at optical
wavelengths, so previous searches using $i$-band photometry
\citep[e.g.,][]{Slesnick:2006ij} were not able to detect them.  Similarly, both
objects lie within the area searched by \citet{Esplin:2014he} using WISE
photometry, but are fainter than the $W1\le14$~mag limit used in that search.
\citet{Luhman:2006cd} would have detected PSO~J077.1+24 in 2MASS photometry, but
its $\jht=0.46\pm0.28$~mag color is bluer than the $\jht\ge0.6$~mag cut used in
that search.\footnote{We note the large error on this color is due to this
  object being near the detection limit of 2MASS.  Our deeper MKO photometry
  (Table 2) shows this object has $(J-H)_{\rm MKO}=0.75\pm0.06$~mag.}
PSO~J060.3+25 lies just outside the \citet{Luhman:2006cd} search area but would
also have been excluded due to its relatively blue \hkt\ color (also having a
large error).  We note also that \citet{Luhman:2006cd} used previously known
low-mass members of Taurus as templates to define color cuts, and many of those
members are reddened by local extinction.  Our discoveries do not appear to be
reddened, as discussed in Sections \ref{taurus.evidence.phot}
and~\ref{taurus.2m0437}.

\subsection{Evidence for Membership}
\label{taurus.evidence}

\subsubsection{Youth}
\label{taurus.evidence.youth}
Both PSO~J060.3+25 and PSO~J077.1+24 have \vlg\ gravity classes
(Section~\ref{results.gravity}).  While more work is needed to calibrate the
ages of \vlg\ objects \citep {Allers:2013hk}, the classification suggests an age
$\lesssim30$~Myr, much younger than the field population.

\subsubsection{Photometry}
\label{taurus.evidence.phot}
\begin{figure}
\begin{center}
  \begin{minipage}[t]{0.49\textwidth}
    \includegraphics[width=1.00\columnwidth, trim = 20mm 0 8mm 0]{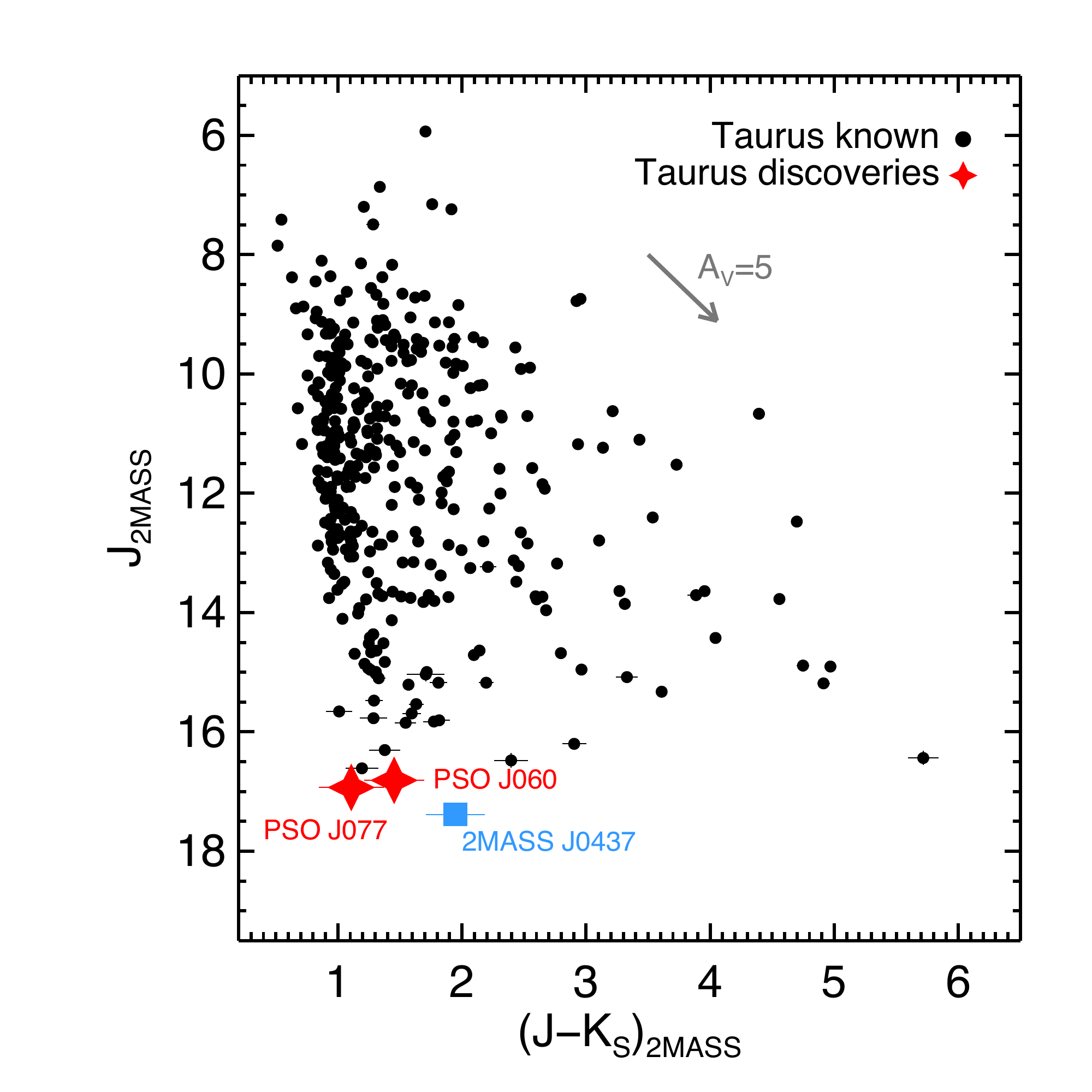}
  \end{minipage}
  \hfill
  \begin{minipage}[t]{0.49\textwidth}
    \includegraphics[width=1.00\columnwidth, trim = 20mm 0 8mm 0]{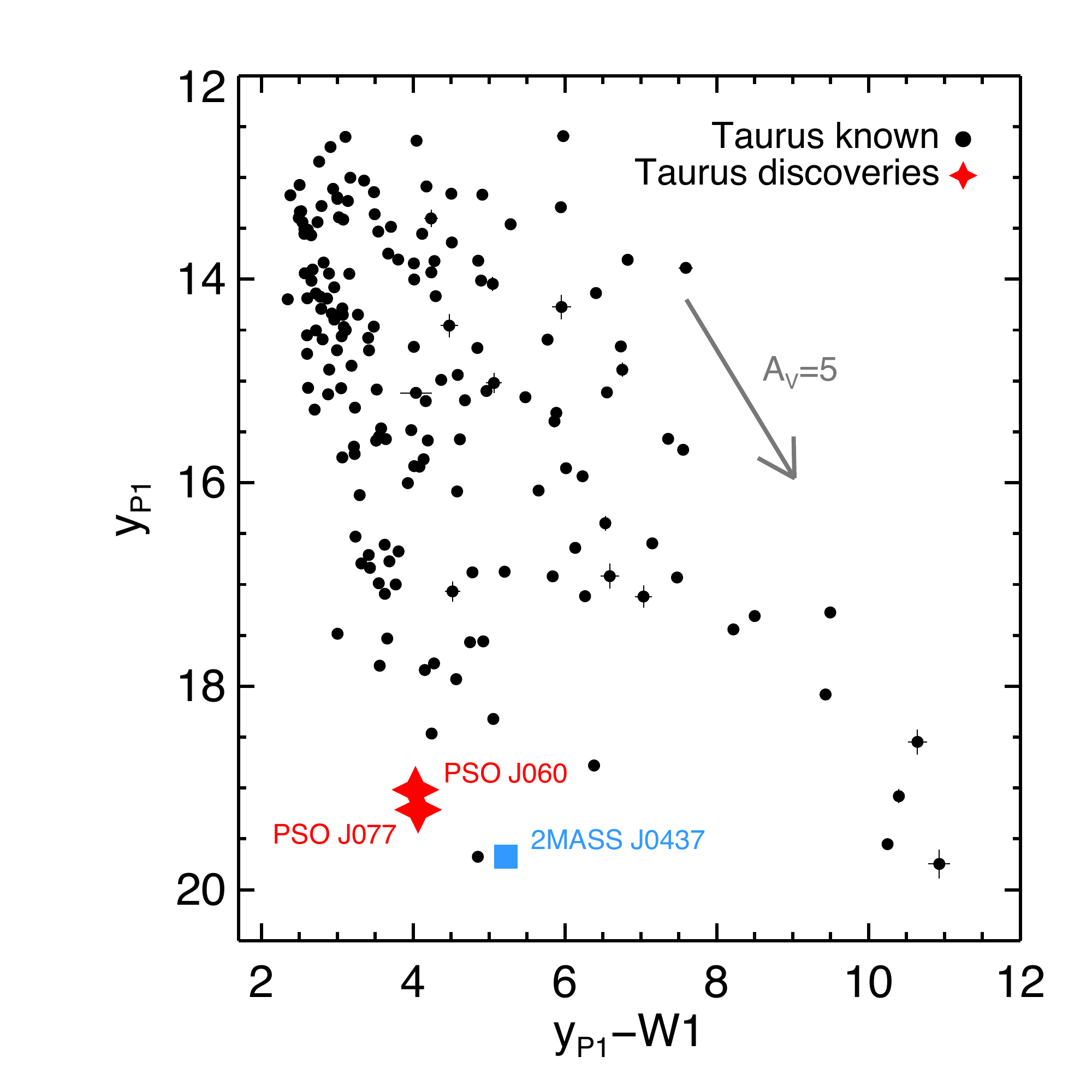}
  \end{minipage}
  \caption{Comparison of the photometry of our discoveries in the Taurus
    star-forming region (red stars) to known Taurus members from
    \citet{Esplin:2014he} (black circles). We also highlight the known Taurus L1
    dwarf 2MASS~J0437+2331 (Section~\ref{taurus.2m0437}, blue square) and
    indicate reddening vectors equivalent to an extinction of $A_V=5$~mag with
    gray arrows.  {\it Left}:~$J$ vs. $J-K_S$ (2MASS) diagram. {\it
      Right}:~\yps\ vs. \ywa\ diagram for Taurus objects not saturated in PS1.
    Both plots show an unreddened cluster sequence on the left, with many
    objects significantly reddened by the Taurus molecular cloud.  Our two
    discoveries lie at the faint end of the cluster sequence and are minimally
    affected by extinction, consistent with their locations on the outer edges
    of Taurus.}
  \label{fig.taurus.cmd}
\end{center}
\end{figure}

Figure~\ref{fig.taurus.cmd} compares the photometry of PSO~J060.3+25 and
PSO~J077.1+24 to that of known Taurus members from \citet{Esplin:2014he}.  The
$J$ vs. $J-K_S$ (2MASS) and \yps\ vs. \ywa\ color-magnitude diagrams for Taurus
make evident the significant reddening in this region of the sky.  We calculated
reddening vectors for these color-magnitude diagrams using the \yps\ coefficient
from \citet[their Table 6, $R_V=3.1$]{Schlafly:2011iu} and the $J/K_S/W1$
coefficients from \citet[their Table 3]{Davenport:2014gm}.  We include these
reddening vectors, scaled to an extinction of $A_V=5$~mag, in
Figure~\ref{fig.taurus.cmd}.  Our two discoveries sit at the faint end of the
unreddened cluster sequence, consistent with their projected locations on the
unobscured outskirts of the region.  While some young early-L dwarfs have
unusually red \jkt\ colors for their spectral types
\citep[e.g.,][]{Gagne:2015dc}, we note that the \jkt\ colors of our Taurus
discoveries ($1.45\pm0.24$~mag for PSO~J060.3+25, $1.11\pm0.26$~mag for
PSO~J077.1+24) are consistent with those of older field dwarfs with the same
spectral types \citep{Schmidt:2010ex,Faherty:2013bc}.  However, both
PSO~J060.3+25 and PSO~J077.1+24 have \wawb\ colors (Figure~\ref{fig.w1w2.yw1})
that are $3\sigma$ redder than those of field early-L dwarfs, a common sign of
low gravity \citep{Gizis:2012kv}.

\subsubsection{Proper Motions}
\label{taurus.evidence.pm}
To assess the kinematic consistency of our discoveries with previously known
members of Taurus, we created a list of proper motions for the Taurus stars and
brown dwarfs from \citet{Esplin:2014he} that are not saturated in PS1.  We
obtained the proper motions from PS1 Processing Version~3.2 (PV3.2), the same
source as the proper motions of our discoveries (Section~\ref{kinematics}).  We
discuss our complete list of proper motions in detail in
Appendix~\ref{appendixa}.  Figure~\ref{fig.taurus.pm} compares the proper
motions of PSO~J060.3+25 and PSO~J077.1+24 with all reliable PS1 Taurus proper
motions.  We calculated a weighted mean proper motion for the known Taurus
members of ($\mua=7.6\pm0.2,\ \mud=-17.4\pm0.2$~\my), with a weighted rms of
4.9~\my\ in R.A. and 6.4~\my\ in Dec.

PSO~J077.1+24 has a proper motion of ($14.1\pm12.5,-27.1\pm12.1$~\my),
consistent with the mean Taurus proper motion.  Because the Taurus region has a
number of distinct subgroups, we also compared the proper motion of
PSO~J077.1+24 to that of the closest subgroup on the sky identified by
\citet{Luhman:2009cn}, L1544.  This subgroup has a median proper motion of
\citep[$0.9\pm1,-17.6\pm1$~\my;][]{Luhman:2009cn}, consistent with
PSO~J077.1+24.

For PSO~J060.3+25, we adopt the proper motion ($14.3\pm3.1,-26.4\pm3.2$~\my)
from \citet{Bouy:2015gl}.  Our PS1 proper motion of
($19.0\pm8.2,-38.1\pm8.2$~\my) is consistent, but the \citet{Bouy:2015gl}
measurement is more precise.  The adopted proper motion is very similar to
PSO~J077.1+24 and consistent with our mean Taurus proper motion.  The closest
Taurus subgroup on the sky to PSO~J060.3+25 identified by \citet{Luhman:2009cn},
B209 ($\approx$4$^{\circ}$~away), has a mean proper motion of
\citep[$6.9\pm1,-22.3\pm1$~\my; ][]{Luhman:2009cn}, consistent within
$1.8\sigma$ in R.A.~and $1.0\sigma$ in Dec.

\begin{figure}
\begin{center}
  \includegraphics[width=1.00\columnwidth]{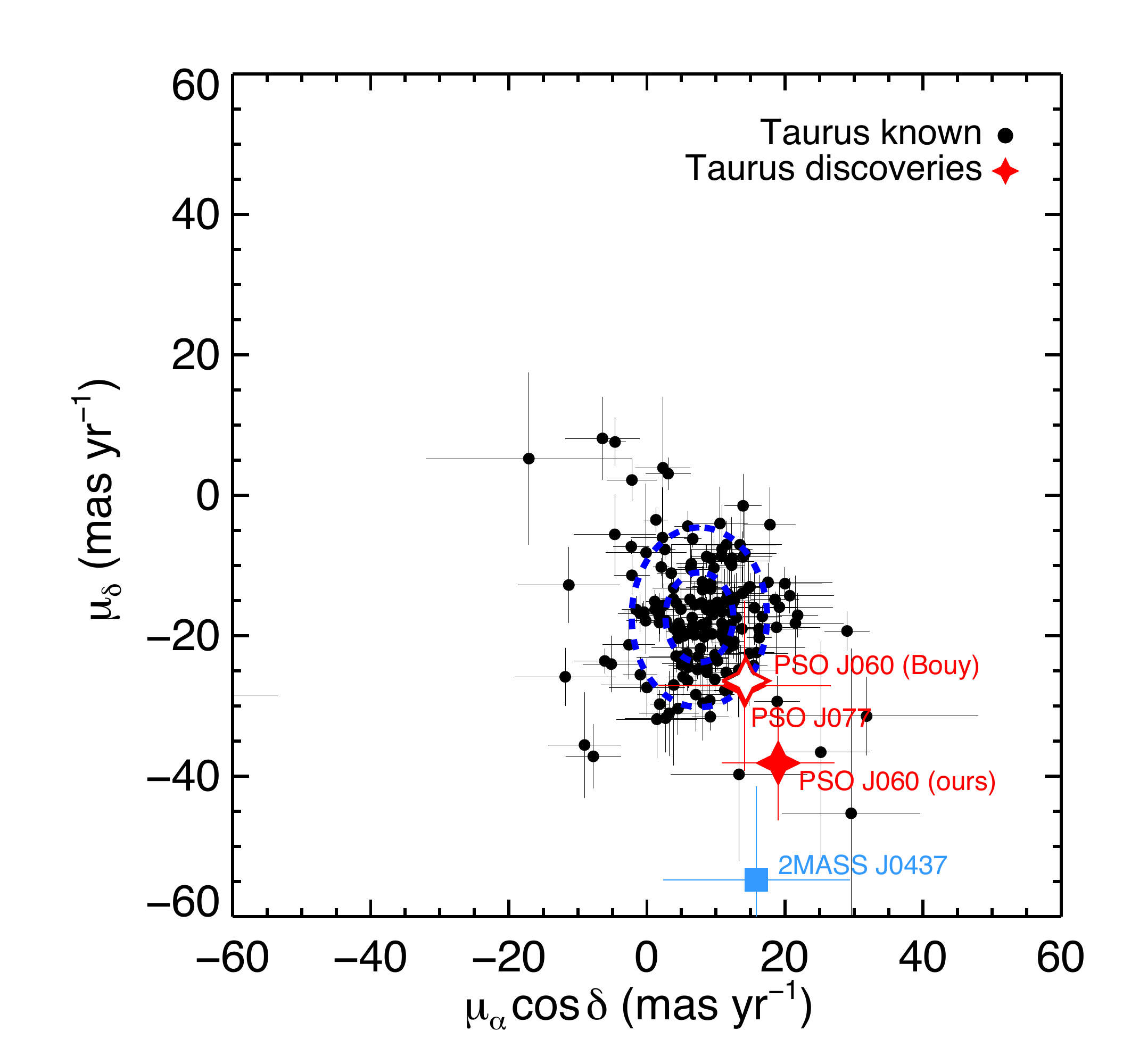}
  \caption{Vector-point diagram showing the proper motions of our discoveries in
    the Taurus star-forming region (red stars) and those of known Taurus members
    from \citet{Esplin:2014he} that are not saturated in PS1 and have reliable
    PS1 proper motion fits (black circles).  We also include the only previously
    known L~dwarf in Taurus, 2MASS~J0437+2331 \citep{Luhman:2009cn}, which we
    classify as L1.  We adopt the proper motion for PSO~J060.3+25 from
    \citet{Bouy:2015gl}, shown with an open red star.  Note that the PS1 proper
    motion for PSO~J077.1+24 and the \citet{Bouy:2015gl} proper motion for
    PSO~J060.3+25 are very similar and their symbols coincide in the figure.
    Both of these adopted proper motions are consistent with the mean Taurus
    proper motion, while the PS1 proper motion for 2MASS~J0437+2331 is
    $\approx$2$\sigma$ discrepant in \mud.}
  \label{fig.taurus.pm}
\end{center}
\end{figure}

\begin{figure}
\begin{center}
  \begin{minipage}[t]{0.49\textwidth}
    \includegraphics[width=1.00\columnwidth, trim = 20mm 0 8mm 0]{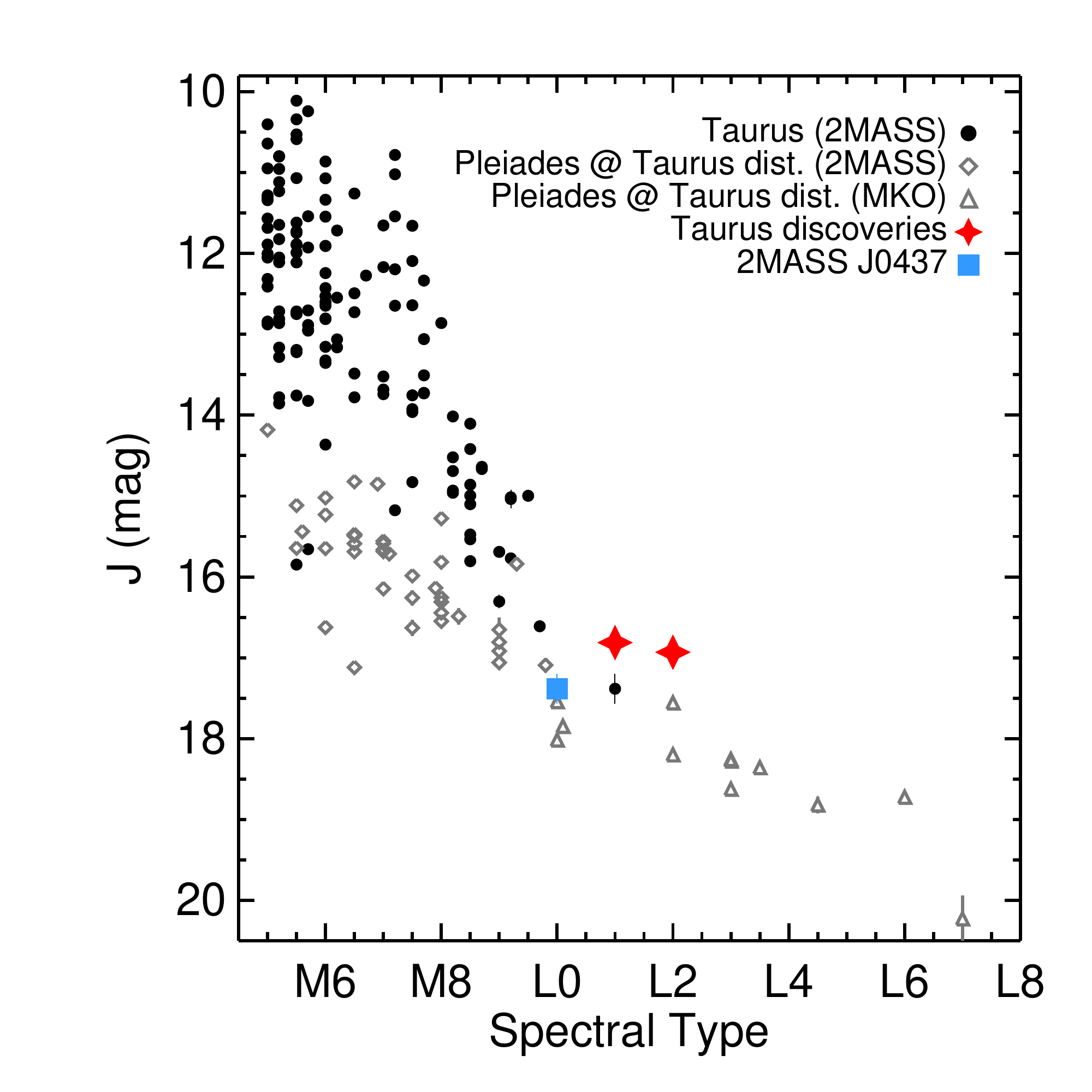}
  \end{minipage}
  \hfill
  \begin{minipage}[t]{0.49\textwidth}
    \includegraphics[width=1.00\columnwidth, trim = 20mm 0 8mm 0]{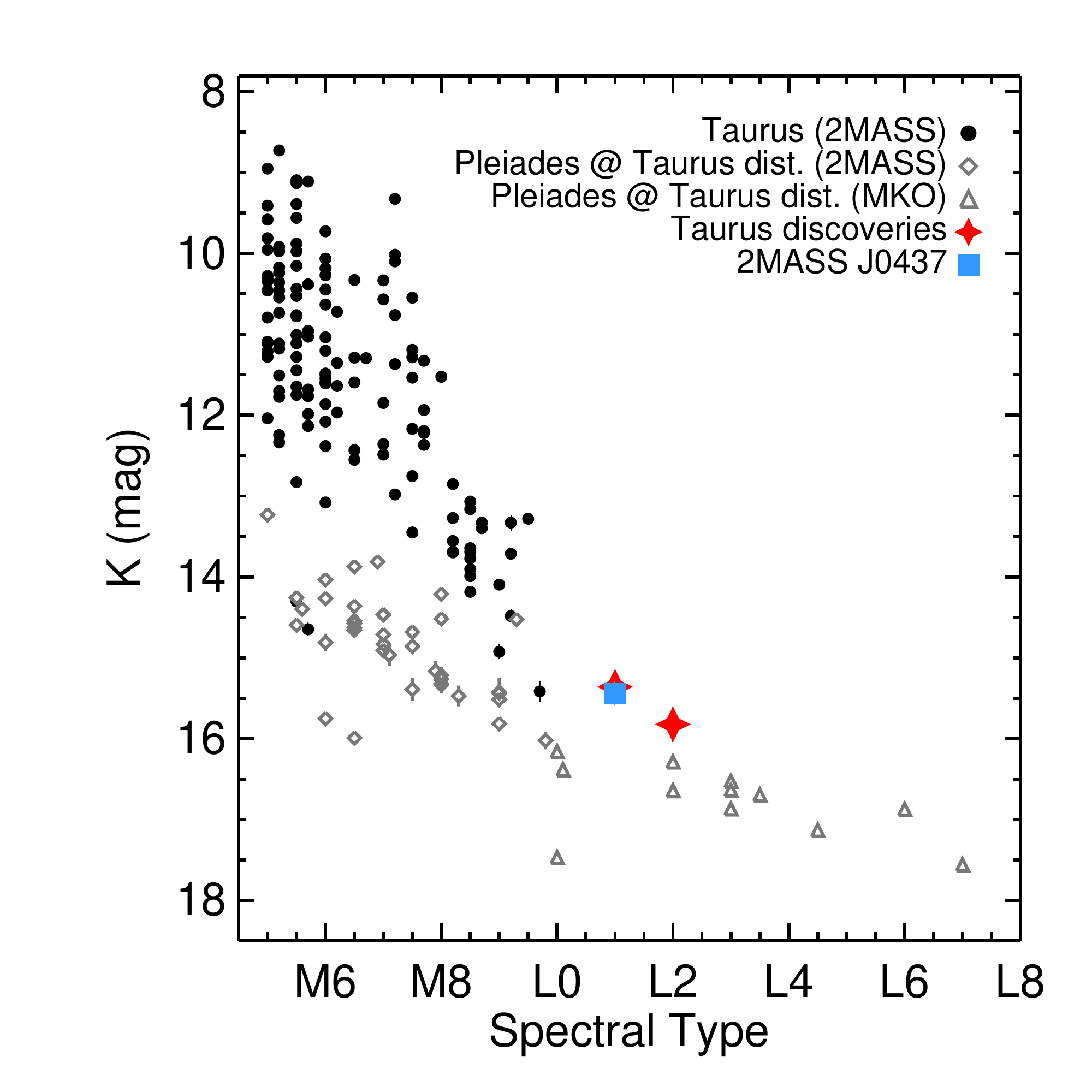}
  \end{minipage}
  \caption{$J$ ({\it left}) and $K$ ({\it right}) apparent magnitudes as a
    function of spectral type for our discoveries in the Taurus star-forming
    region (red stars) compared with known members of Taurus \citep[black
    circles]{Esplin:2014he} and the Pleiades (grey open symbols).  Pleiades
    magnitudes have been adjusted by +0.136~mag to place the objects at the
    distance of Taurus (145~pc).  We use 2MASS photometry
    \citep[diamonds]{Skrutskie:2006hl} for the brighter Pleiades members, and
    MKO photometry from the UKIDSS Galactic Clusters Survey
    \citep[triangles]{Lawrence:2012wh} for members too faint to be detected by
    2MASS.  We also highlight the previously coolest known member of Taurus,
    2MASS~J0437+2331 (blue square), which we classify as L1 on the AL13 system.
    Our discoveries lie $\gtrsim$1~mag above the Pleiades sequence but are
    consistent with an extension of the Taurus sequence, supporting membership
    in the younger Taurus region.  References for Pleiades spectral types:
    \citet{Bihain:2006ck,Bihain:2010gm,Festin:1998cr,Martin:1996gm,Martin:1998iz,Martin:1998hr,Martin:2000jp,Pinfield:2003di,Stauffer:1998kg,Stauffer:1998kt,Steele:1995bi,ZapateroOsorio:1997ce,ZapateroOsorio:2014dr}.}
  \label{fig.taurus.pleiades}
\end{center}
\end{figure}

PSO~J060.3+25 (a.k.a.\ DANCe~J040116.80+255752.2) was also previously identified
by \citet{Sarro:2014ci} and \citet{Bouy:2015gl} as a high-probability (93\%)
member of the Pleiades cluster, based on $z\,YJHK_S$ photometry and astrometry.
The Pleiades lie at a mean distance of $136.2\pm1.2$~pc \citep{Melis:2014id},
commensurate with the $\approx$145~pc distance to Taurus.  The projected center
of the Pleiades lies $\approx$3.5$^\circ$ away from PSO~J060.3+25, or
$\approx$8~pc at the distance of the Pleiades, so it is possible that
PSO~J060.3+25 falls within the $9.5\pm0.5$~pc tidal radius of the Pleiades
\citep{Danilov:2015dz}.  The mean proper motion for low-mass brown dwarfs
($0.012-0.025$~\msun) in the Pleiades is ($21.6,-47.6$~\my) with a dispersion of
$\sigma_\mu=7.5\pm6.1$~\my\ \citep{ZapateroOsorio:2014em}, so the proper motion
of PSO~J060.3+25 is intermediate between the bulk motions of Taurus and the
Pleiades, and consistent within $2\sigma$ with both groups.  Our \vlg\
classification for PSO~J060.3+25 suggests a younger age ($\lesssim30$~Myr) than
for the Pleiades \citep [$\approx$125~Myr;][]{Stauffer:1998kt}, although the age
range of \vlg\ objects has not been firmly established
\citep[e.g.,][]{Liu:2016co}.  We note that \citet{Allers:2013vv} classified
several Pleiades members as \vlg, but they used spectra with mostly lower
resolution ($R\approx50$) and S/N ($\lesssim$20) than the prism spectra we
present here for our discoveries, so we regard those classifications as
provisional.  In Figure~\ref{fig.taurus.pleiades} we also compare the $J$ and
$K$-band photometry of known Taurus members and our discoveries, as a function
of spectral type, to known VLM members of the Pleiades.  The published spectral
types for the Pleiades members are derived from multiple sources and methods and
are therefore heterogenous, but classification of these objects using the AL13
system shows a consistent result \citep{Allers:2013vv}.
Figure~\ref{fig.taurus.pleiades} shows that our discoveries have $J$ magnitudes
more consistent with the younger, brighter members of Taurus.  We therefore find
it more likely that PSO~J060.3+25 is a member of the 1--2~Myr old Taurus region.
A radial velocity measurement would help to further assess the Taurus membership
of PSO~J060.3+25.

\subsubsection{Likelihood of Field Contamination}
\label{taurus.foreground}
We investigated the possibility that PSO~J060.3+25 or PSO~J077.1+24 could be a
foreground or background field object in the direction of Taurus by estimating
the number of such contaminating field objects from our search.  For this
estimate, we generously defined the boundaries of the Taurus region to be
$4^{\rm h}00^{\rm m}\le\alpha\le5^{\rm h}15^{\rm m}$ and
$14^\circ\le\delta\le32^\circ$ (see Figure~\ref{disc.map}), covering
309.4~deg$^2$.  Our search covered the entire sky between declinations
$-30^\circ$ and $+70^\circ$ except for locations within $3^\circ$ of the
Galactic plane (Paper II), an area totaling 28,070 deg$^2$.  (We noted in
Section~\ref{photometry} that our search also avoided reddened regions
identified by \citealt{Cruz:2003fi}, but PSO~J060.3+25 actually lies within one
of these reddened regions, so we include those regions in this estimate.)

We found a total of fourteen \vlg\ L0--L2 dwarfs in our search, including two
previously known objects and three discoveries that we consider to be strong
\vlg\ candidates.  We would therefore expect our search to find 0.15 \vlg\
L0--L2 dwarfs in an arbitrary Taurus-sized area of sky.

In addition, we assessed the likelihood that an early-L dwarf observed in the
direction of Taurus would have a proper motion consistent with members of Taurus
(as do PSO~J060.3+25 and PSO~J077.1+24).  We used the Besan\c{c}on Galactic
model\footnote{http://model.obs-besancon.fr} \citep{Robin:2003jk} to generate a
synthetic population of field M dwarfs in a volume spanning our Taurus
boundaries between 50~pc and 240~pc.  We assigned uncertainties to the synthetic
proper motions using an astrometric error vs.\ $K_S$ relationship derived from
our Taurus proper motions (Section~\ref{taurus.evidence.pm}).  Assuming that
early-L dwarfs have the same kinematics as M dwarfs in the field, we used Monte
Carlo trials to determine that $20.3\pm0.2$\% of early-L dwarfs in the
direction of Taurus will have proper motions within $3\sigma$ of the mean Taurus
\mua\ and \mud.

We would therefore expect our search to find $(3.13\pm0.03)\times10^{-2}$ \vlg\
dwarfs within our Taurus boundaries having proper motions consistent with Taurus
membership.  Poisson statistics give us a probability of 96.9\% that neither
PSO~J060.3+25 nor PSO~J077.1+24 is an interloping field object in the direction
of Taurus, and a negligible $5\times10^{-4}$ probability that both objects are
contaminants.

\subsubsection{Comparison with 2MASS J04373705+2331080}
\label{taurus.2m0437}
Prior to our discoveries, 2MASS~J0437+2331 was the only known free-floating L
dwarf member of Taurus, discovered and classified as L0 by
\citet{Luhman:2009cn}.  We used the SpeX Prism spectrum for 2MASS~J0437+2331
from \citet{Bowler:2014dk} to assign a near-infrared spectral type of L$1\pm1$,
based on a calculated type of L$0.9\pm0.9$ using the AL13 indices
(Section~\ref{results.indices}).

\citet{Luhman:2009cn} claimed membership in Taurus for 2MASS~J0437+2331 based on
weaker \nai\ and \ki\ absorption features in its red-optical spectrum, which are
recognized signs of youth \citep{Kirkpatrick:2006hb}, along with its central
projected location in Taurus.  Similarly to two of our Sco-Cen discoveries, the
spectrum for 2MASS~J0437+2331 does not have a high enough S/N for us to
confidently assign a gravity class, but Figure~\ref{fig.gravplots.2m0437} shows
that it more closely resembles the L1 \vlg\ standard than the field standard,
confirming its young age.  Figure~\ref{fig.taurus.cmd} shows that the
\yps$JK_SW1$ photometry of 2MASS~J0437+2331 is also similar to that of
PSO~J060.3+25 and PSO~J077.1+24, and consistent with being a slightly reddened
member of Taurus.  2MASS~J0437+2331 has a proper motion of
($15.9\pm13.5,-54.8\pm13.4$~\my), consistent with our mean motion of Taurus in
R.A. but nearly $2\sigma$ different in Dec.  We note that 2MASS~J0437+2331
satisfies all the criteria for our search (Section~\ref{photometry}), except
that it lies well within the excluded reddened region of \citet{Cruz:2003fi}.

\begin{figure}
\begin{center}
  \vspace{-20pt}
  \includegraphics[width=1.00\columnwidth]{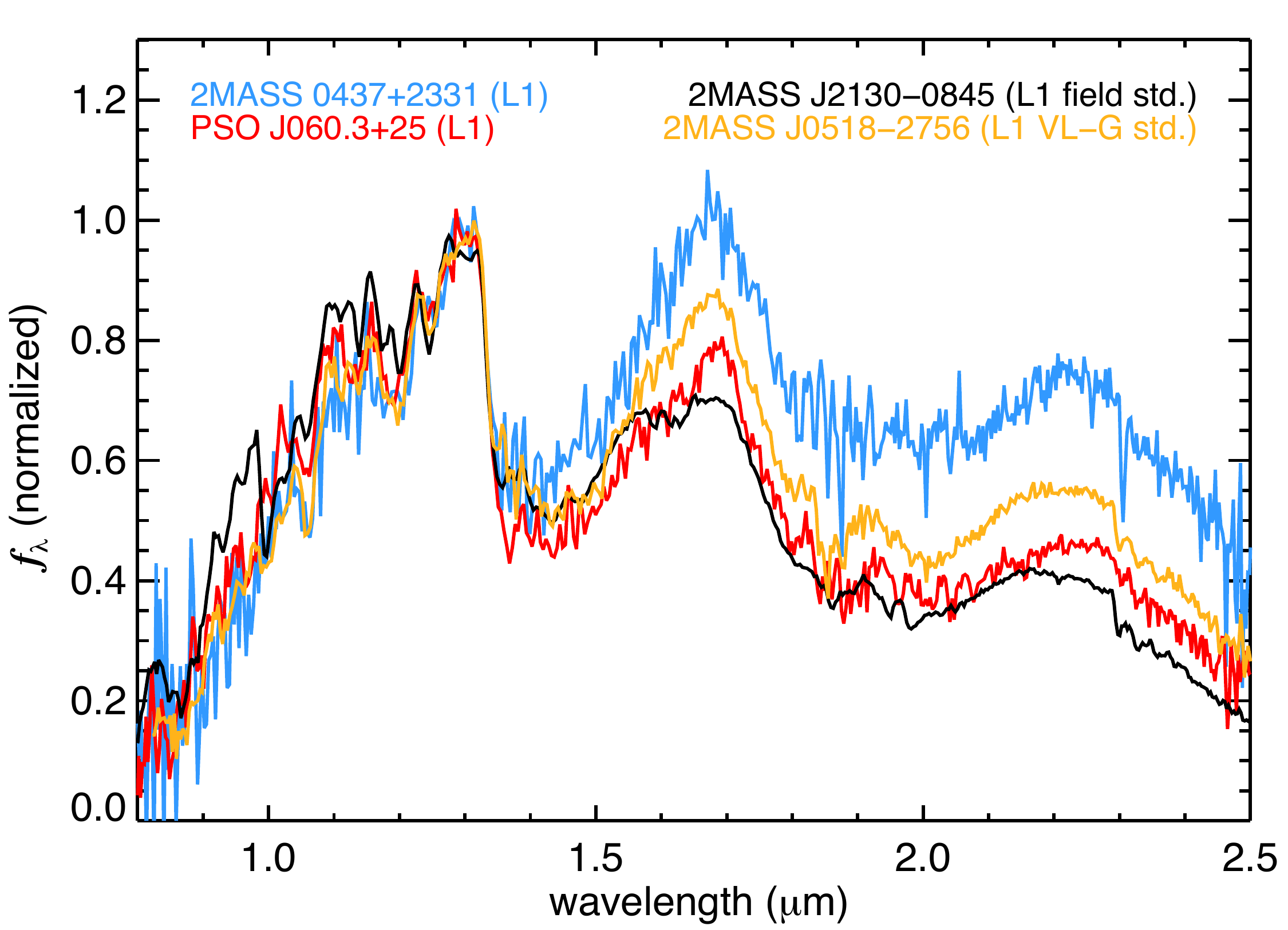}
  \includegraphics[width=1.00\columnwidth]{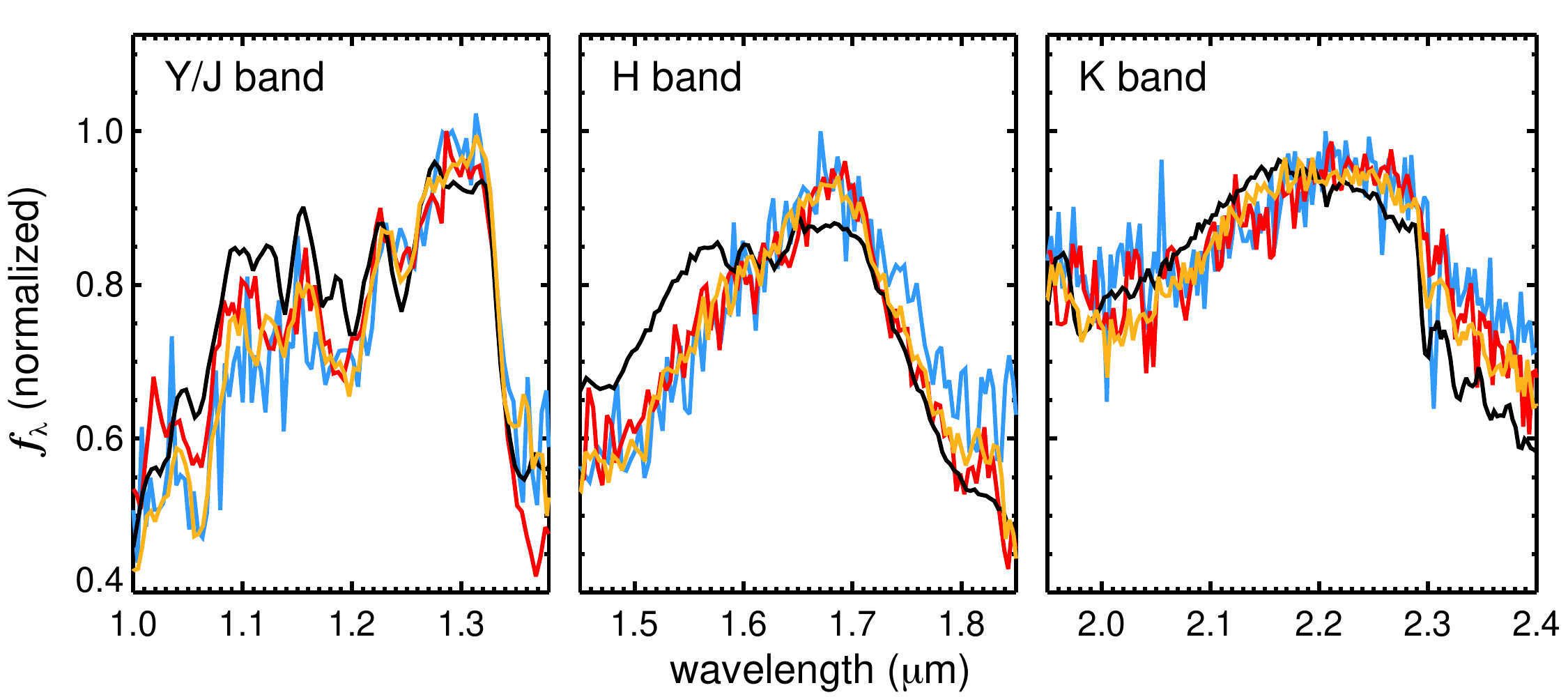}
  \caption{\textit{Top}: The SpeX Prism spectrum of 2MASS~J0437+2331 (blue) from
    \citet{Bowler:2014dk}, overplotted with PSO~J060.3+25 (red) along with the
    field \citep[black;][]{Kirkpatrick:2010dc} and \vlg\
    \citep[orange;][]{Allers:2013hk} standards of the same spectral type (L1) as
    PSO~J060.3+25.  All four spectra are normalized at the $J$-band peak.
    2MASS~J0437+2331 is notably redder than PSO~J060.3+25 as well the L1 \vlg\
    standard, while PSO~J060.3+25 has colors similar to those of the field
    standard.  \textit{Bottom}: The same four spectra plotted separately for
    Y/J, H, and K~bands, normalized separately for each band to compare the
    spectral shapes in each band.  The two young Taurus objects and the L1 \vlg\
    standard have similar shapes in all bands, distinct from the older field
    standard.}
  \label{fig.2m0437.060.spectra}
\end{center}
\end{figure}

\begin{figure}
\begin{center}
  \includegraphics[width=1.00\columnwidth]{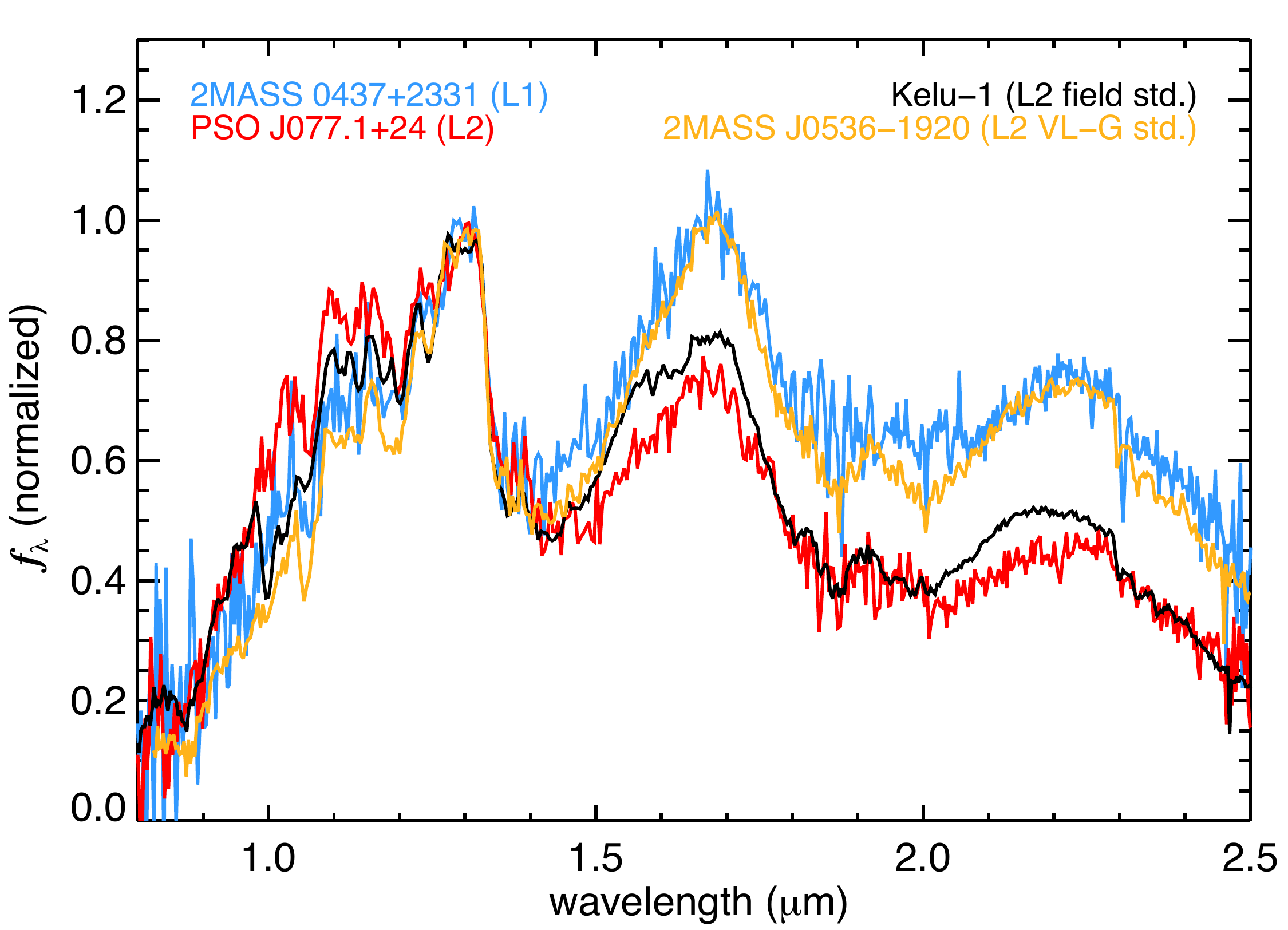}
  \includegraphics[width=1.00\columnwidth]{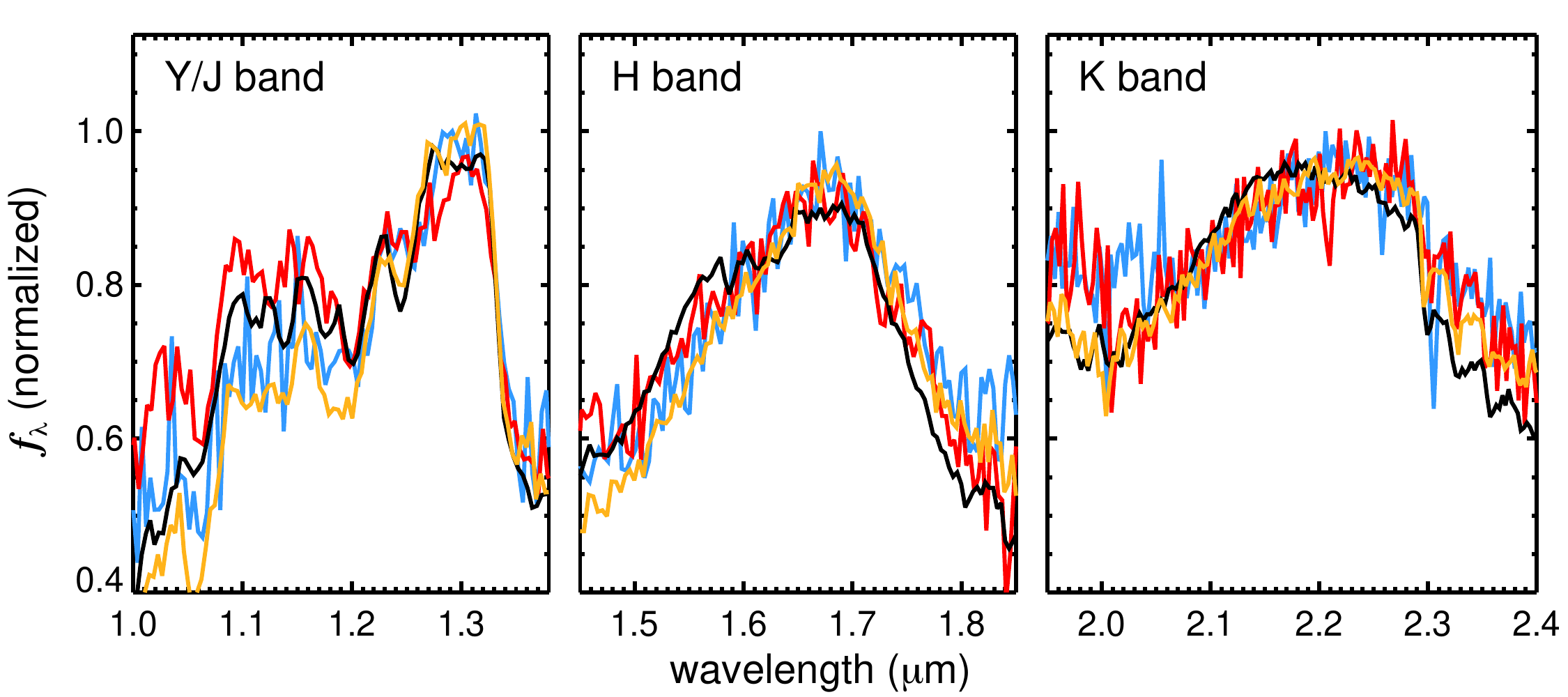}
  \caption{The SpeX Prism spectrum of 2MASS~J0437+2331 (blue) from
    \citet{Bowler:2014dk}, overplotted with PSO~J077.1+24 (red) along with the
    field \citep[black;][]{Kirkpatrick:2010dc} and \vlg\
    \citep[orange;][]{Allers:2013hk} standards of the same spectral type (L2) as
    PSO~J077.1+24, using the same format as Figure~\ref{fig.2m0437.060.spectra}.
    2MASS~J0437+2331 is significantly redder than PSO~J077.1+24 but has similar
    colors to the L2 \vlg\ standard, while PSO~J077.1+24 has colors more similar
    to those of the field standard.}
\label{fig.2m0437.077.spectra}
\end{center}
\end{figure}

\begin{figure}
\begin{center}
  \epsscale{0.9}
  \plotone{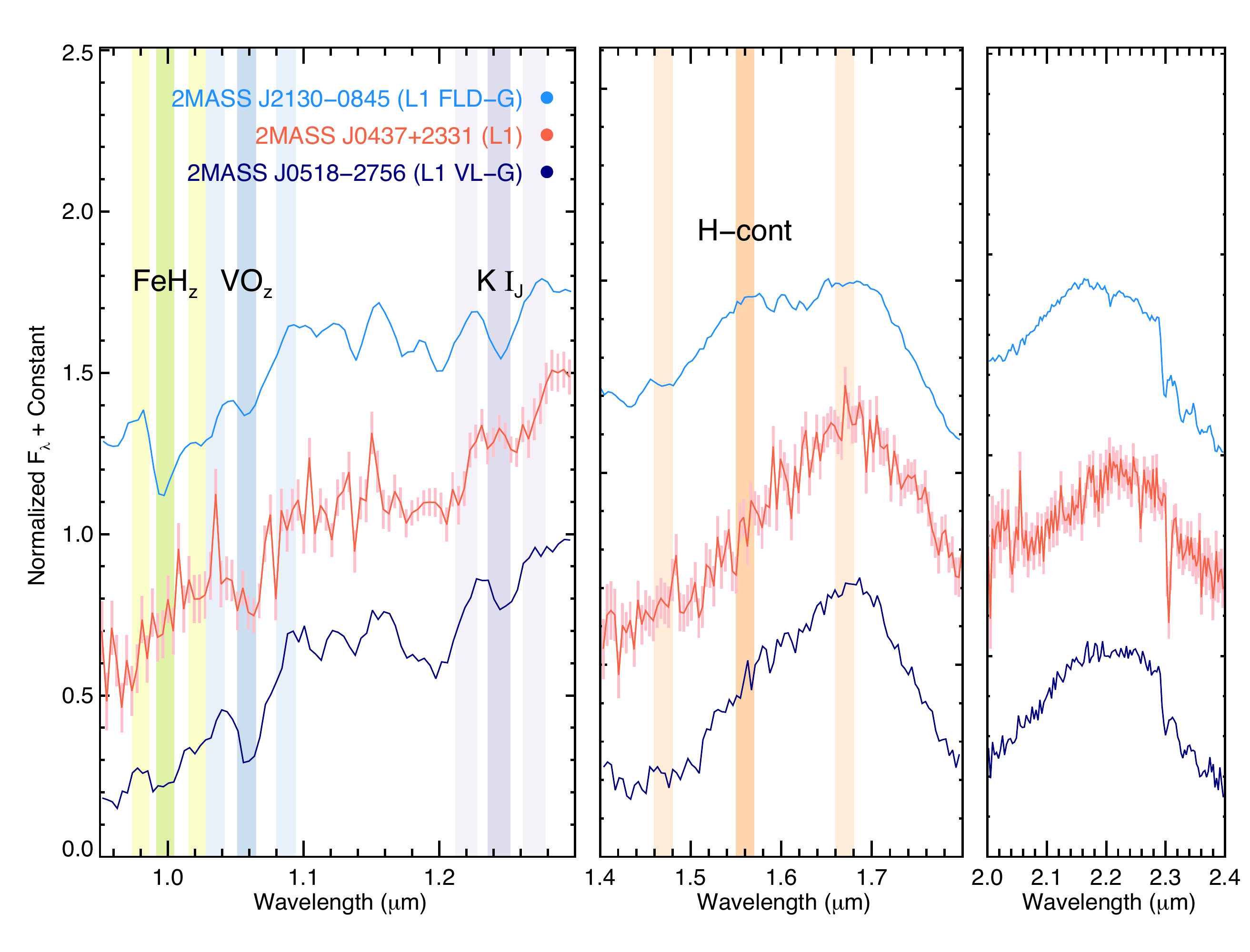}
  \caption{Same as Figure~\ref{fig.gravplots} but showing the SpeX Prism
    spectrum for 2MASS~J0437+2331 \citep{Luhman:2009cn,Bowler:2014dk}.  We did
    not determine a gravity class for 2MASS~J0437+2331 due to low spectral S/N,
    but its \fehz, \voz, \kij, and H-cont features do resemble those of the L1
    \vlg\ standard \citep{Allers:2013hk}.}
  \label{fig.gravplots.2m0437}
\end{center}
\end{figure}

We compare the spectrum of 2MASS~J0437+2331 to those of PSO~J060.3+25 and
PSO~J077.1+24 in Figures \ref{fig.2m0437.060.spectra}
and~\ref{fig.2m0437.077.spectra}, along with the appropriate field
\citep{Kirkpatrick:2010dc} and \vlg\ (AL13) standards.  Interestingly, the
near-IR spectrum of 2MASS~J0437+2331 is notably redder than those of our
discoveries, and its position in the color-magnitude diagrams in
Figure~\ref{fig.taurus.cmd} is consistent with an extinction of
$A_V\approx2-4$~mag.  This redness was also noted by
\citet{AlvesdeOliveira:2013dx}, who calculated an extinction of
$A_V=2.1-3.3$~mag for 2MASS~J0437+2331 based on 2MASS photometry and comparison
to the near-IR spectra of other young M and L dwarfs.  However,
\citet{Luhman:2009cn} found an extinction of $A_J=0$~mag for 2MASS~J0437+2331
using an optical spectrum, which is more sensitive to dust-induced reddening
than longer near-IR wavelengths.  Our near-IR spectrum also closely resembles
the L2 \vlg\ standard in color.  It therefore appears that the red near-IR
colors of 2MASS~J0437+2331 are photospheric in nature.

Overall, we find supporting evidence that 2MASS~J0437+2331 is a member of
Taurus, and we find the photometric and spectral qualities of PSO~J060.3+25 and
PSO~J077.1+24 to be very similar to those of 2MASS~J0437+2331.  The only notable
difference is the redder overall near-infrared spectral slope of
2MASS~J0437+2331, which does not appear to be due to interstellar extinction
\citep{Luhman:2009cn}.

\subsubsection{Membership in Taurus}
\label{taurus.bonafide}
Because of the locations of PSO~J060.3+25 and PSO~J077.1+24 on Taurus
color-magnitude diagrams, their plausibly consistent proper motions, their \vlg\
gravity classifications, their photometric and spectral similarity to the known
Taurus L1 dwarf 2MASS~J0437+2331, and the low probability of contamination by
field objects, we consider PSO~J060.3+25 and PSO~J077.1+24 to be bona fide
members of Taurus.  Their near-infrared colors, consistent with those of field
early-L dwarfs but bluer than 2MASS~J0437+2331, confirm that the near-infrared
redness observed in some low-gravity early-L dwarfs (e.g.,
\citealt{Gizis:2012kv,Faherty:2013bc}; see also \citealt{Aller:2016kg}) is not a
universal feature even for very young (1--2~Myr) L1 and L2 dwarfs.

\subsection{Luminosities and Masses}
\label{taurus.masses}
To estimate the masses of our Taurus discoveries, we assumed a distance of
$145\pm15$~pc and an age of 1--2~Myr \citep{Kraus:2009dh}.  We first calculated
bolometric luminosities for each object using our spectral types, the
$K_{\rm MKO}$ bolometric corrections of \citet[their Table 6]{Liu:2010cw}, and
the distance to Taurus.  We then used the Lyon/DUSTY evolutionary
models\footnote{The more recent BHAC15 models \citep{Baraffe:2015fw} do not
  extend to masses below $0.01\msun\ (\approx10\mjup)$.}
\citep{Chabrier:2000hq} and our $L_{\rm bol}$ values to interpolate masses at
the age of Taurus.  We propagated the uncertainties on our spectral types
($\pm1$~subtype), $K_{\rm MKO}$ magnitudes, bolometric correction, distance, and
age into our mass determinations using Monte Carlo simulations, and we quote
68\% confidence limits.  We used normal distributions for each uncertainty
except for age, for which we used a uniform distribution spanning 1--2~Myr to
avoid unreasonably young ages.  We estimate masses of $6.0^{+0.9}_{-0.8}$~\mjup\
for PSO~J060.3+25 and $5.9^{+0.9}_{-0.8}$~\mjup\ for PSO~J077.1+24
(Table~\ref{tbl.kin.mass}).  We also applied this method to 2MASS~J0437+2331
using $K_{\rm MKO}=15.20\pm0.02$~mag from the UKIDSS Galactic Clusters Survey
\citep{Lawrence:2007hu,Lawrence:2013wf}.  We estimate log($L_{\rm bol}$/\lsun)
$=-3.17^{+0.9}_{-1.0}$~dex and a mass of $7.1^{+1.1}_{-1.0}$~\mjup\ for
2MASS~J0437+2331, consistent with the masses of our discoveries and the 4--7
\mjup\ estimate of \citet{Luhman:2009cn}.

With no evidence of companionship to any nearby star or of unresolved binarity
(Section~\ref{results.binaries}), our discoveries are among the lowest mass
free-floating substellar objects ever discovered, similar to 2MASS~J0437+2331,
the young $\beta$ Pictoris Moving Group L~dwarf PSO~J318.5338$-$22.8603
\citep[$8.3\pm0.5$~\mjup;][]{Liu:2013gy,Allers:2016gz}, the young TW~Hydrae
Associaton L~dwarfs 2MASS~J11193254$-$1137466
\citep[4.3--7.6~\mjup;][]{Kellogg:2016fo} and WISEA~J114724.10$-$204021.3
\citep[5--13~\mjup;][]{Schneider:2016iq}, the AB Doradus Moving Group T~dwarf
SDSS~J111010.01+011613.1 \citep[$\approx$10--12~\mjup;][]{Gagne:2015kf}, and the
field Y~dwarf WISE~J085510.83$-$071442.5
\citep[3--10~\mjup;][]{Luhman:2014jd,Leggett:2015dn}.  They provide significant
evidence that free-floating planetary-mass objects can form as part of normal
star-formation processes.

For comparison, we also converted spectral types into effective temperatures,
and then used the DUSTY models and our \teff\ values to estimate masses.  No
empirically calibrated conversion of spectral type to \teff\ has been determined
for very young L dwarfs, so we extrapolated the scale of
\citet{Luhman:2003gd,Luhman:2008ik}, arriving at 2000~K for the L1 dwarf and
1800~K for the L2 dwarf.  We assumed an error of $\pm100$~K for each object.
With this distance-independent approach, we estimate masses of
$7.1^{+1.4}_{-1.1}$~\mjup\ for PSO~J060.3+25 and $5.2^{+0.9}_{-0.8}$~\mjup\ for
PSO~J077.1+24.  If instead we use the field dwarf (i.e., not young)
SpT-to-\teff\ conversion of \citet[Eq. 3]{Stephens:2009cc}, we find temperatures
of $2112\pm100$~K for the L1 dwarf and $1971\pm100$~K for the L2 dwarf,
resulting in masses of $8.6^{+2.0}_{-1.6}$~\mjup\ for PSO~J060.3+25 and
$6.8^{+1.3}_{-1.1}$~\mjup\ for PSO~J077.1+24.

We note also that a recent study by \citet{Daemgen:2015fp} identified evidence
suggesting an older sub-population of Taurus with an age of $\approx$20~Myr.  If
confirmed, and our discoveries are in fact members of this sub-population, the
older age would lead to a factor of $\approx$3 increase in our mass estimates.

\subsection{Evidence for Circumstellar Disks}
\label{taurus.disks}
Many low-mass stellar members of Taurus are known to host circumstellar disks
\citep[e.g.,][]{Andrews:2013ku,Esplin:2014he}.  We searched for evidence of
elevated fluxes at mid-infrared wavelengths that would indicate the presence of
circumstellar disks around our Taurus discoveries.  We fit the BT-Settl grid of
synthetic spectra from \citet{Baraffe:2015fw} to our 0.85--2.45~\um\ prism
spectra of PSO~J060.3+25, PSO~J077.1+24, and 2MASS~J0437+2331 following the
method of \citet{Bowler:2011gw}.  In summary, the models are smoothed to the
resolving power of the data and resampled to the same wavelength grid.  The
1.60--1.65~\um\ and 1.8--1.95~\um\ regions are ignored to avoid incomplete
methane line lists and low S/N portions of our spectra.  The spectra are
flux-calibrated to each object's $J$-band photometry.  A scale factor, equal to
the square of the object's radius divided by its distance, is calculated by
minimizing the $\chi^2$ value following \citet{Cushing:2008kb}.  Assuming a
distance of $145\pm15$~pc to Taurus allows us to simultaneously infer the radius
at each grid point.

The results of the fits are shown in Figure~\ref{fig.taurus.sed}.  The best-fit
synthetic spectra ($T_{\mathrm{eff}}$=1800~K, log~$g$=5.5~dex [cgs] for both of
our discoveries and $T_{\mathrm{eff}}$=1600~K, log~$g$=5.5~dex for
2MASS~J0437+2331) offer relatively poor fits to the data, largely failing to
reproduce the observed $H$ and $K$-band spectral shapes.  The best-fit models
have field-age surface gravities, contrary to the \vlg\ classes indicated by the
observed spectra, so we include synthetic spectra with the same \teff\ as the
best-fit models but with log~$g$=3.5~dex (roughly the gravity expected for \vlg\
objects) in Figure~\ref{fig.taurus.sed} for comparison.  The inferred radii for
our discoveries are all $\ge$2~\rjup, consistent with expectations for very
young objects still undergoing gravitational contraction
\citep[e.g.,][]{Burrows:1997jq}.  Synthetic photometry of the best fitting
models is generally consistent with the observed photometry from Pan-STARRS,
UKIRT, and the $W1$ (3.4~\um) filter from WISE, but is significantly lower for
the $W2$ (4.6~\um) channel at the 7--9$\sigma$ level.  This may represent
evidence of thermal excess from a disk around both objects, but we note that the
observed $W2$ photometry is much more consistent with the log~$g$=3.5~dex model
spectra.  The discrepancy at $W2$ is therefore likely to be a consequence of the
poor model fits, or possibly a result of a systematic error in the model
atmospheres, for example from imperfect opacity sources.

\begin{figure}
\begin{center}
  \includegraphics[width=1.00\columnwidth, trim = 0 20mm 10mm 35mm]{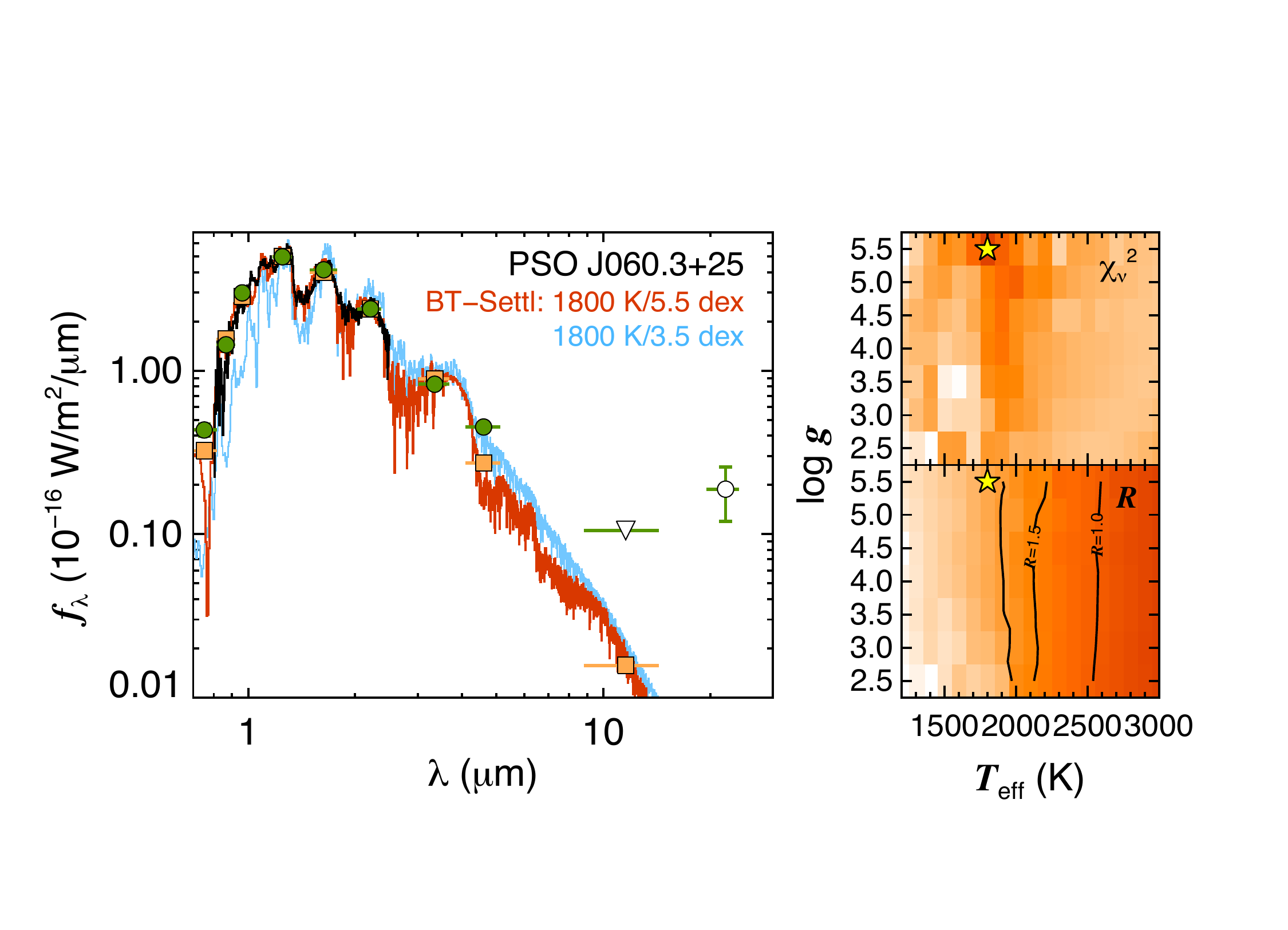}
  \includegraphics[width=1.00\columnwidth, trim = 0 20mm 10mm 35mm]{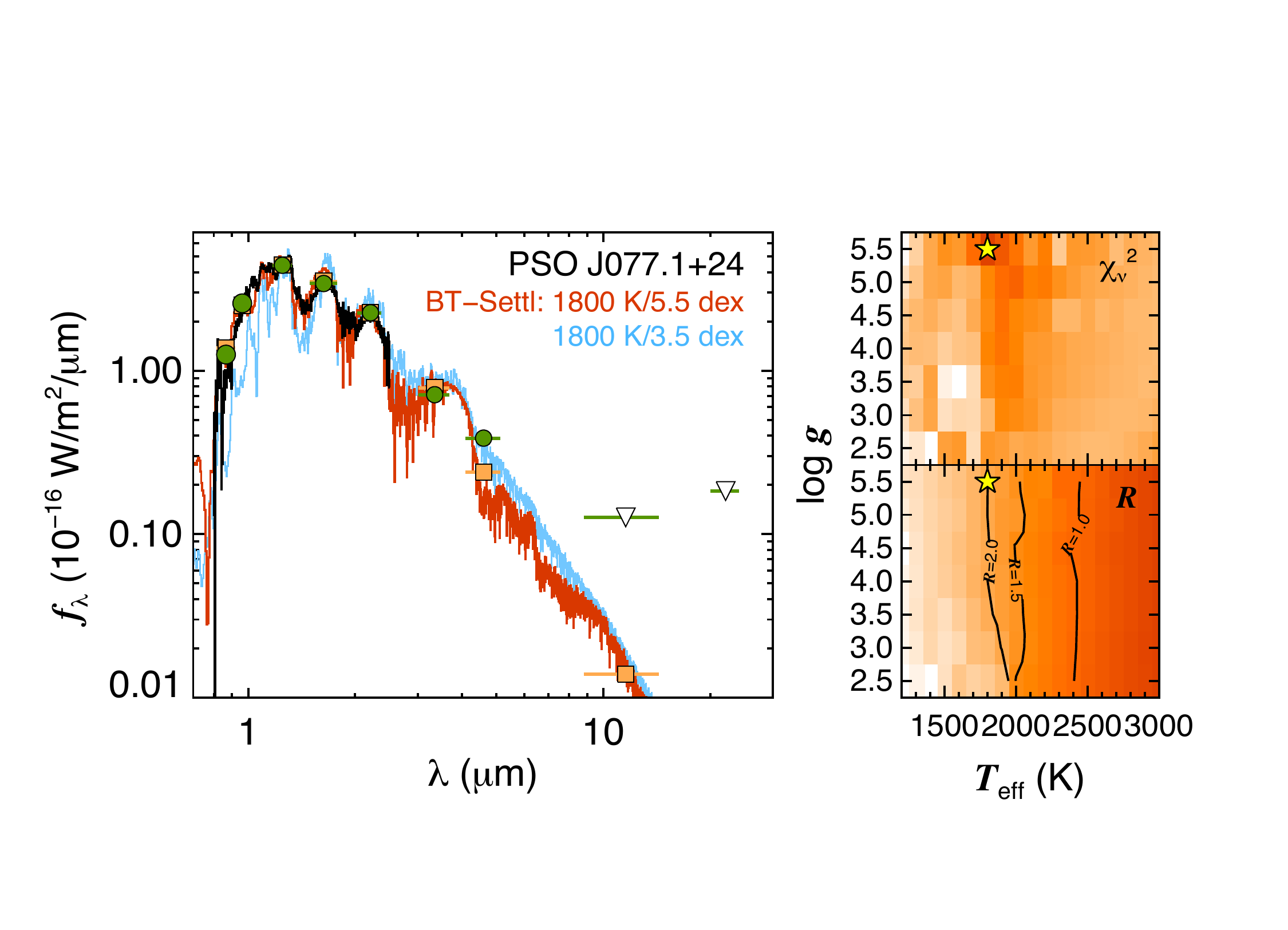}
  \caption{Best-fit BT-Settl model spectra \citep{Baraffe:2015fw} for our two
    Taurus discoveries and 2MASS~J0437+2331.  For each object, the left-hand
    plot includes our prism spectrum (black), the best-fit model spectrum (red),
    and the synthetic spectrum with the same \teff\ as the best-fit model but
    with log~$g=3.5$ (roughly that expected for \vlg\ objects) in blue.  In
    addition, we plot observed PS1/MKO/AllWISE photometry (green circles), and
    synthetic photometry for the best-fit model (orange squares).  Upper limits
    for $W3$ (12~\um) and $W4$ (22~\um) are plotted with open triangles; the
    $W4$ detection for PSO~J060.3+25 (open circle) is marginal at $2.6\sigma$.
    The right-hand plots show the $\chi^2$ surface for the model
    ($T_{\mathrm{eff}}$, log~$g$) fits (\textit{top}) and the inferred radius in
    units of \rjup\ (\textit{bottom}).  The best-fit models match the observed
    spectra fairly poorly, particular in the $H$ and $K$-band morphology.  The
    observed excess flux relative to the best-fit models at $W2$ (4.6~\um) in
    all three objects may indicate the presence of a disk, but the excess is not
    seen relative to the low-gravity model spectra, and may therefore be the
    result of a systematic error in the model atmospheres.}
  \figurenum{fig.taurus.sed.1}
\end{center}
\end{figure}

\begin{figure}[h]
\begin{center}
  \includegraphics[width=1.00\columnwidth, trim = 0 20mm 10mm 35mm]{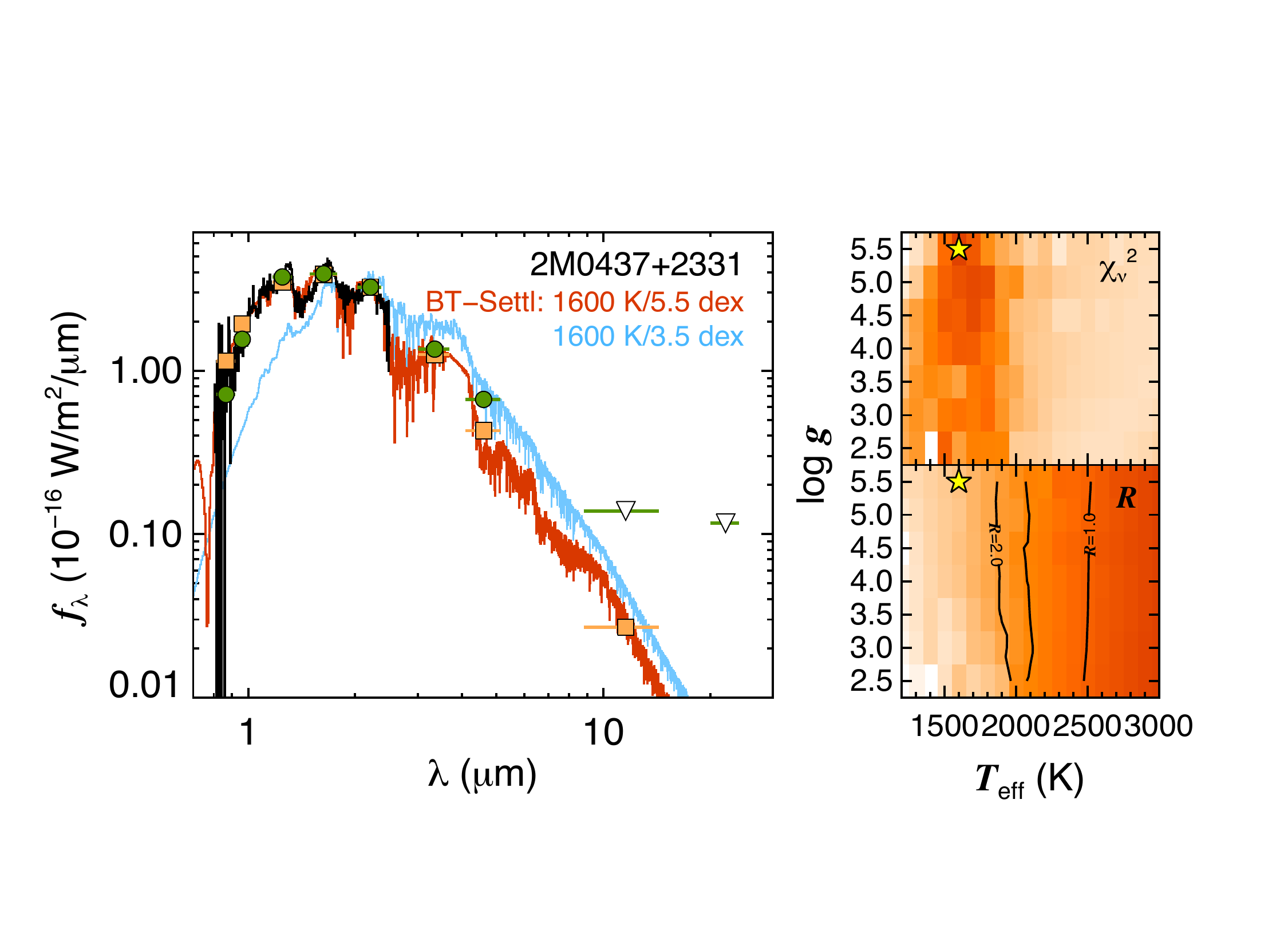}
  \caption{continued.}
  \label{fig.taurus.sed}
\end{center}
\end{figure}

The Taurus objects have photometric upper limits in WISE for the $W3$ (12~\um)
and $W4$ (22~\um) bands, with one exception.  PSO~J060.3+25 has a reported
$2.6\sigma$ detection in $W4$ that is significantly brighter than the synthetic
model photometry.  We visually inspected the WISE images of this object and
could not confirm that the $W4$ detection is distinct from noise.  A clear
excess at 22~\um\ would indicate the presence of a disk, but this marginal
detection requires confirmation by deeper imaging.

We note that \citet{Luhman:2009cn} also found no excess at mid-IR wavelengths in
{\it Spitzer} photometry that would indicate the presence of a disk around
2MASS~J0437+2331.

\section{Scorpius-Centaurus Discoveries}
\label{scocen}
The Scorpius-Centaurus Association is the nearest OB association to the Sun.  We
have discovered six new late-M and early-L members of Sco-Cen.  Using the
boundaries defined by \citet{deZeeuw:1999fe}, PSO~J237.1$-$23 (M7) and
PSO~J239.7$-$23 (L0) lie within the Upper Scorpius subgroup (hereinafter Upper
Sco), while PSO~J228.6$-$29 (L1), PSO~J229.2$-$26 (L0), PSO~J231.7$-$26 (L0),
and PSO~J231.8$-$29 (L0) sit on the northern outskirts of the Upper Centaurus
Lupus subgroup (hereinafter UCL).  Figure~\ref{disc.map} shows the sky locations
of our discoveries.  Upper Sco and UCL are among the reddened regions on the sky
identified by \citet{Cruz:2003fi} that we excluded from our search, and our
discoveries lie just outside the excluded areas.

Like Taurus, the Upper Sco region has been searched multiple times for brown
dwarfs.  Unlike in Taurus, more than a dozen L0--L2 dwarfs have previously been
confirmed in Upper Sco, in particular by \citet{Lodieu:2008hm} using early
release data from the UKIDSS Galactic Clusters Survey (GCS).  Our discoveries
were not found by that search, nor by \citet{Lodieu:2006di,Lodieu:2011df},
because of incomplete coverage of the region by the early version of GCS.  They
also remained undetected by searches using optical data as part of the selection
process \citep{Martin:2004ki,Slesnick:2006ij,Slesnick:2008ci}.  The objects were
detected in GCS Data Release 8 and later versions, but were only observed in $H$
and $K$ bands, and so were not included in the search of
\citet{Dawson:2011jq,Dawson:2013ex} who required $Z$ and $J$ photometry in their
selection process.  PSO~J237.1$-$23 was identified as a candidate member of
Upper Sco by \citet{Lodieu:2013eo} as part of their ``$HK$-only sample'', but
PSO~J239.7$-$23 is too faint to qualify for this sample.  Searches for young
ultracool dwarfs have not focused on UCL, so our four discoveries lie outside
regions targeted by previous efforts.

In this Section we follow the structure of Section~\ref{taurus}, presenting the
evidence that our discoveries are members of Sco-Cen
(Section~\ref{scocen.evidence}), estimating their masses
(Section~\ref{scocen.masses}), and comparing their SEDs to model atmospheres to
look for evidence of circumstellar disks (Section~\ref{scocen.disks}).

\subsection{Evidence for Membership}
\label{scocen.evidence}

\subsubsection{Youth}
\label{scocen.evidence.youth}
Four of our six Sco-Cen discoveries have \vlg\ gravity classes, suggesting ages
$\lesssim30$~Myr.  The other two, PSO~J228.6$-$29 and PSO~J229.2$-$26, have
lower-S/N spectra that do not permit robust calculation of the AL13
gravity-sensitive indices but nevertheless show clear visual indications of low
gravity (Section~\ref{results.gravity}).

\subsubsection{Photometry}
\label{scocen.evidence.phot}
\begin{figure}
\begin{center}
  \begin{minipage}[t]{0.49\textwidth}
    \includegraphics[width=1.00\columnwidth, trim = 20mm 0 8mm 0]{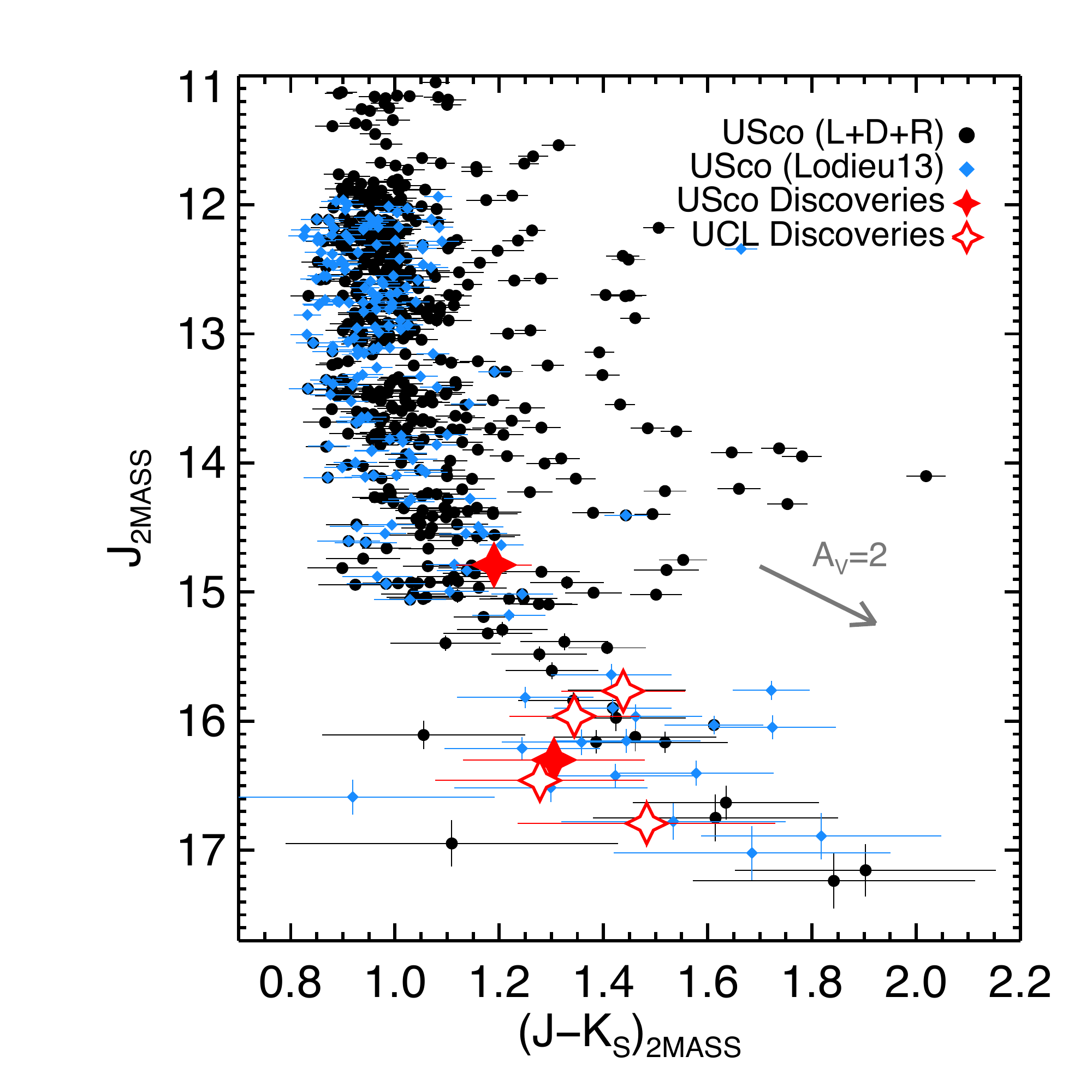}
  \end{minipage}
  \hfill
  \begin{minipage}[t]{0.49\textwidth}
    \includegraphics[width=1.00\columnwidth, trim = 20mm 0 8mm 0]{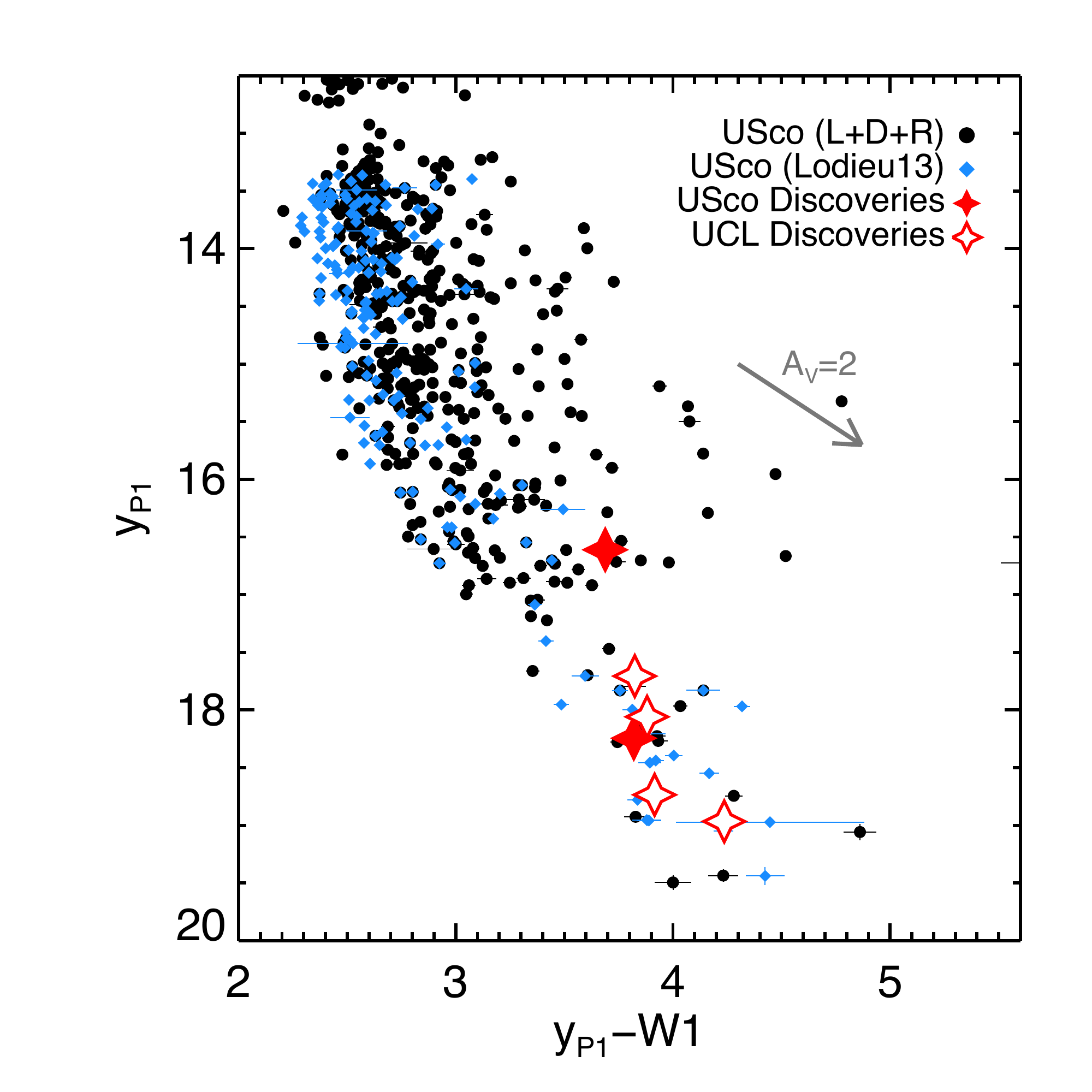}
  \end{minipage}
  \caption{Comparison of the photometry of our discoveries in Upper Sco (USco,
    filled red stars) and Upper Centaurus Lupus (UCL, open red stars) to known
    members of Upper Sco from \citet{Luhman:2012hj}, \citet{Dawson:2014hl}, and
    \citet{Rizzuto:2015bs} (black circles, labeled ''L+D+R'' in the legend) and
    known and candidate Upper Sco members from UKIDSS GCS \citep[blue
    diamonds]{Lodieu:2013eo}. {\it Left}:~$J$ vs. $J-K$ (MKO) diagram. {\it
      Right}:~\yps\ vs. \ywa\ diagram for non-saturated objects in PS1.  We
    include reddening vectors (gray arrows) scaled to an extinction of
    $A_V=2$~mag.  All six of our discoveries have photometry lying along the
    cluster sequences.  The brighter Upper Sco discovery, PSO~J237.1$-$23, has a
    redder \ywa\ color suggesting the presence of a circumstellar disk
    (Section~\ref{scocen.disks}).}
  \label{fig.scocen.cmd}
\end{center}
\end{figure}

Figure~\ref{fig.scocen.cmd} demonstrates the consistency of our six discoveries'
photometry with that of known Upper Sco members from \citet[hereinafter
LM12]{Luhman:2012hj}, \citet{Dawson:2014hl}, and \citet{Rizzuto:2015bs}, and
with photometric/astrometric candidates from unreddened regions in the UKIDSS
GCS Data Release 10 \citep[their Table A1]{Lodieu:2013eo}.  The LM12 catalog
contains a handful of objects that are reddened by interstellar extinction, and
we include reddening vectors (Section~\ref{taurus.evidence.phot}) scaled to an
extinction of $A_V=2$~mag in Figure~\ref{fig.scocen.cmd}.  Most of our
discoveries lie along the unreddened cluster sequence of the Upper Sco $J$
vs. $J-K$ and \yps\ vs. \ywa\ color-magnitude diagrams, fully consistent with
the reddening-free sample of \citet{Lodieu:2013eo}.  PSO~J237.1$-$23 is redder
than the cluster sequence in \ywa, likely evidence for a circumstellar disk
(Section~\ref{scocen.disks}).  As with PSO~J060.3+25 and PSO~J077.1+24 in
Taurus, the Sco-Cen objects have \jkt\ colors that are consistent with field
L0--L1 dwarfs but have \wawb\ colors that are 1--3$\sigma$ redder than the field
population \citep{Gizis:2012kv,Faherty:2013bc}.

\subsubsection{Proper Motion}
\label{scocen.evidence.pm}
\begin{figure}
\begin{center}
  \begin{minipage}[t]{0.49\textwidth}
    \includegraphics[width=1.00\columnwidth, trim = 20mm 0 8mm 0]{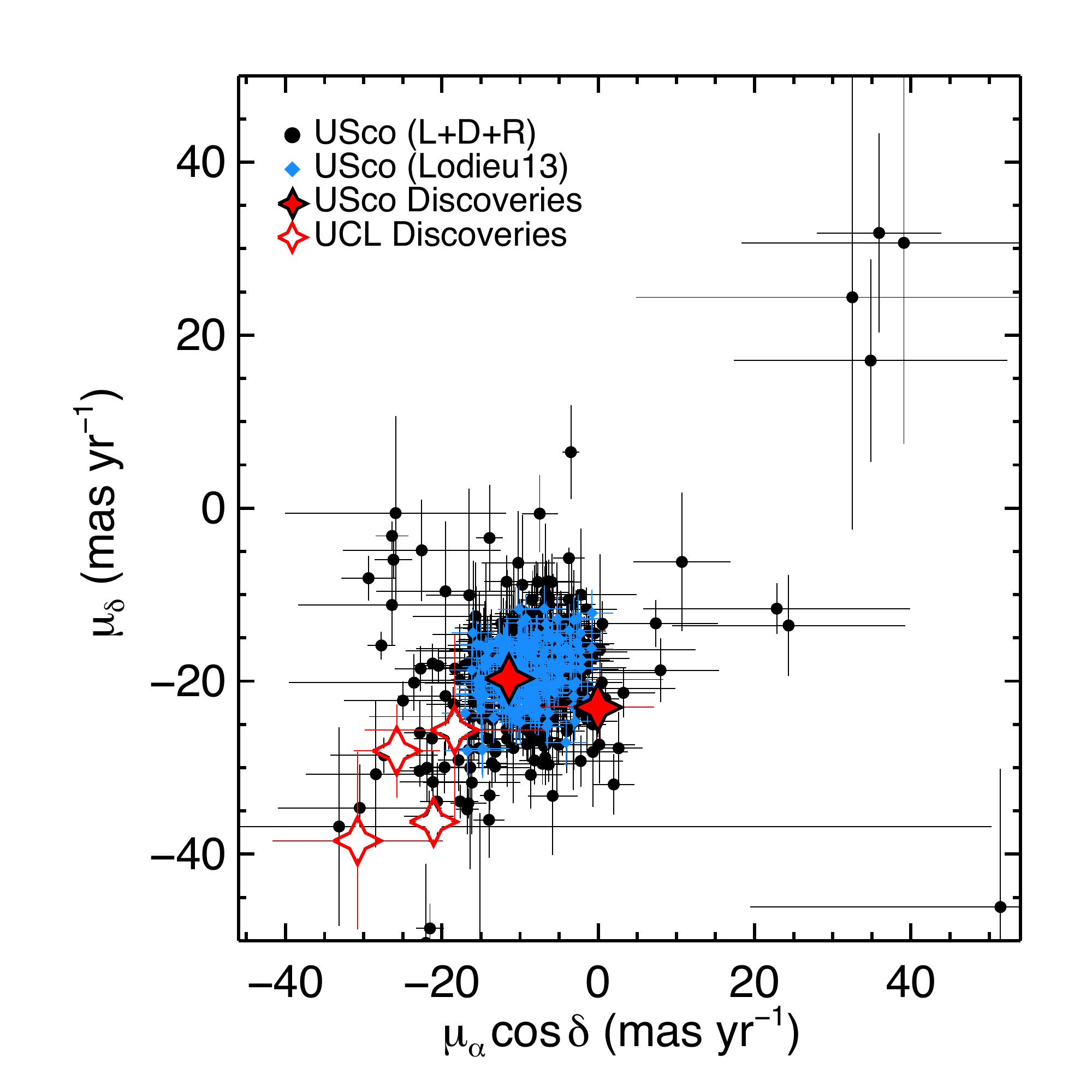}
  \end{minipage}
  \hfill
  \begin{minipage}[t]{0.49\textwidth}
    \includegraphics[width=1.00\columnwidth, trim = 20mm 0 8mm 0]{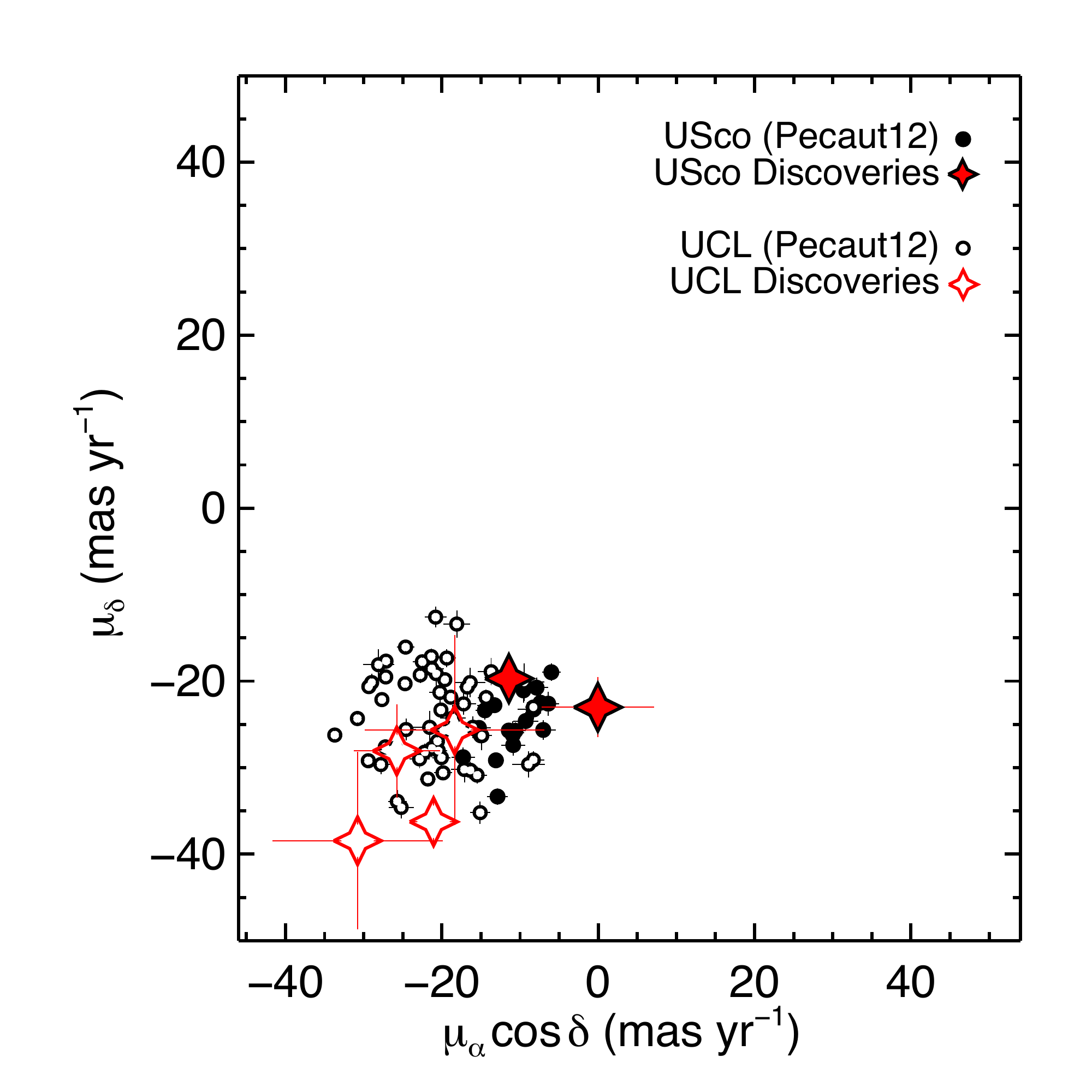}
  \end{minipage}
  \caption{Vector-point diagrams comparing the proper motions of our discoveries
    in Upper Scorpius (USco, filled red stars) and Upper Centaurus Lupus (UCL,
    open red stars) to those of Sco-Cen objects from the literature.  {\it
      Left}: We show our proper motions for objects in the Upper Sco lists of
    LM12, \citet{Dawson:2014hl}, and \citet{Rizzuto:2015bs} (black circles,
    labeled ''L+D+R'' in the legend) that are not saturated in PS1 and have
    reliable proper motion fits, and the proper motions of known and candidate
    Upper Sco members from UKIDSS GCS \citep[blue diamonds]{Lodieu:2013eo}.
    {\it Right}: Proper motions of F-type stars in Upper Sco (black filled
    circles) and UCL (black open circles) from \citet{Pecaut:2012gp}. All six of
    our Sco-Cen discoveries have proper motions consistent with all literature
    sources.}
  \label{fig.scocen.pm}
\end{center}
\end{figure}

We compare the proper motions of our Sco-Cen discoveries to the proper motions
of several literature sources in Figure~\ref{fig.scocen.pm}.
\citet{Pecaut:2012gp} calculated proper motions for F-type stars in Upper Sco
and UCL, and \citet{Lodieu:2013eo} calculated proper motions for a list of
unreddened photometric/astrometric members and candidates in UKIDSS GCS.  LM12
do not quote proper motions for their catalog of Upper Sco members, so we
obtained PS1 Processing Version~3.2 (PV3.2) proper motions for these objects as
well as those from \citet{Dawson:2014hl} and \citet{Rizzuto:2015bs}.  From these
PS1 proper motions we calculated a weighted mean proper motion for known Upper
Sco members of ($\mua=-8.5\pm0.1,\ \mud=-19.6\pm0.1$~\my), with a weighted rms
of 4.3~\my\ in R.A. and 5.6~\my\ in Dec.  Our complete list of PS1 proper
motions for objects from this combined catalog that are not saturated in PS1 is
described in Appendix~\ref{appendixb}.  Figure~\ref{fig.scocen.pm} demonstrates
the consistency of the proper motions of our discoveries with all literature
sources.

\subsubsection{Likelihood of Field Contamination}
\label{scocen.foreground}
We estimated the likelihood that any of our Sco-Cen discoveries could be
interloping foreground or background field objects, using the same approach as
in Section~\ref{taurus.foreground}.  Upper Sco and UCL are distinct regions with
different ages and bulk proper motions \citep{Pecaut:2012gp}, so we considered
them separately.

We defined the boundaries of Upper Sco to be $343^\circ\le l\le360^\circ$ and
$10^\circ\le b\le30^\circ$ (Figure~\ref{disc.map}).  The portion of this region
surveyed by PS1 (i.e., north of $\delta=-30^\circ$) covers 281.0~deg$^2$.  We
found that $28.42\pm0.11$\% of a synthetic population (Besan\c{c}on Galactic
model) of field M dwarfs in the direction of Upper Sco will have proper motions
within $3\sigma$ of the mean Upper Sco \mua\ and \mud\
(Section~\ref{scocen.evidence.pm}).  We estimate that our search would find
$(3.99\pm0.02)\times10^{-2}$ field \vlg\ dwarfs within the projected boundaries
of Upper Sco having proper motions consistent with Upper Sco membership.  We
find a probability of 96.1\% that both PSO~J237.1$-$23 and PSO~J239.7$-$23 are
members of Upper Sco.

For UCL, we used the data for F stars from \citet{Pecaut:2012gp} to calculate a
weighted mean proper motion of ($\mua=-23.0\pm0.1,\ \mud=-23.8\pm0.1$~\my) with
a weighted rms of 5.7~\my\ in R.A. and 4.9~\my\ in Dec.  Only 0.51\% of our
synthetic M dwarf population have proper motions consistent within $3\sigma$
with the mean UCL motion.  PSO~J228.6$-$29, PSO~J229.2$-$26, PSO~J231.7$-$26,
and PSO~J231.8$-$29 all have proper motions consistent with UCL, so the
probability that any of them is a field interloper is negligible based on proper
motions alone.

\subsubsection{Membership in Scorpius-Centaurus}
\label{scocen.bonafide}
All six of our discoveries in the direction of Sco-Cen have \vlg\ gravity
classifications or clear spectral indications of low gravity.  Their positions
in color-magnitude diagrams, proper motions, and very low probability of
contamination by field objects confirm that they are members of Sco-Cen.

\subsection{Luminosities and Masses}
\label{scocen.masses}
We calculated bolometric luminosities and estimated the masses of our Sco-Cen
discoveries using the method described in Section~\ref{taurus.masses}.  We
adopted a distance of $145\pm15$~pc \citep{deZeeuw:1999fe,Preibisch:1999jk} and
an age of $11\pm2$~Myr \citep{Pecaut:2012gp} for Upper Sco.  The mean distance
to UCL is $140\pm2$~pc \citep{deZeeuw:1999fe}.  \citet{deBruijne:1999bh} found a
substantial depth of $50\pm20$~pc for UCL, but all four of our UCL discoveries
have photometry consistent with members of Upper Sco
(Figure~\ref{fig.scocen.cmd}), so we used the same distance uncertainty as Upper
Sco and adopted a UCL distance of $140\pm15$~pc, along with an age of
$16\pm1$~Myr \citep{Pecaut:2012gp}.  We used normal distributions for the age
and distance uncertainties in our Monte Carlo simulations.  Our luminosity and
mass estimates for our Sco-Cen discoveries are listed in
Table~\ref{tbl.kin.mass}.  The masses span $15-36$~\mjup, near the low-mass end
of the brown dwarf regime and comparable to the lowest mass members known in
these regions \citep{Lodieu:2011df,Aller:2013bc}.

\subsection{A Candidate Circumstellar Disk}
\label{scocen.disks}
At ages $\gtrsim$10~Myr, our discoveries in Sco-Cen are less likely to harbor
circumstellar disks than are the $\approx$1--2~Myr Taurus objects
\citep[e.g.,][]{Mathews:2012bw}.  However, LM12 have demonstrated that
$\approx$25\% of M5--L0 objects in Upper Sco have disks detectable at
mid-infrared wavelengths.  They developed color vs. spectral type relationships
to identify stars and brown dwarfs with candidate circumstellar disks, using
colors including $K_S-W2$, $K_S-W3$, and $K_S-W4$.  Our L dwarf discoveries were
not detected in $W3$ or $W4$, and LM12 caution that the $K_S-W2$ colors do not
reliably discriminate between excess flux from a disk and rapidly reddening
photospheres beyond spectral type M8.5.  We nevertheless checked the $K_S-W2$
colors of our L dwarfs (including the Taurus discoveries), and none are redder
than the typical colors of Upper Sco M9--L1 dwarfs (LM12), so we cannot identify
any candidate disk hosts among these objects.

PSO~J237.1$-$23, the lone M dwarf (M7) among our discoveries, has significantly
mid-infrared redder colors than those of late-M dwarfs lacking disks, strongly
suggesting the presence of a circumstellar disk.  The $K_S-W2$ and $K_S-W3$
colors both satisfy the LM12 criteria by $\gtrsim5\sigma$.  PSO~J237.1$-$23 also
has a marginal $W4$ detection at $8.75\pm0.45$~mag, which if real would give the
object a $K_S-W4$ color over 3.5~mag redder than the LM12 limit for disk-hosting
M7~dwarfs.  We therefore consider PSO~J237.1$-$23 to be a clear candidate
circumstellar disk host, joining over a dozen other candidates in Upper Sco with
spectral types M7 or later (LM12).

We also looked for evidence of excess mid-infrared fluxes using the method
described in Section~\ref{taurus.disks}, fitting the BT-Settl model spectra
\citep{Baraffe:2015fw} to the prism spectra of our Sco-Cen discoveries.  The
results are shown in Figure~\ref{fig.scocen.sed}.  As with our Taurus
discoveries, the best-fit synthetic spectra for our Sco-Cen objects have
field-age gravities inconsistent with the low-gravity features in the observed
spectra, fit the emprical $JHK$-band morphologies fairly poorly, and have radii
$\ge2$~\rjup\ (consistent with models).  Synthetic photometry from the models is
generally consistent with the observed photometry.  The significantly lower
synthetic flux at $W2$ seen in our Taurus discoveries
(Figure~\ref{fig.taurus.sed}) is present here for two of the six Sco-Cen
objects, although again this is likely to be an consequence of poor model fits.
Our disk candidate, PSO~J237.1$-$23, does show an excess in observed flux
relative to the best-fit and low-gravity models at all four \WISE\ bands,
although the model synthetic fluxes are higher than observations in the optical
bands, indicating that the fit has failed to correctly capture the observed SED.

\begin{figure}
\begin{center}
  \includegraphics[width=1.00\columnwidth, trim = 0 20mm 10mm 35mm]{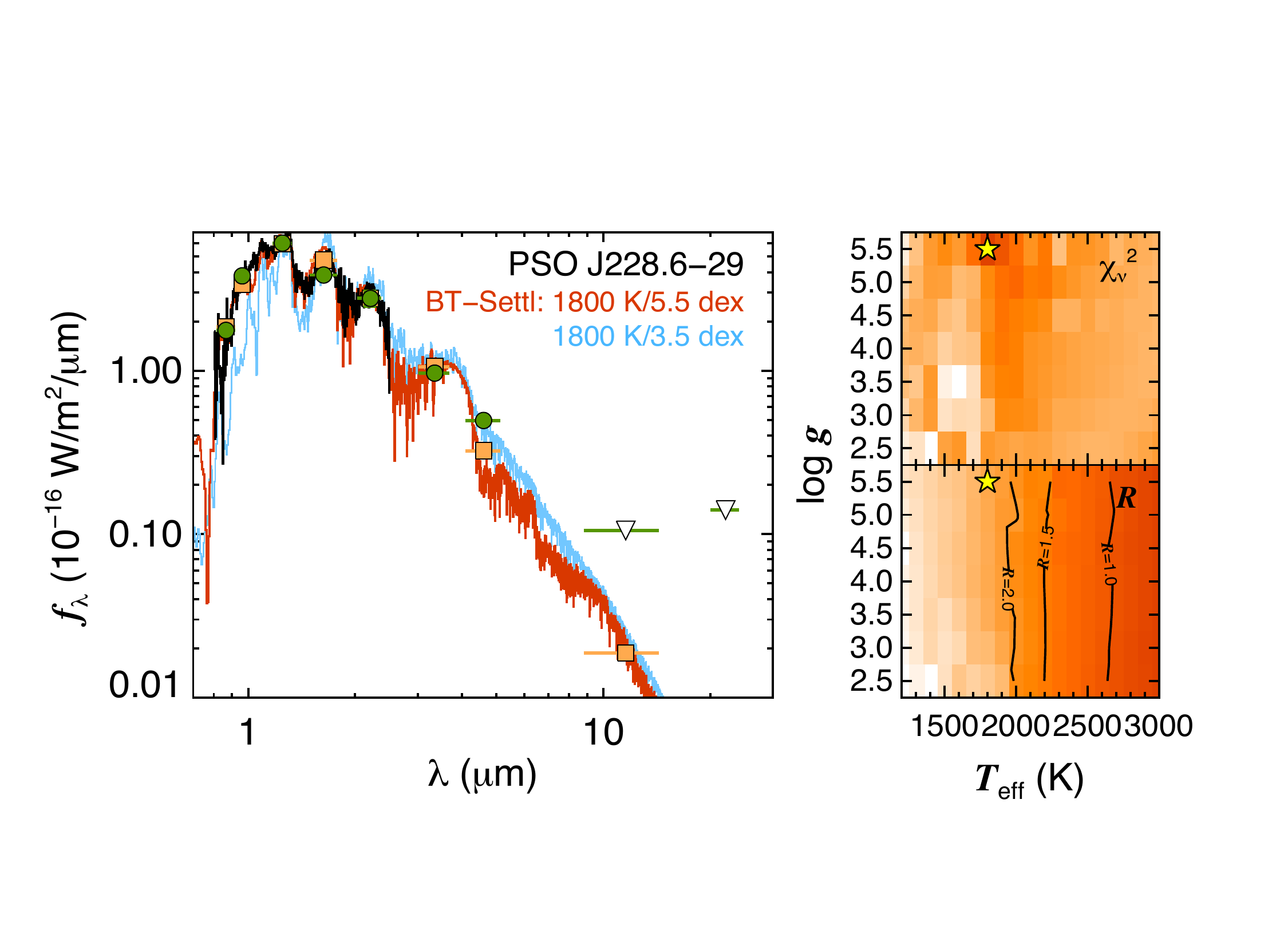}
  \includegraphics[width=1.00\columnwidth, trim = 0 20mm 10mm 35mm]{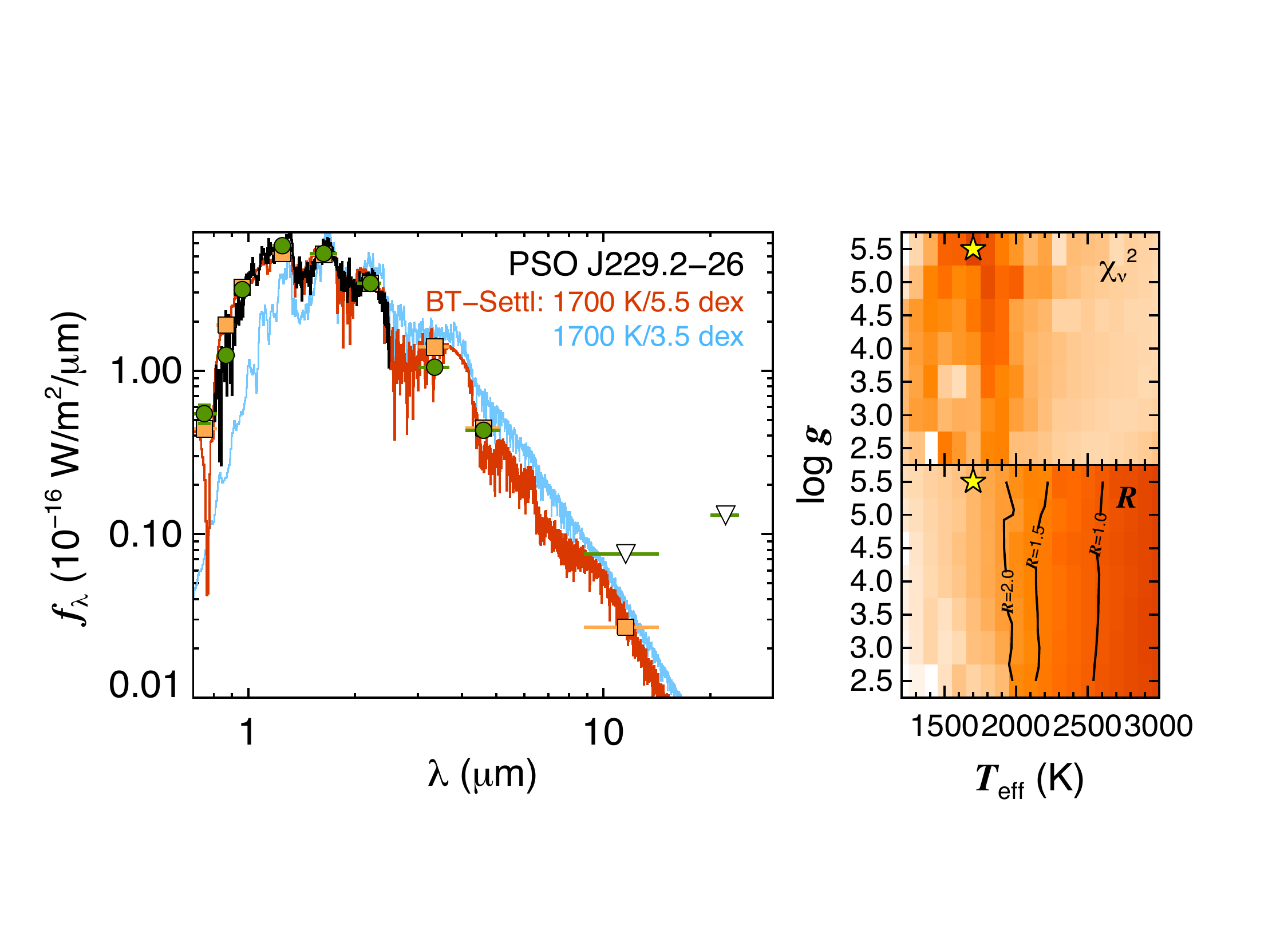}
  \caption{Same as Figure~\ref{fig.taurus.sed}, but showing the best-fit
    BT-Settl model spectra \citep{Baraffe:2015fw} to our Sco-Cen discoveries.
    The best-fit models again match the observed spectra fairly poorly,
    particularly in the $H$ and $K$-band morphology.  The observed excess flux
    at $W2$ (4.6~\um) seen in Figure~\ref{fig.taurus.sed} is seen here in three
    of the six objects, but the excess again disappears in two cases
    (PSO~J228.6$-$29 and PSO~J239.7$-$23) when compared to the low-gravity
    models expected for \vlg\ objects.  However, PSO~J237.1$-$23 does show a
    clear excess in flux relative to the models at all four \WISE\ bands
    ($\ge3.4$~\um), implying the presence of a circumstellar disk.}
\figurenum{fig.scocen.sed.1}
\end{center}
\end{figure}

\begin{figure}
\begin{center}
  \includegraphics[width=1.00\columnwidth, trim = 0 20mm 10mm 35mm]{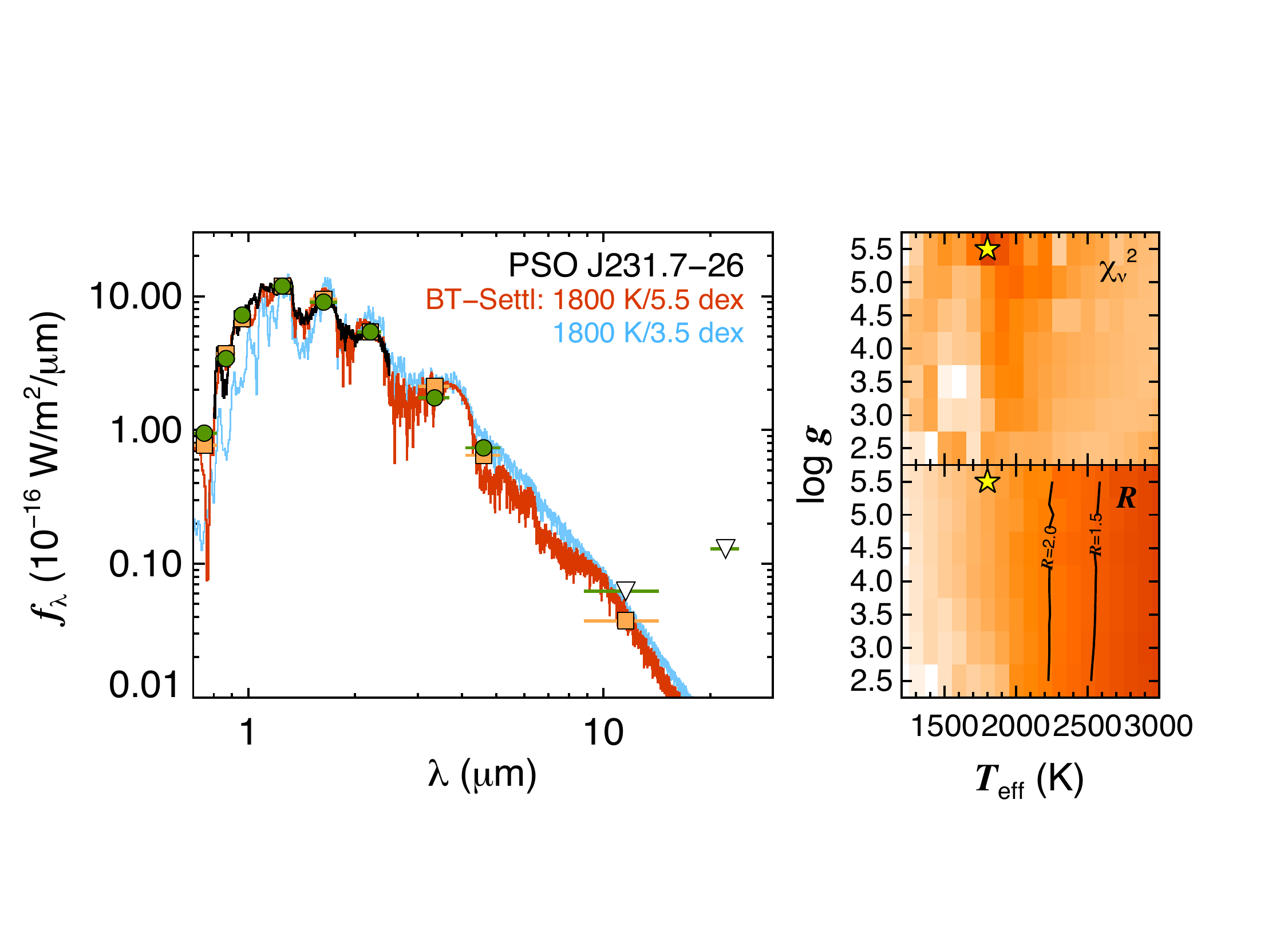}
  \includegraphics[width=1.00\columnwidth, trim = 0 20mm 10mm 35mm]{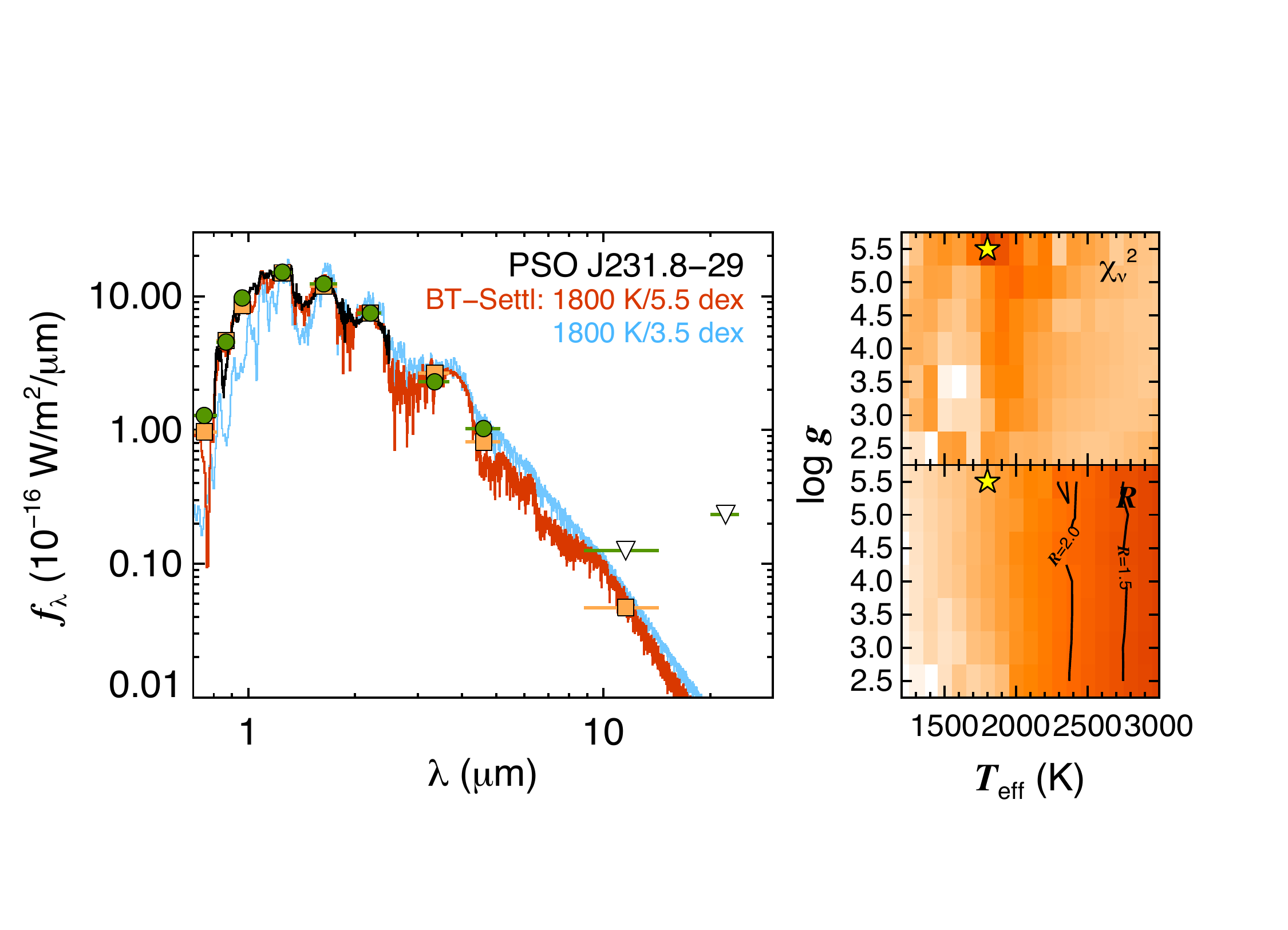}
  \caption{continued.}
\figurenum{fig.scocen.sed.2}
\end{center}
\end{figure}

\begin{figure}
\begin{center}
  \includegraphics[width=1.00\columnwidth, trim = 0 20mm 10mm 35mm]{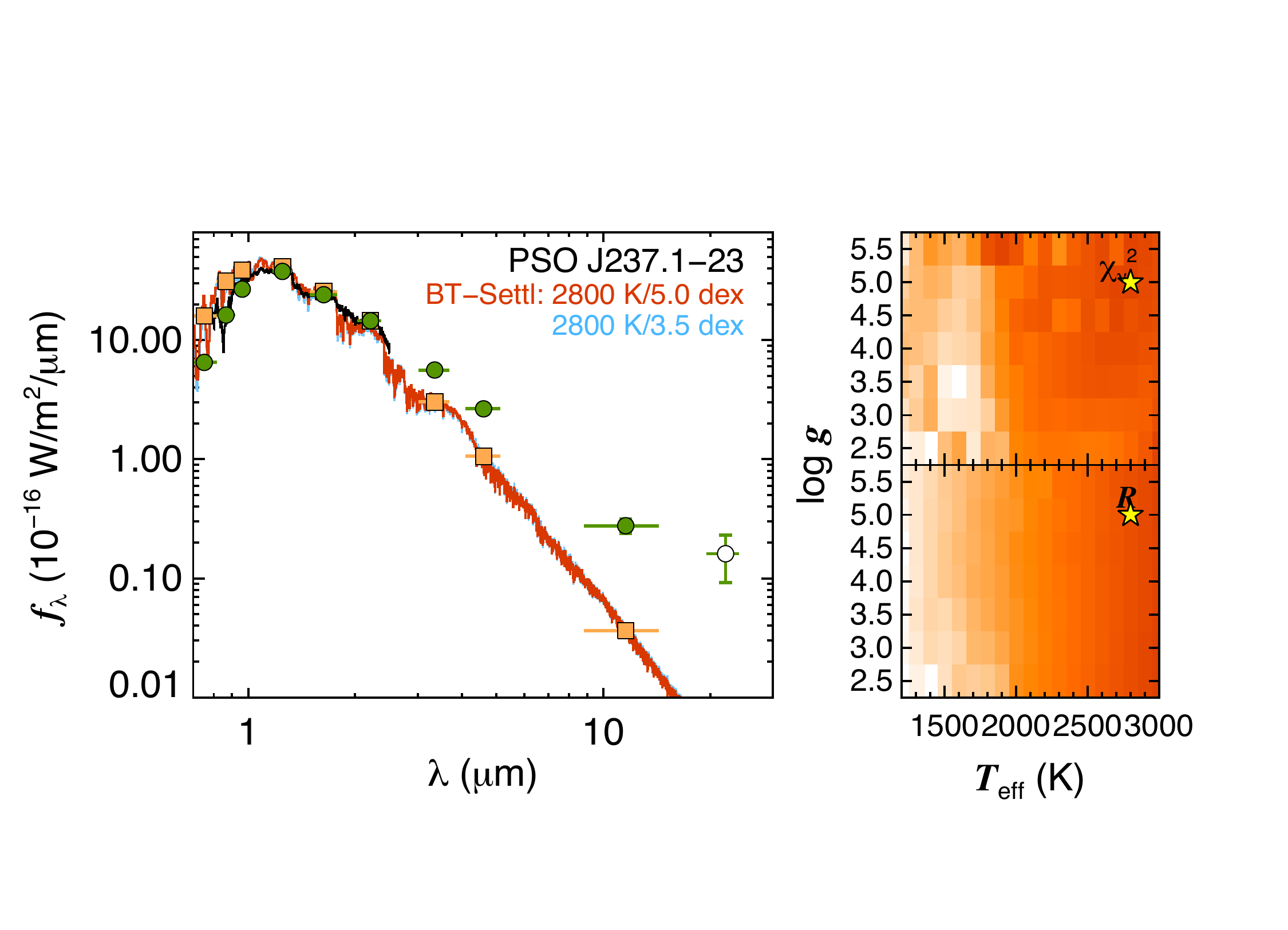}
  \includegraphics[width=1.00\columnwidth, trim = 0 20mm 10mm 35mm]{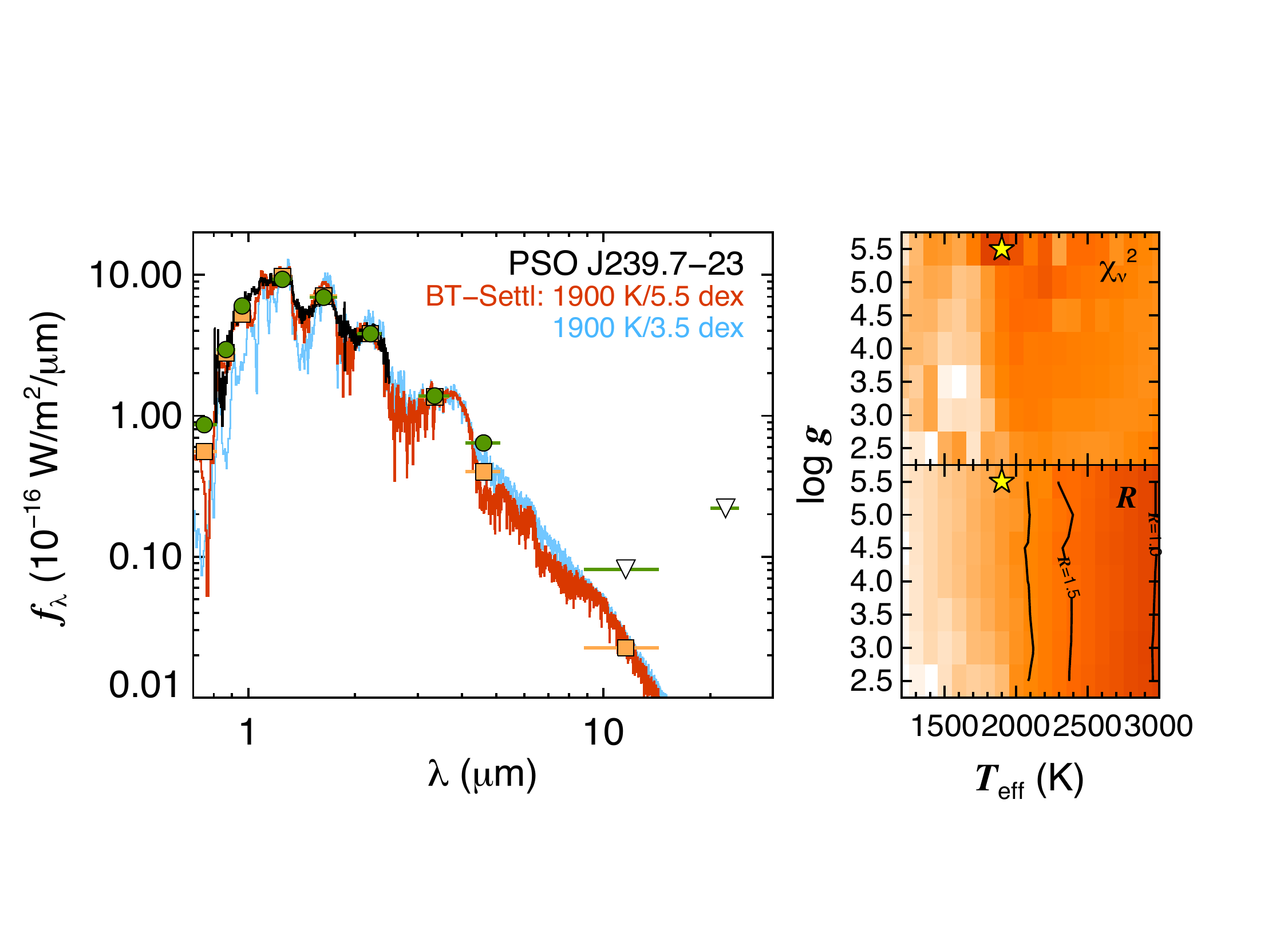}
  \caption{continued.}
  \label{fig.scocen.sed}
\end{center}
\end{figure}

\section{Summary}
\label{summary}
As part of a wide-field search for L/T transition dwarfs using the \PS\ and
\WISE\ surveys, we have serendipitously discovered eight young late-M and
early-L dwarfs in the nearby Taurus and Scorpius-Centaurus star-forming regions.
PSO~J060.3+25 (spectral type L1) and PSO~J077.1+24 (L2) are members of Taurus.
Both have \vlg\ gravity classifications indicating ages $\lesssim30$~Myr,
photometry consistent with previously known ultracool members of Taurus, and
proper motions consistent with the Taurus population.  We estimate the
probability that neither object is a foreground (or background) field dwarf to
be 97\%.  The spectral and photometric properties of our two discoveries are
also similar to the only previously known free-floating L~dwarf in Taurus,
2MASS~J0437+2331 \citep{Luhman:2009cn}.  At the young ($\approx$1--2~Myr) age of
Taurus, PSO~J060.3+25 and PSO~J077.1+24 have estimated masses of
$\approx$6~\mjup, and they join 2MASS~J0437+2331 ($\approx$7~\mjup, spectral
type L1) as the lowest-mass isolated known members of Taurus.  PSO~J077.1+24 is
additionally the coolest known free-floating object discovered in Taurus to
date.

PSO~J060.3+25 was previously identified by \citet{Sarro:2014ci} and
\citet{Bouy:2015gl} as DANCe~J040116.80+255752.2, a likely ultracool member of
the Pleiades (age~$\approx125$~Myr) based on its photometry and astrometry.  Our
spectrum confirms the late spectral type of PSO~J060.3+25, but its \vlg\ gravity
class implies an age ($\lesssim$30~Myr) consistent with the much younger Taurus
star-forming region, and its near-infrared photometry is more consistent with
other VLM members of Taurus.

The other six M7--L1 dwarf discoveries lie on the outskirts of the Upper
Scorpius and Upper Centaurus-Lupus associations (ages $\approx11$--16~Myr), with
estimated masses $\approx15-36$~\mjup.  Four have \vlg\ gravity classifications;
our spectra for the other two did not have enough S/N for confident gravity
classification, but visual inspection finds they have clear spectral signatures
of low gravity.  The photometry and proper motions of all six objects are fully
consistent with membership in Scorpius-Centaurus.  \citet{Lodieu:2013eo}
previously identified PSO~J237.1$-$23 as an astrometric and photometric
candidate member of Upper Sco, which we confirm with our independent discovery
and spectroscopy.

We found no spectral indications that any of our discoveries have unresolved
companions, nor did we find any comoving objects nearby.  The Taurus objects
represent strong evidence that normal star formation processes can produce
isolated objects with masses as low as $\approx$6~\mjup.

The $J-K$ colors of all seven young L~dwarf discoveries are consistent with
those of older field L0--L2 dwarfs.  This contrasts with the redder $J-K$ colors
of some previously discovered young early-L dwarfs, and confirms that
near-infrared redness is not a universal feature of very young (1--2~Myr) brown
dwarfs.  Our discoveries do have \wawb\ colors that are redder than those of
early-L field objects, which we identify as the primary reason we discovered
these objects during a search for L/T transition dwarfs.

We fit BT-Settl synthetic spectra \citep{Baraffe:2015fw} to our observed spectra
and found that the best-fit models reproduce our spectra relatively poorly in
the near-infrared.  At 4.6~\um\ (\WISE\ $W2$ band), all three Taurus objects and
two Sco-Cen objects show a significant observed excess flux over the model
predictions.  These elevated fluxes are suggestive of the presence of a
circumstellar disk but may also indicate a source of systematic error in the
model atmospheres.  The M7 dwarf PSO~J237.1$-$23 shows strong excess fluxes at
$W2$, $W3$, and possibly at $W4$, making it a likely host for a circumstellar
disk.

Our discovery of these eight young brown dwarfs in well-searched regions of the
sky, while looking for older objects with cooler spectral types, has a few
important implications:
\begin{itemize}
  \item The combination of PS1 and \WISE\ photometry is a powerful tool for
    identifying young ultracool dwarfs (see also Paper II).
  \item Unusually red \wawb\ colors in late-M and early-L dwarfs may indicate
    the objects are young (Figure~\ref{fig.w1w2.yw1}), providing leverage for
    searches for young M/L dwarfs.
  \item There are likely to be more young planetary-mass brown dwarfs that could
    be discovered with focused searches in even well-studied star-forming regions. 
\end{itemize}

We thank the anonymous referee for a helpful critique that improved the quality
of this paper.  We thank Katelyn Allers, Brian Cabreira, Dave Griep, Tony
Matulonis, and Eric Volqardsen for assisting with IRTF observations.  The \PS\
Surveys (PS1) have been made possible through contributions of the Institute for
Astronomy, the University of Hawaii, the Pan-STARRS Project Office, the
Max-Planck Society and its participating institutes, the Max Planck Institute
for Astronomy, Heidelberg and the Max Planck Institute for Extraterrestrial
Physics, Garching, The Johns Hopkins University, Durham University, the
University of Edinburgh, Queen's University Belfast, the Harvard-Smithsonian
Center for Astrophysics, the Las Cumbres Observatory Global Telescope Network
Incorporated, the National Central University of Taiwan, the Space Telescope
Science Institute, the National Aeronautics and Space Administration under Grant
No. NNX08AR22G issued through the Planetary Science Division of the NASA Science
Mission Directorate, the National Science Foundation under Grant
No. AST-1238877, the University of Maryland, Eotvos Lorand University (ELTE),
and the Los Alamos National Laboratory.  This project makes use of data products
from the Wide-field Infrared Survey Explorer, which is a joint project of the
University of California, Los Angeles, and the Jet Propulsion
Laboratory/California Institute of Technology, funded by the National
Aeronautics and Space Administration.  This work has made use of data from the
European Space Agency (ESA) mission {\it Gaia}
(\url{http://www.cosmos.esa.int/gaia}), processed by the {\it Gaia} Data
Processing and Analysis Consortium (DPAC,
\url{http://www.cosmos.esa.int/web/gaia/dpac/consortium}). Funding for the DPAC
has been provided by national institutions, in particular the institutions
participating in the {\it Gaia} Multilateral Agreement.  This research has made
use of the 2MASS data products; the UKIDSS data products; the VISTA data
products; NASA's Astrophysical Data System; the SIMBAD and Vizier databases
operated at CDS, Strasbourg, France, and the Database of Ultracool Parallaxes,
maintained by Trent Dupuy at https://www.cfa.harvard.edu/$\sim$tdupuy/plx.  WMJB
received support from NSF grant AST09-09222.  WMBJ, MCL, and EAM received
support from NSF grant AST-1313455.  Finally, the authors wish to recognize and
acknowledge the very significant cultural role and reverence that the summit of
Mauna Kea has always held within the indigenous Hawaiian community. We are most
fortunate to have the opportunity to conduct observations from this mountain.

{\it Facilities:} \facility{IRTF (SpeX)}, \facility{PS1}, \facility{UKIRT (WFCAM)}

\appendix

\section{Proper Motions of Known Low-Mass Taurus Members}
\label{appendixa}

\subsection{\PS\ Proper Motions}
\label{appendixa.method}
We compiled a catalog of proper motions for low-mass members of Taurus using the
\PS\ $3\pi$ (PS1) Survey, Processing Version~3.2 (PV3.2).  Photometry and
positions from PV3.2 were publically released as part of PS1~DR1 (K. Chambers et
al., 2017, in prep), with proper motions and parallaxes planned for a future PS1
release.  PS1 astrometry includes \PS\ observations from November 2009 to March
2014, as well as detections from 2MASS (October 1997 to November 2000) and
\textit{Gaia}~DR1 \citep[Epoch
2015.0;][]{GaiaCollaboration:2016cu,Lindegren:2016gr} lying within 1'' of the
mean PS1 position.  PS1 astrometry, including proper motions, is calibrated to
the \textit{Gaia}~DR1 reference frame.

A full description of the proper motion calculations can be found in E. Magnier
et al. (2017, in prep).  Briefly, all PS1, 2MASS, and \textit{Gaia} detections
for an object were fit simultaneously for position, parallax, and proper motion
using iteratively-reweighted least squares regression with outlier clipping.
Errors were estimated for each object using a bootstrapping approach, drawing
random samples in a Monte Carlo fashion (allowing duplicates) from the set of
detections not rejected in the astrometric fit.

\subsection{Catalog}
\label{appendixa.catalog}
To create our catalog, we began with the list of 414 Taurus members from
\citet{Esplin:2014he}.  Using the 2MASS positions (or WISE positions for the
seven objects with no 2MASS detection), we cross-matched this list with the PS1
database using a $3''$ matching radius, and found 363 matches with a proper
motion measured by PS1.  We supplemented these matches with our two Taurus
discoveries presented in the paper.  We verified that none of the PS1 sources
were identified as quasars, transients, periodic variables, or solar system
objects in the PS1 database, and we excluded any objects with poor PSF fits
({$\tt psf\_qf<0.85$}).  To avoid saturation in PS1 we also excluded objects
having $\ips<14.5$~mag or $\yps<12.5$~mag (corresponding roughly to a spectral
type of M3--M4).  This left us with 187 members of Taurus having proper motions
measured by PS1.  Almost every object in the catalog of \citet{Esplin:2014he}
was detected by 2MASS, so the PS1 proper motions have time baselines of
$\approx$14--17 years.  Our catalog includes 27 objects with no previously
published proper motion and 93 measurements that improve on the best available
literature values drawn from NOMAD \citep{Zacharias:2005tx}, PPMXL
\citep{Roeser:2010cr}, SDSS DR9 \citep{Ahn:2012ih}, UCAC4
\citep{Zacharias:2013cf}, UKIDSS GCS DR9 \citep{Lawrence:2013wf}, URAT1
\citep{Finch:2016us}, USNO-B \citep{Monet:2003bw}, and \citet{Riaz:2013fg}. It
is the largest catalog to date for proper motions of low-mass (spectral
types~$\gtrsim$~M3) members of Taurus.

\begin{figure}
\begin{center}
    \includegraphics[width=1.00\columnwidth]{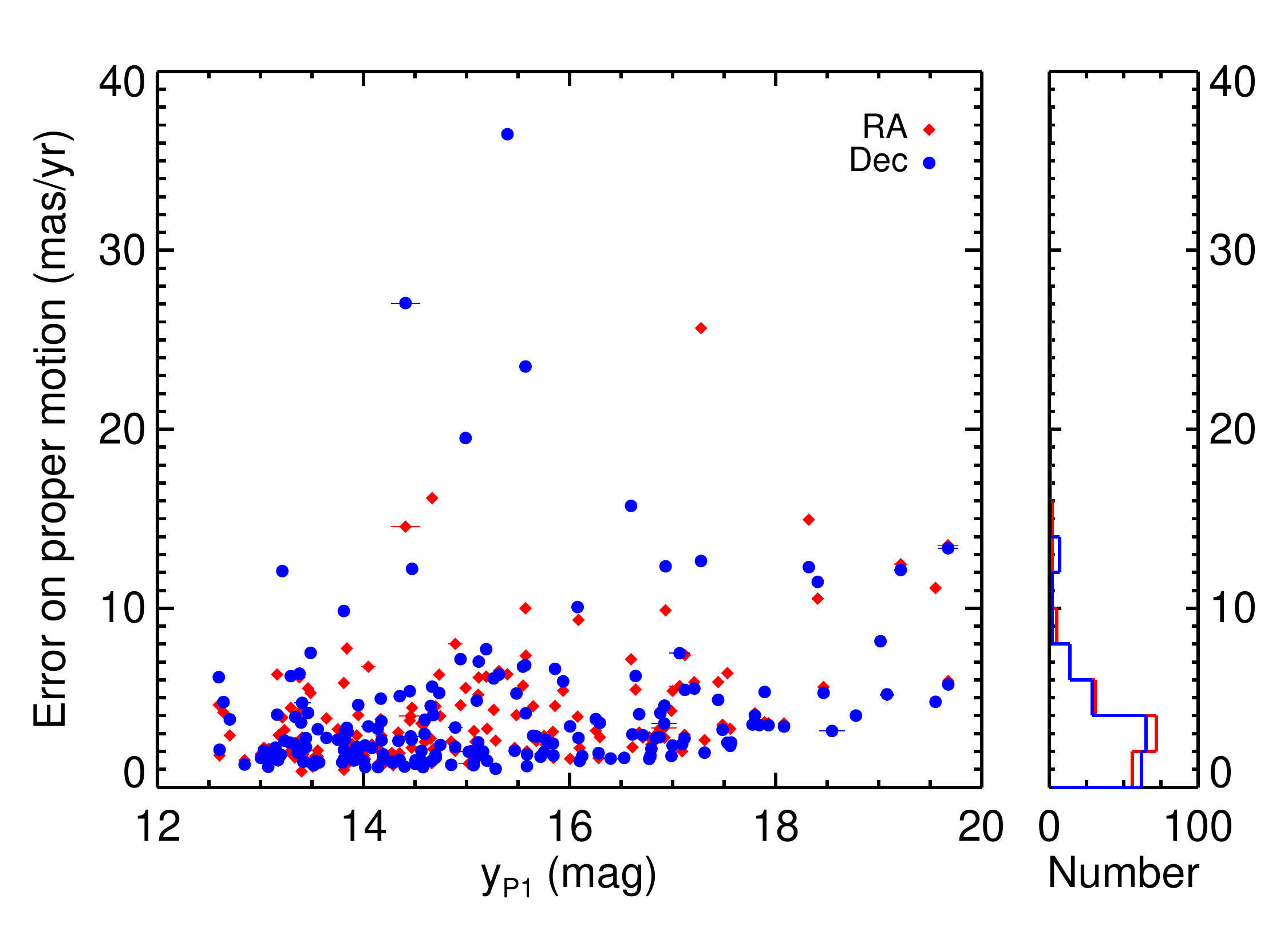}
    \caption{Errors on our \mua\ (red diamonds) and \mud\ (blue dots) as a
      function of \yps\ for known Taurus members that are not saturated in PS1.
      The histogram on the right shows the distributions of the errors.  Most
      errors are $\lesssim$5~\my.}
  \label{fig.taurus.pm.err}
\end{center}
\end{figure}

We list our proper motions in Table~\ref{tbl.taurus.allpm}, along with the \ips\
and \yps\ photometry, the number of epochs used, the reduced $\chi^2$, and the
time baseline for each proper motion fit.  We adopt a photometric precision
floor of 0.01~mag for the PS1 photometry, following the analysis of
\citet{Schlafly:2012da}.  The errors reported in the PS1 database are formal
errors that do not include systematics, and are often smaller than 0.01~mag.  We
do not report photometry with errors larger than 0.2~mag.
Figure~\ref{fig.taurus.pm.err} shows the proper motion errors as a function of
the \yps\ magnitude of each source, and indicates that most of the errors are
$\lesssim$5~\my.
 
\begin{figure}
\begin{center}
    \includegraphics[width=1.00\columnwidth]{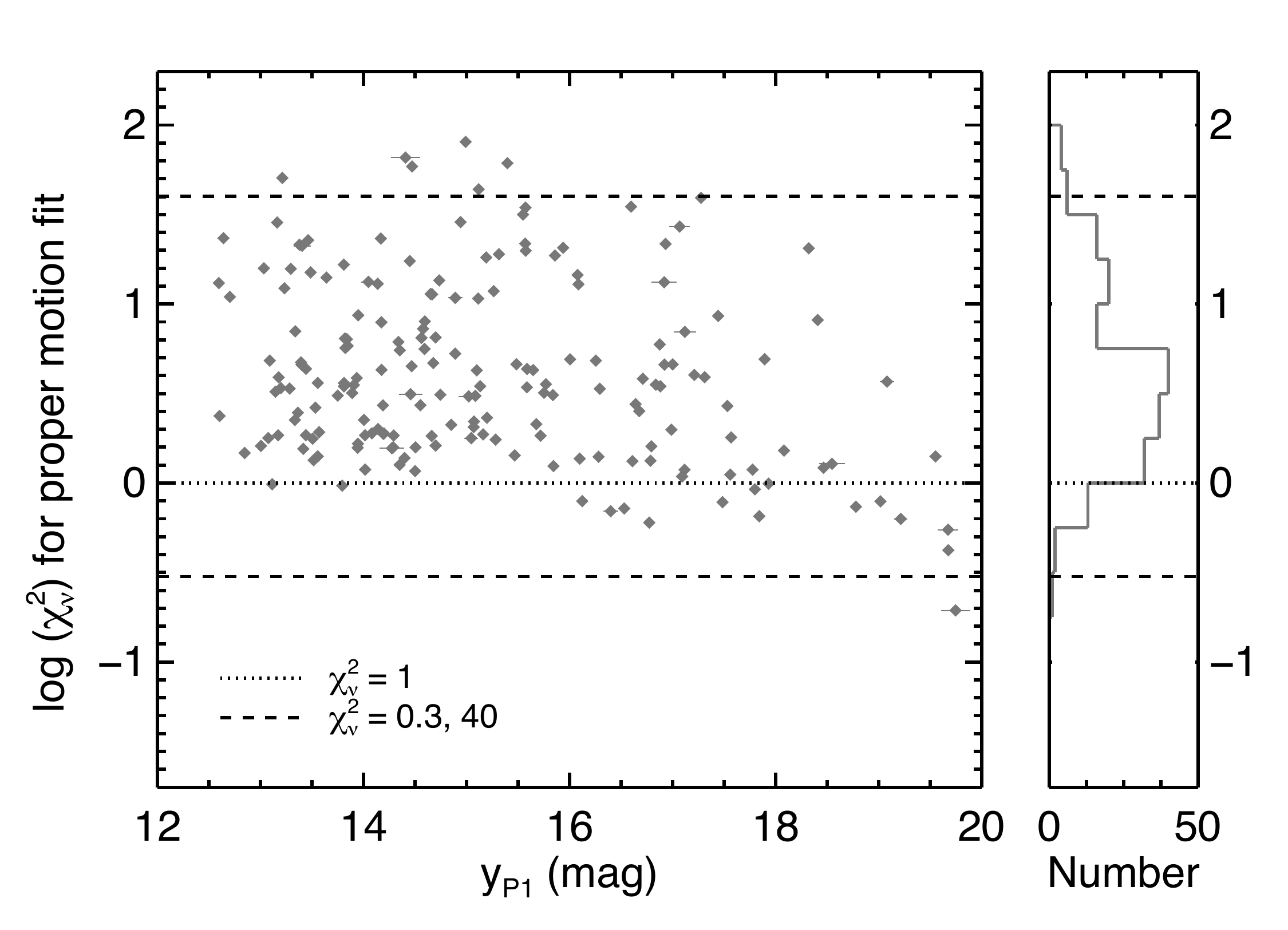}
    \caption{Reduced $\chi^2$ for our proper motions as a function of \yps\ for
      known Taurus members that are not saturated in PS1.  The histogram on the
      right shows the distribution of \rchi. The two dashed lines mark
      $\rchi=0.3$ and $\rchi=40$, values between which we regard our proper
      motion fits and errors as reliable (Figure~\ref{fig.taurus.reliable}).
      The dotted line marks $\rchi=1$. The fact that most of the fits have
      $\rchi>1$ suggests that our estimates for the PS1 astrometric errors are
      small compared to the scatter in positions between epochs.}
  \label{fig.taurus.pm.chisq}
\end{center}
\end{figure}

In Figure~\ref{fig.taurus.pm.chisq} we plot the reduced $\chi^2$ for the proper
motion fits as a function of \yps.  Most of our proper motion fits have
$\rchi>1$, suggesting that the astrometric uncertainties of the individual PS1
epochs are small compared to the scatter in R.A.~and Dec.~of the epochs.  The
proper motion errors may therefore be underestimated (by a factor of
$\approx$2).

\begin{figure}
\begin{center}
    \includegraphics[width=1.00\columnwidth]{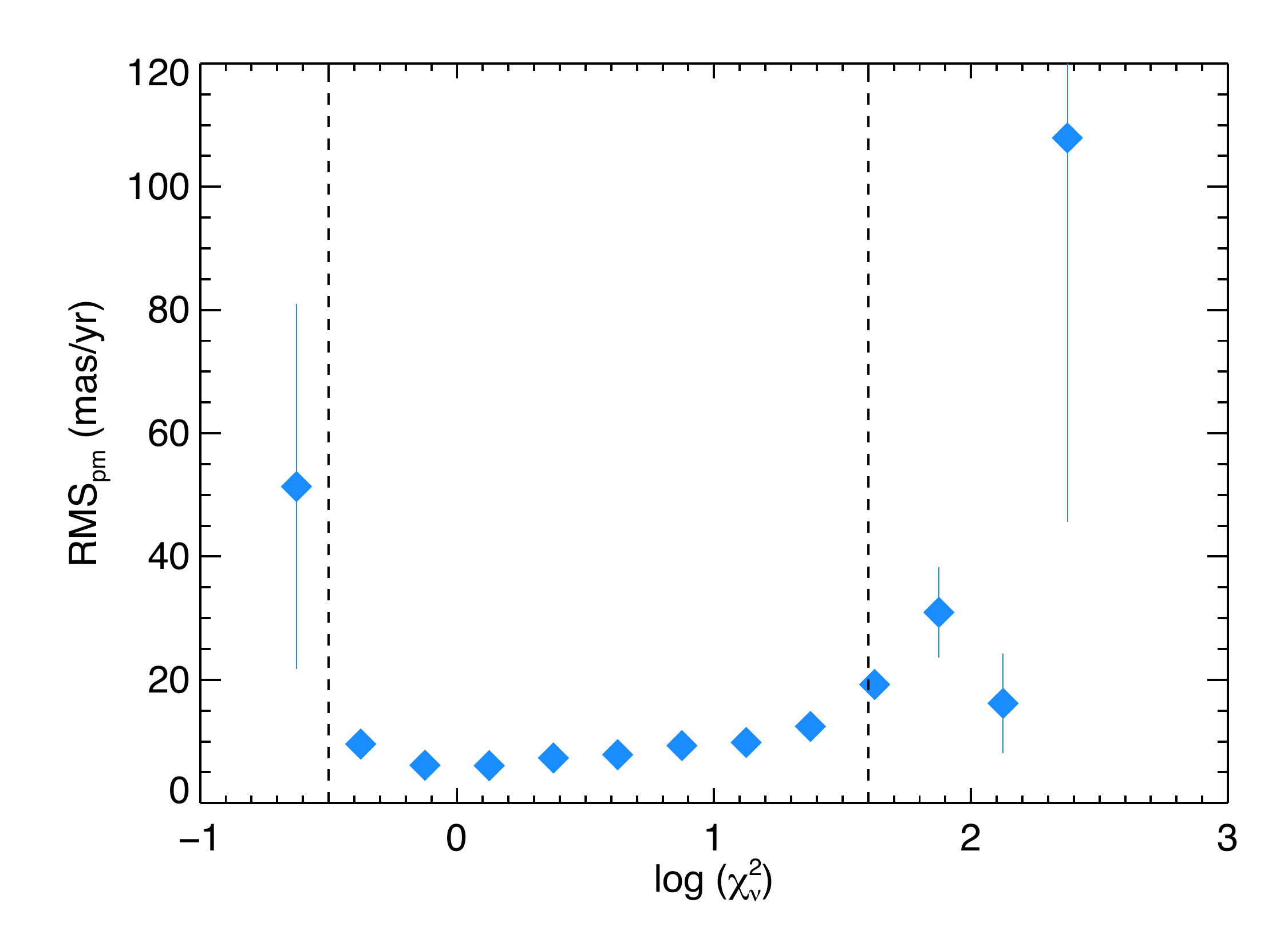}
    \caption{The rms of PS1 proper motions of well-detected objects in a
      0.5~deg$^2$ patch of sky near Taurus, for bins of 0.25 in log(\rchi).
      More than 97\% of the proper motions in this patch of sky are less than
      50~\my.  The low rms for proper motions with $-0.5<{\rm log} (\rchi)<1.6$
      (between the vertical dashed black lines), i.e., $0.3<\rchi<40$, implies
      that those proper motion measurements are reliable, while measurements
      with larger or smaller \rchi\ are less reliable.}
  \label{fig.taurus.reliable}
\end{center}
\end{figure}

To assess the reliability of our proper motions, we calculated proper motions
and \rchi\ for objects in a 0.5~deg$^2$ patch of sky near Taurus
($80^\circ<\alpha<81^\circ$, $5.5^\circ<\delta<6.0^\circ$), in which more than
97\% of objects have proper motions less than 50~\my.  We used all 8,962 objects
in this patch meeting the same criteria as our Taurus catalog and also having
$\yps<19.75$~mag to match the faintness limit of our Taurus catalog.  Because
the proper motions of this sample are small, the rms of the proper motions gives
us an estimate of the quality of the measurements.
Figure~\ref{fig.taurus.reliable} shows that proper motions measurements with
$0.3<\rchi<40$ had rms $\approx5-15$~\my, while proper motions with larger or
smaller \rchi\ had significantly greater spread.  We therefore adopted these
\rchi\ values as the limits between which we regard our proper motions as
reliable measurements.  Table~\ref{tbl.taurus.allpm} separates our Taurus proper
motions into those we regard as reliable (all but 5 of the objects) and
unreliable.  We report all of our proper motions for completeness, but when
$\rchi\le0.3$ or $\rchi\ge40$ the proper motions should be treated with caution.

Using the reliable proper motions and inverse variance weighting, we calculate a
weighted mean proper motion for Taurus of
($\mua=7.6\pm0.2,\ \mud=-17.4\pm0.2$~\my), with a weighted rms of 4.9~\my\ in
R.A. and 6.4~\my\ in Dec.  We compare this to the catalog of brighter Taurus
members compiled by \citet[][hereinafter D05]{Ducourant:2005iu}, whose proper
motions were similarly calculated using optical and 2MASS data and whose time
baselines range from 10 to more than 100 years.  Using the proper motions from
the full D05 Taurus catalog, we calculate a weighted mean proper motion of
($7.9\pm1.1,-20.5\pm1.0$~\my).  We note a $\approx$3~\my\ discrepancy with the
mean proper motion for our Taurus list, which is likely due to differences in
the astrometric reference frames used for the two samples.  The \PS~PV3.2
astrometry is tied to the \textit{Gaia}~DR1 reference frame, while the D05
proper motions use data from many sources with different astrometric reference
frames.  Only ten Taurus objects in the D05 catalog are not saturated in PS1 and
have reliable proper motions, preventing a robust object-by-object comparison of
the PS1 proper motions with those from D05.  The ten shared objects mostly have
proper motions from the two catalogs that are consistent within errors.

\section{Proper Motions of Known Upper Scorpius Members}
\label{appendixb}
We calculated proper motions for 482 members of Upper Sco from the lists of
\citet{Luhman:2012hj}, \citet{Dawson:2014hl}, \citet{Rizzuto:2015bs}, and this
paper, that met the same selection criteria we used for Taurus members in
Appendix~\ref{appendixa.catalog}.  Our catalog comprises the largest set of
proper motions for low-mass (spectral types~$\gtrsim$~M3) members of Upper Sco
published to date.  It includes 40 objects for which no proper motion has
previously been published, and 266 that improve on existing literature values,
which were drawn from NOMAD \citep{Zacharias:2005tx}, PPMXL
\citep{Roeser:2010cr}, SDSS DR9 \citep{Ahn:2012ih}, UCAC4
\citep{Zacharias:2013cf}, UKIDSS GCS DR9 \citep{Lawrence:2013wf}, USNO-B
\citep{Monet:2003bw}, \citet{Dawson:2011jq},
\citet{Lodieu:2007if,Lodieu:2013cj}, and \citet{Lodieu:2013eo}.  We list our
proper motions, number of epochs used, reduced $\chi^2$, and time baseline for
each proper motion fit in Table~\ref{tbl.uppersco.allpm}.

\begin{figure}
\begin{center}
    \includegraphics[width=1.00\columnwidth]{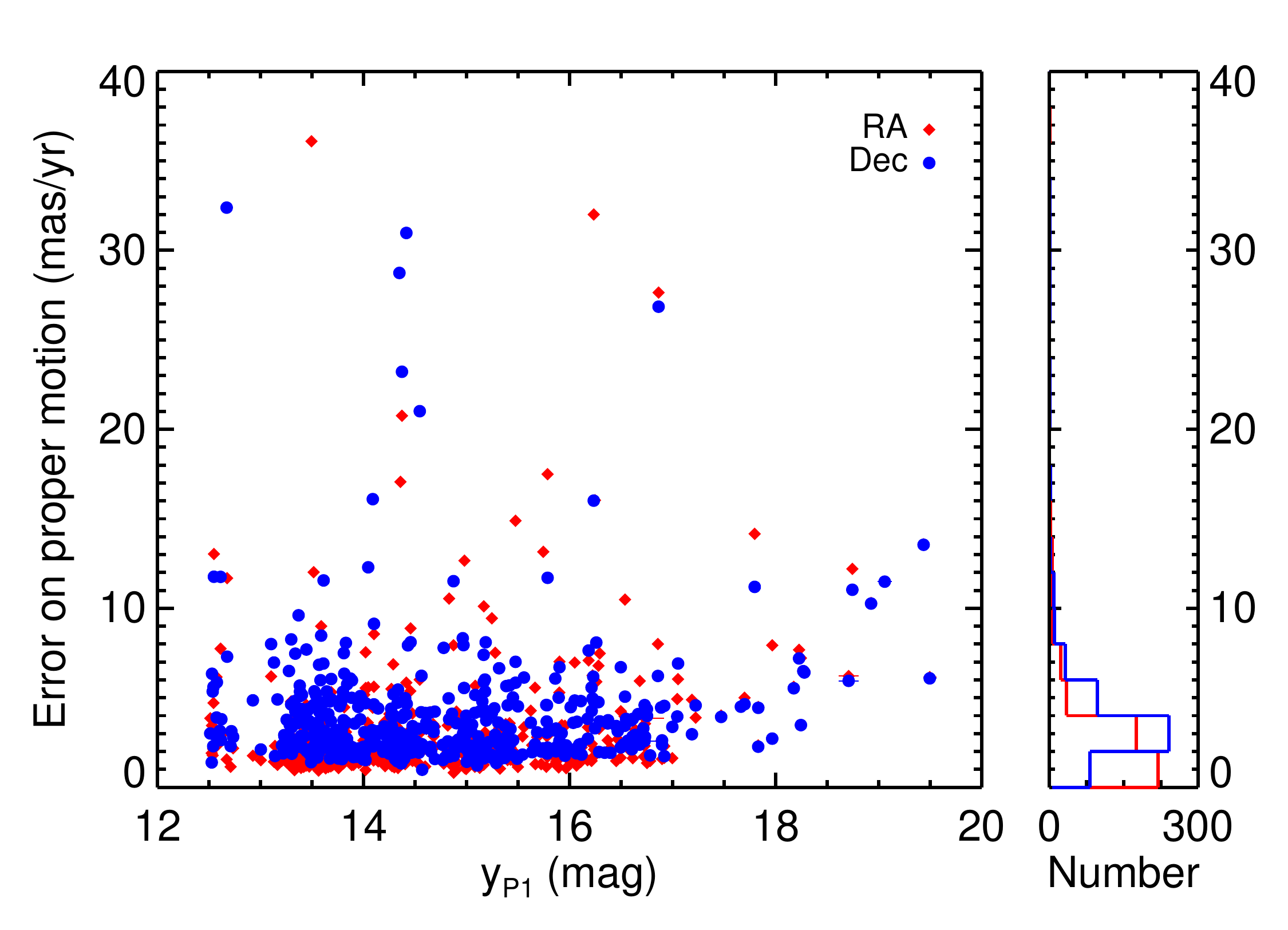}
    \caption{Errors on our \mua\ and \mud\ as a function of \yps\ for known
      Upper Sco members that are not saturated in PS1, using the same format as
      Figure~\ref{fig.taurus.pm.err}.  Most errors are $\lesssim$5~\my.}
  \label{fig.scocen.pm.err}
\end{center}
\end{figure}

\begin{figure}
\begin{center}
    \includegraphics[width=1.00\columnwidth]{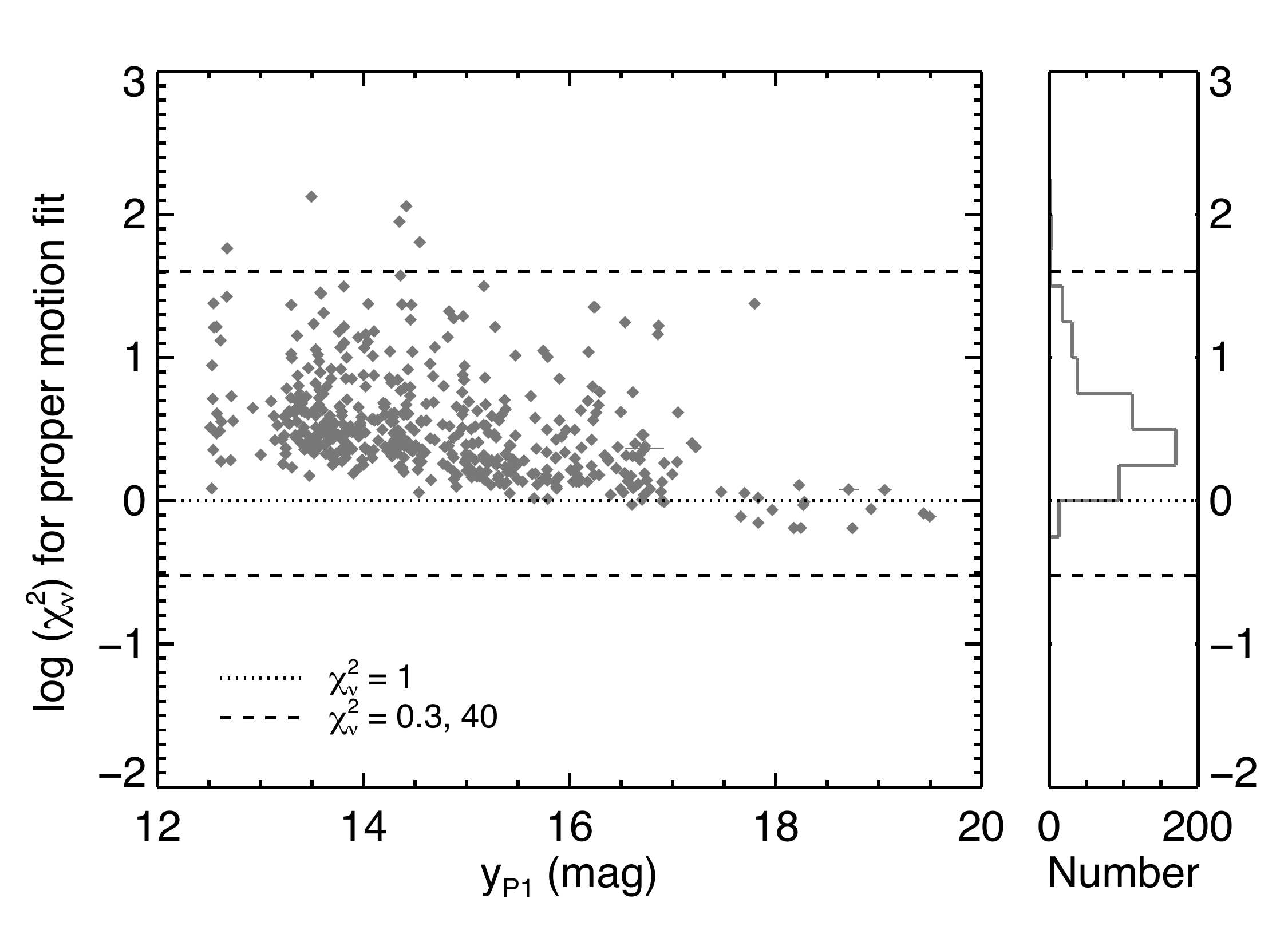}
    \caption{Reduced $\chi^2$ for our proper motion fits as a function of \yps\
      for known Upper Sco members that are not saturated in PS1, using the same
      format as Figure~\ref{fig.taurus.pm.chisq}.  The two dashed lines mark
      $\rchi=0.3$ and $\rchi=40$, values between which we regard our proper
      motion fits and errors as reliable (Appendix~\ref{appendixa.catalog}).
      The dotted line marks $\rchi=1$. As in Taurus, most of the proper motion
      fits have $\rchi>1$.}
  \label{fig.scocen.pm.chisq}
\end{center}
\end{figure}

Figure~\ref{fig.scocen.pm.err} shows the distribution of errors for our Upper
Sco proper motions as a function of the \yps\ magnitude of each source.  As with
our Taurus sample (Appendix~\ref{appendixa}), most of the errors are
$\lesssim$5~\my.  Figure~\ref{fig.scocen.pm.chisq} shows the reduced $\chi^2$
for the proper motion fits as a function of \yps.  Again like Taurus, most of
our proper motion fits have $\rchi>1$, implying that the proper motion errors
may be underestimated (by a factor of $\approx$2).

Using the reliable fits from our Upper Sco list, we calculate a weighted mean
proper motion for Upper Sco of ($\mua=-8.5\pm0.1,\ \mud=-19.6\pm0.1$~\my), with
a weighted rms of 4.3~\my\ in R.A. and 5.6~\my\ in Dec.  We compare our proper
motions to those listed in \citet[hereinafter P12]{Pecaut:2012gp} for F stars in
Upper Sco, which were determined by Hipparcos \citep{vanLeeuwen:2007dc} or
Tycho-2 \citep{Hog:2000wk}.  These stars are all saturated in PS1, so we are not
able to measure accurate proper motions using PS1 data and compare them directly
to the P12 proper motions.  For the P12 Upper Sco catalog, we calculate a
weighted mean proper motion of ($-11.5\pm0.3,-25.0\pm0.2$~\my).  We also
cross-matched the P12 objects with the UCAC4 catalog \citep{Zacharias:2013cf},
and using those proper motions obtained a weighted mean of
($-9.9\pm0.2,-21.6\pm0.2$~\my).  The source of the discrepancy between these two
mean proper motion vectors for Upper Sco and our determination from PS1~PV3.2 is
unclear, and may indicate a difference in the bulk motions of higher mass stars
(from P12) and low mass stars and brown dwarfs (our catalog) in Upper Sco.

\section{A New SpeX Prism Spectrum for the L0 Field Standard}
\label{appendixc}
\begin{figure}
\begin{center}
 \vspace{-20pt}
  \includegraphics[width=1.00\columnwidth]{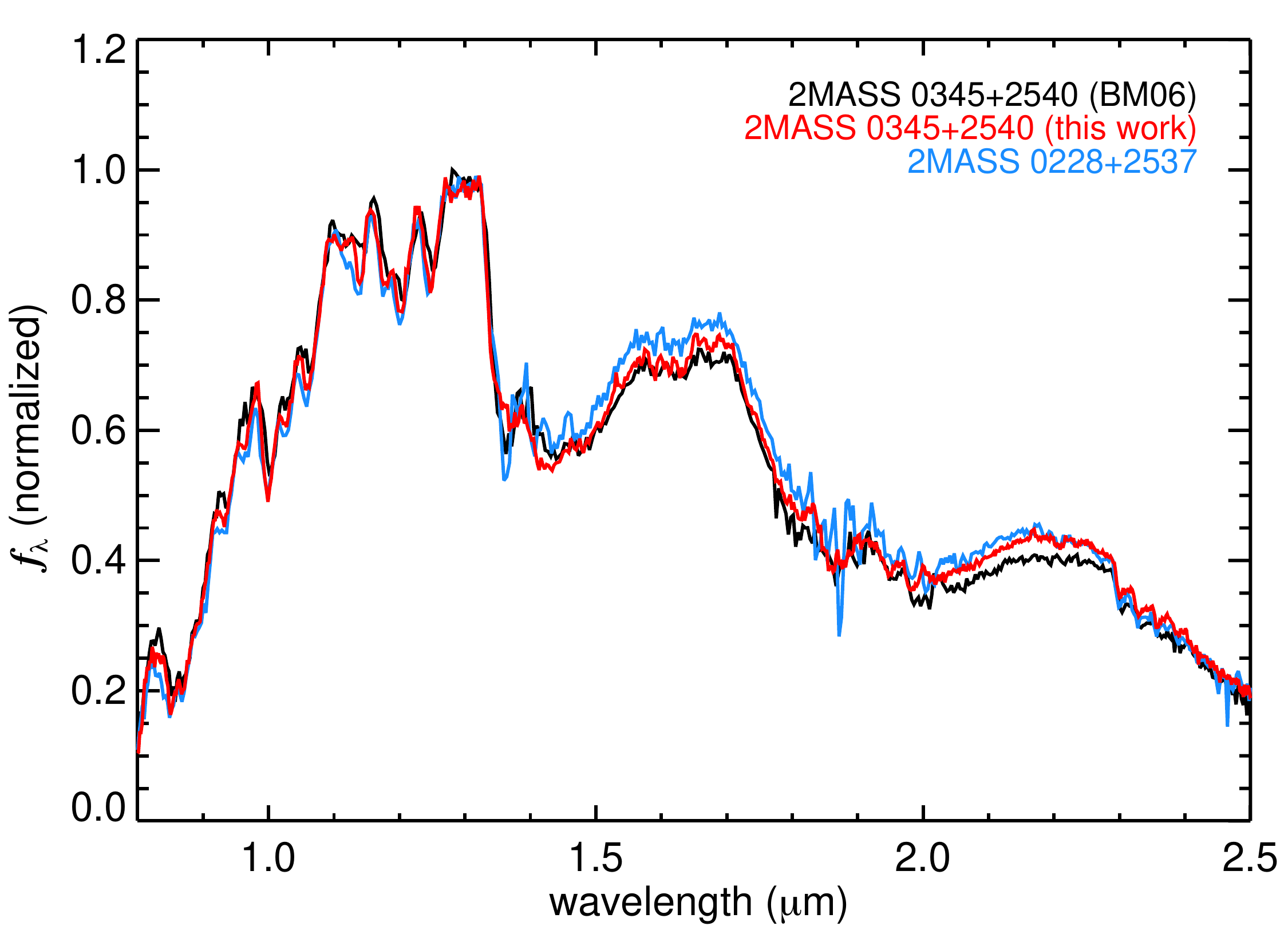}
  \includegraphics[width=1.00\columnwidth]{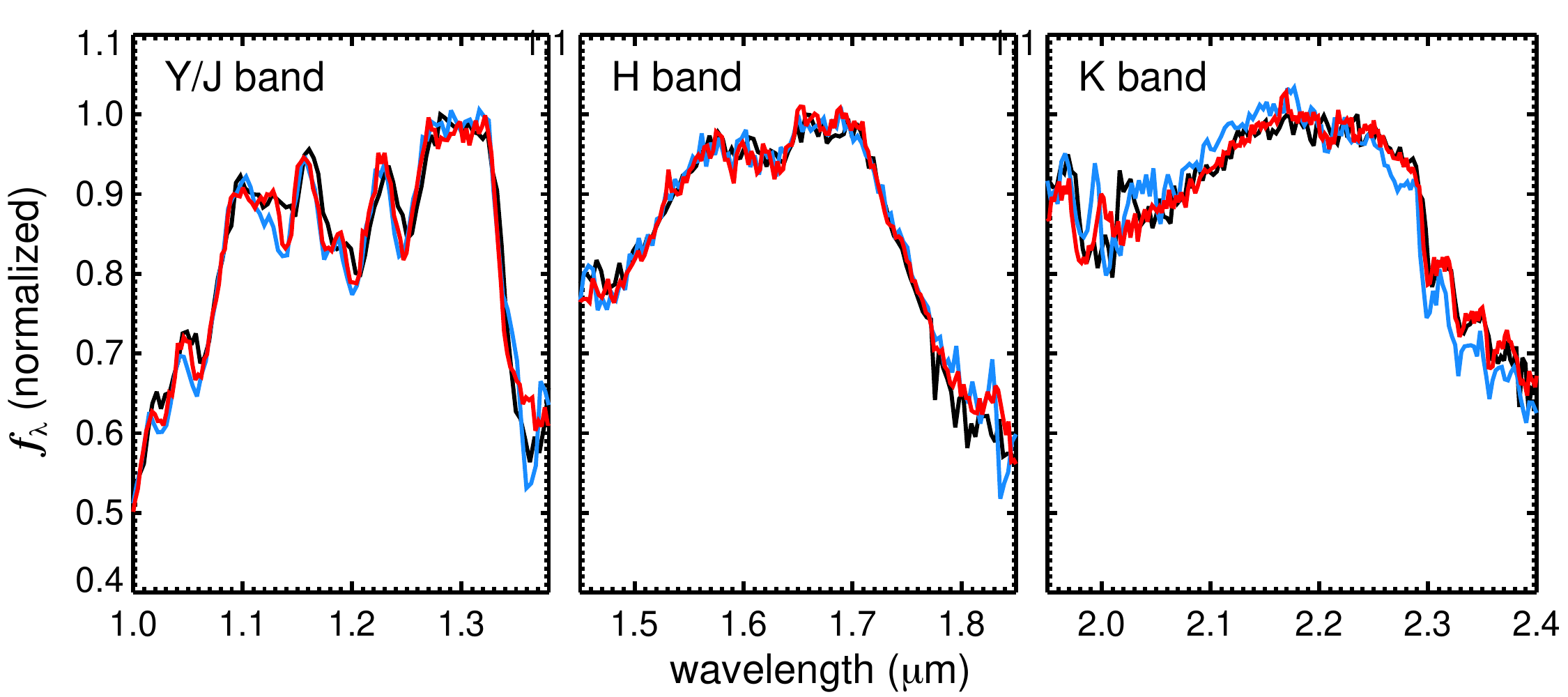}
  \caption{\textit{Top}: The SpeX Prism Library spectra for the L0 field
    standard 2MASS~J0345+2540 \citep[BM06, black]{Burgasser:2006jj} and the L0
    dwarf 2MASS~J0228+2537 \citep[blue]{Burgasser:2008cj}, compared with our new
    SpeX prism spectrum for 2MASS~J0345+2540 (red).  \textit{Bottom}: The same
    three spectra normalized and plotted separately for Y/J, H, and K~bands to
    compare the spectral shapes in each band.  The offset of the BM06
    2MASS~J0345+2540 spectrum towards longer wavelengths is evident in the
    $J$-band absorption features.  We use our new spectrum for 2MASS~J0345+2540
    for analysis in this paper.}
  \label{fig.2m0345.spectra}
\end{center}
\end{figure}

We identified a wavelength offset in the spectrum of 2MASS~J03454316+2540233
(hereinafter 2MASS~J0345+2540) publically available from the SpeX Prism
Library\footnote{http://pono.ucsd.edu/$\sim$adam/browndwarfs/spexprism}.
2MASS~J0345+2540 is the field L0 spectral standard for both optical
\citep{Kirkpatrick:1999ev} and near-infrared \citep{Kirkpatrick:2010dc}
wavelengths.  The spectrum, first published in \citet[hereinafter
BM06]{Burgasser:2006jj}, is shifted $\approx$0.01~\um\ toward longer wavelengths
(Figure~\ref{fig.2m0345.spectra}).  The offset is insignificant when visually
compared to other spectra over the full 0.8--2.5~\um\ range of SpeX prism
spectra, but is large enough to impact calculations of the \citet{Allers:2013hk}
gravity-sensitive spectral indices that use $\approx$0.02~\um-wide $J$-band
absorption features (Section~\ref{results.gravity}).  The offset is equivalent
to a velocity of $\approx$2,400~\kms, two orders of magnitude larger than the
radial velocities typical of nearby late-M and early-L dwarfs
\citep{Burgasser:2015dx}, so the offset is almost certainly due to a wavelength
calibration error.

To obtain a spectrum for 2MASS~J0345+2540 with an accurate wavelength
calibration, we observed the object on 2016 February 03 UT with IRTF/SpeX in
prism mode, using the 0.5'' slit.  Conditions were clear.  Observations were
made at an airmass of 1.01 and comprised six exposures of 120~sec using an ABBA
nodding pattern.  Immediately after we observed the A0V star HD~19600 for
telluric calibration.  We reduced the 2MASS~J0345+2540 spectrum using Spextool
v.~4.0 in standard fashion.  The final spectrum has a mean S/N of 115 in $J$
band (1.20--1.31~\um).  Figure~\ref{fig.2m0345.spectra} shows our spectrum
compared with the BM06 spectrum for 2MASS~J0345+2540 and the SpeX Prism Library
spectrum for the L0 dwarf 2MASS~J02281101+2537380 \citep{Burgasser:2008cj}.  The
redward offset on the BM06 spectrum is evident in the $J$-band absorption
features.  We therefore used our new spectrum for 2MASS~J0345+2540 in our
analysis (Sections \ref{results.indices} and~\ref{results.gravity}).

\bibliography{\string~/Astro/LaTeX/willastro}

\input{tbl1_ps1phot}

\begin{deluxetable}{lccccrccc}
\tablecolumns{9}
\tablewidth{0pc}
\tabletypesize{\tiny}
\tablecaption{Near-Infrared Photometry\label{tbl.nir}}
\tablehead{   % column headings
  \colhead{} &
  \multicolumn{4}{c}{2MASS Photometry} &
  \colhead{} &
  \multicolumn{3}{c}{MKO Photometry} \\
  \cline{2-5} \cline{7-9} \\
  \colhead{Name} &
  \colhead{2MASS Name} &
  \colhead{\jtwo} &
  \colhead{\htwo} &
  \colhead{\ktwo} &
  \colhead{} &
  \colhead{$J_{\rm MKO}$} &
  \colhead{$H_{\rm MKO}$} &
  \colhead{$K_{\rm MKO}$} \\
  \colhead{} &
  \colhead{} &
  \colhead{(mag)} &
  \colhead{(mag)} &
  \colhead{(mag)} &
  \colhead{} &
  \colhead{(mag)} &
  \colhead{(mag)} &
  \colhead{(mag)}
}
\startdata

\cutinhead{Taurus Discoveries}
PSO~J060.3+25 & J04011678+2557527 & $16.81\pm0.17$ & $15.73\pm0.14$ & $15.36\pm0.17$ & & $16.93\pm0.04$ & $16.10\pm0.03$ & $[15.53\pm0.07]$ \\
PSO~J077.1+24 & J05082480+2422518 & $16.93\pm0.14$ & $16.47\pm0.25$ & $15.82\pm0.22$ & & $17.06\pm0.04$ & $16.31\pm0.04$ & $15.59\pm0.03$ \\

\cutinhead{Taurus (previously known)}
2MASS~J0437+2331 & J04373705+2331080 & $17.38\pm0.19$ & $16.13\pm0.15$ & $15.44\pm0.15$ & & $17.24\pm0.02$\tablenotemark{a} & $16.16\pm0.01$\tablenotemark{a} & $15.20\pm0.02$\tablenotemark{b} \\

\cutinhead{Scorpius-Centaurus Discoveries}
PSO~J228.6$-$29 & J15144258$-$2942315 & $16.79\pm0.19$ & $15.95\pm0.15$ & $15.31\pm0.16$ & & $16.72\pm0.05$ & $16.18\pm0.05$ & $[15.37\pm0.11]$ \\
PSO~J229.2$-$26 & J15165651$-$2640251 & $16.46\pm0.13$ & $15.98\pm0.18$ & $15.18\pm0.15$ & & $16.76\pm0.03$ & $15.85\pm0.02$ & $[15.14\pm0.08]$ \\
PSO~J231.7$-$26 & J15270961$-$2626574 & $15.96\pm0.08$ & $15.21\pm0.11$ & $14.62\pm0.10$ & & $15.98\pm0.02$ & $15.25\pm0.03$ & $[14.64\pm0.05]$ \\
PSO~J231.8$-$29 & J15273464$-$2903354 & $15.77\pm0.09$ & $14.83\pm0.07$ & $14.33\pm0.08$ & & $[[15.72\pm0.09]]$ & $[[14.91\pm0.08]]$ & $[[14.29\pm0.08]]$ \\
PSO~J237.1$-$23 & J15483530$-$2308557 & $14.79\pm0.05$ & $14.13\pm0.07$ & $13.60\pm0.05$ & & $[[14.73\pm0.06]]$ & $[[14.19\pm0.07]]$ & $[[13.57\pm0.06]]$ \\
\enddata
\tablecomments{All 2MASS photometry is from the 2MASS Point Source Catalog
  \citep{Cutri:2003vr}. MKO photometry was obtained using UKIRT/WFCAM
  \citep{Best:2015em} unless enclosed in brackets or otherwise indicated.  $H_{\rm MKO}$ and $K_{\rm
    MKO}$ magnitudes enclosed in single brackets were synthesized using our
  observed $J_{\rm MKO}$ magnitudes and our low-resolution spectra
  (Section~\ref{obser}). MKO magnitudes enclosed in double brackets were
  synthesized using the 2MASS magnitudes for the corresponding filters and our
  low-resolution spectra.}
\tablenotetext{a}{From the UKIDSS Galactic Plane Survey \citep[DR6;][]{Lucas:2012wf}.}
\tablenotetext{b}{From the UKIDSS Galactic Clusters Survey \citep[DR9;][]{Lawrence:2013wf}.}
\end{deluxetable}

\begin{deluxetable}{lcccccc}
\tablecolumns{7}
\tablewidth{0pc}
\tabletypesize{\scriptsize}
\tablecaption{IRTF/SpeX Observations \label{tbl.obslog}}
\tablehead{   % column headings
  \colhead{Object} &
  \colhead{Date} &
  \colhead{Slit} &
  \colhead{R} & 
  \colhead{$t_\mathrm{int}$} & 
  \colhead{S/N\tablenotemark{a}} &
  \colhead{A0 V Standard} \\
  \colhead{} &
  \colhead{(UT)} &
  \colhead{('')} &
  \colhead{($\lambda/\Delta\lambda$)} &
  \colhead{(s)} &
  \colhead{} &
  \colhead{} 
}
\startdata

\vspace{-6pt}  & 2013 Jan 25 & 0.8 & 75 & 1920 &  & HD 27761 \\ \vspace{-6pt}
PSO~J060.3+25\tablenotemark{b} &  &  &  &  & 37 & \\ 
 & 2013 Nov 23 & 0.8 & 75 & 960 & & HD 31069 \\

\vspace{-6pt}  & 2013 Jan 26 & 0.8 & 75 & 840 &  & HD 34977 \\ \vspace{-6pt}
PSO~J077.1+24\tablenotemark{b} &  &  &  &  & 34 & \\
 & 2013 Nov 23 & 0.8 & 75 & 1320 &  & HD 38245 \\

PSO~J228.6$-$29 & 2013 Apr 18 & 0.8 & 75 & 720 & 24 & HD 146880 \\
PSO~J229.2$-$26 & 2013 Apr 17 & 0.8 & 75 & 2880 & 26 & HD 143822 \\
PSO~J231.7$-$26 & 2015 May 18 & 0.5 & 120 & 1440 & 49 & HD 133466 \\
PSO~J231.8$-$29 & 2015 May 29 & 0.5 & 120 & 3600 & 84 & HD 146606 \\
PSO~J237.1$-$23 & 2015 May 18 & 0.5 & 120 & 720 & 79 & HD 133466 \\
PSO~J239.7$-$23 & 2013 Apr 18 & 0.8 & 75 & 720 & 33 & HD 147013 \\

\enddata
\tablenotetext{a}{Mean $J$-band (1.20--1.31~\um) S/N per resolution element.}
\tablenotetext{b}{We present a combined spectrum and S/N from the two observations (Section~\ref{obser}).}
\end{deluxetable}

\begin{deluxetable}{lccccccc}
\tablecaption{Index-Based Spectral Types from \citet{Allers:2013hk}
  \label{tbl.indices.allers}}
\tablecolumns{8}
\tabletypesize{\scriptsize}
\tablewidth{0pt}
\tablehead{   % column headings
  \colhead{} &
  \multicolumn{5}{c}{Index-Derived Spectral Types} &
  \multicolumn{2}{c}{} \\
  \cline{2-6} &
  \multicolumn{4}{c}{} &
  \colhead{Avg.} &
  \colhead{Visual} &
  \colhead{Adopted} \\
  \colhead{Name} &
  \colhead{\hto} &
  \colhead{\htod} &
  \colhead{\htoa} &
  \colhead{\htob} &
  \colhead{SpT\tablenotemark{a}} &
  \colhead{SpT\tablenotemark{b}} &
  \colhead{SpT\tablenotemark{c}}
}
\startdata

\cutinhead{Taurus Discoveries}
PSO~J060.3+25 & L1.3$_{-1.4}^{+1.4}$ & \nodata & L0.8$_{-1.2}^{+1.2}$ & L0.5$_{-1.2}^{+1.1}$ & L0.9$_{-0.9}^{+0.9}$ & L0.5 & L1 \\
PSO~J077.1+24 & L2.0$_{-1.2}^{+1.2}$ & \nodata & L0.1$_{-1.2}^{+1.2}$ & L1.5$_{-1.0}^{+0.8}$ & L1.7$_{-0.9}^{+0.9}$ & L1 & L2 \\

\cutinhead{Taurus (previously known)}
2MASS~J0437+2331\tablenotemark{d} & L1.5$_{-1.4}^{+1.3}$ & \nodata & L0.1$_{-1.3}^{+1.3}$ & L0.0$_{-1.2}^{+1.2}$ & L0.9$_{-0.9}^{+0.9}$ & L1.5 & L1 \\

\cutinhead{Scorpius-Centaurus Discoveries}
PSO~J228.6$-$29 & L1.4$_{-2.2}^{+1.9}$ & L1.7$_{-1.4}^{+1.4}$ & L1.1$_{-1.4}^{+1.4}$ & M9.5$_{-1.9}^{+1.7}$ & L0.8$_{-1.3}^{+1.1}$ & L0.5 & L1 \\
PSO~J229.2$-$26 & \nodata & \nodata & M9.8$_{-1.3}^{+1.3}$ & \nodata & M9.8$_{-1.3}^{+1.3}$ & L0 & L0 \\
PSO~J231.7$-$26 & M9.9$_{-0.9}^{+0.9}$ & \nodata & M9.6$_{-1.2}^{+1.1}$ & L0.0$_{-0.9}^{+0.9}$ & M9.9$_{-0.6}^{+0.6}$ & L0 & L0 \\
PSO~J231.8$-$29 & L0.3$_{-0.6}^{+0.6}$ & \nodata & L1.0$_{-1.1}^{+1.1}$ & M9.9$_{-0.6}^{+0.6}$ & L0.2$_{-0.4}^{+0.4}$ & L0 & L0 \\
PSO~J237.1$-$23 & M7.1$_{-0.6}^{+0.5}$ & \nodata & M7.3$_{-1.2}^{+1.2}$ & M7.1$_{-0.6}^{+0.6}$ & M7.1$_{-0.4}^{+0.4}$ & M7.5 & M7 \\
\enddata
\tablenotetext{a}{Calculated as the weighted average of the spectral types from
  Monte Carlo trials for all indices, excluding those that fell outside the
  valid range for each index.}
\tablenotetext{b}{Spectral types determined by visual comparison with the \vlg\
  spectral standards proposed by \citet{Allers:2013hk}.  Uncertainties for these
  visual spectral types are $\pm1$~subtype for all objects.}
\tablenotetext{c}{The index-based average spectral types rounded to the nearest
  subtype, which we adopt as the final spectral types for our discoveries with
  uncertainties of $\pm1$~subtype.}
\tablenotetext{d}{Classified as L0 by \citep{Luhman:2009cn} using an optical spectrum.}
\end{deluxetable}

\begin{deluxetable}{lccccccc}
\tablecolumns{8}
\tablewidth{0pc}
\tabletypesize{\scriptsize}
\tablecaption{Index-Based Spectral Types from \citet{Burgasser:2006cf} for L dwarfs\label{tbl.indices.burg}}
\tablehead{   % column headings
  \colhead{} &
  \multicolumn{6}{c}{Index Values (Derived Spectral Types)\tablenotemark{a}} &
  \colhead{} \\
  \cline{2-7} &
  \multicolumn{5}{c}{} &
  \colhead{Avg.} &
  \colhead{Adopted} \\
  \colhead{Name} &
  \colhead{H$_2$O-$J$} &
  \colhead{CH$_4$-$J$} &
  \colhead{H$_2$O-$H$} &
  \colhead{CH$_4$-$H$} &
  \colhead{CH$_4$-$K$} &
  \colhead{SpT} &
  \colhead{SpT\tablenotemark{b}}
}
\startdata

\cutinhead{Taurus Discoveries}
PSO~J060.3+25 & 0.867 (L2.3) & 0.933 ($<$T0) & 0.780 (L3.4) & 1.118 ($<$T1) & 1.122 ($<$L0) & L2.8$\pm$0.8 & L1 \\
PSO~J077.1+24 & 0.946 (L0.4) & 0.924 ($<$T0) & 0.803 (L2.5) & 1.167 ($<$T1) & 1.137 ($<$L0) & L1.5$\pm$1.4 & L2 \\

\cutinhead{Taurus (previously known)}
2MASS~J0437+2331\tablenotemark{d} & 0.810 (L3.9) & 0.942 ($<$T0) & 0.788 (L3.1) & 1.190 ($<$T1) & 1.147 ($<$L0) & L3.5$\pm$0.6 & L1 \\

\cutinhead{Scorpius-Centaurus Discoveries}
PSO~J228.6$-$29 & 0.852 (L2.7) & 0.837 ($<$T0) & 0.735 (L5.2) & 1.126 ($<$T1) & 1.146 ($<$L0) & L3.9$\pm$1.8 & L1 \\
PSO~J229.2$-$26 & 0.918 (L1.0) & 0.969 ($<$T0) & 0.755 (L4.4) & 1.119 ($<$T1) & 1.111 ($<$L0) & L2.7$\pm$2.4 & L0 \\
PSO~J231.7$-$26 & 0.920 (L1.0) & 0.927 ($<$T0) & 0.841 (L1.0) & 1.167 ($<$T1) & 1.135 ($<$L0) & L1.0$\pm$0.5\tablenotemark{c} & L0 \\
PSO~J231.8$-$29 & 0.923 (L0.9) & 0.905 ($<$T0) & 0.821 (L1.7) & 1.150 ($<$T1) & 1.136 ($<$L0) & L1.3$\pm$0.6 & L0 \\
PSO~J237.1$-$23 & 0.978 ($<$L0) & 0.918 ($<$T0) & 0.929 ($<$L0) & 1.053 ($<$T1) & 1.026 ($<$L0) & \nodata & M7 \\
\enddata
\tablenotetext{a}{Spectral types calculated using the polynomials defined in
  \citet{Burgasser:2007fl}.}
\tablenotetext{b}{For description see Section~\ref{results.indices} and Table~\ref{tbl.indices.allers}.}
\tablenotetext{c}{The two contributing index-based spectral types are both L1.0,
  so we adopt an uncertainty of 0.5~subtypes.}
\tablenotetext{d}{Classified as L0 by \citet{Luhman:2009cn} using an optical spectrum.}
\end{deluxetable}

\begin{deluxetable}{lcccccccc}
\tablecaption{Low-Resolution Gravity Indices from
  \citet{Allers:2013hk} \label{tbl.gravity.allers}}
\tablecolumns{9}
\tabletypesize{\scriptsize}
\tablewidth{0pt}
\tablehead{
  \multicolumn{5}{c}{} &
  \colhead{Index} &
  \colhead{Gravity} &
  \colhead{Gravity} &
  \colhead{Adopted} \\
  \colhead{Name} &
  \colhead{FeH$_{z}$} &
  \colhead{VO$_{z}$} &
  \colhead{KI$_{J}$} & 
  \colhead{H-cont} &
   \colhead{Scores\tablenotemark{a}} &
  \colhead{Score\tablenotemark{b}} &
  \colhead{Class} &
  \colhead{SpT}
}
\startdata

\cutinhead{Taurus Discoveries}
PSO~J060.3+25 & 1.006$_{-0.036}^{+0.035}$ & 1.203$_{-0.034}^{+0.032}$ & 1.038$_{-0.016}^{+0.016}$ & 0.971$_{-0.023}^{+0.022}$ & 2122 (2122) & 2.0$_{-0.5}^{+0.0}$ & \vlg & L1 \\
PSO~J077.1+24 & 1.043$_{-0.042}^{+0.039}$ & 1.196$_{-0.030}^{+0.029}$ & 0.994$_{-0.017}^{+0.017}$ & 0.977$_{-0.026}^{+0.025}$ & 2122 (2122) & 2.0$_{-0.5}^{+0.0}$ & \vlg & L2 \\

\cutinhead{Scorpius-Centaurus Discoveries}
PSO~J231.7$-$26 & 1.075$_{-0.021}^{+0.020}$ & 1.214$_{-0.015}^{+0.015}$ & 1.032$_{-0.011}^{+0.011}$ & 0.993$_{-0.015}^{+0.015}$ & 2n22 (2n22) & 2.0$_{-0.0}^{+0.0}$ & \vlg & L0 \\
PSO~J231.8$-$29 & 1.075$_{-0.011}^{+0.011}$ & 1.239$_{-0.009}^{+0.009}$ & 1.061$_{-0.006}^{+0.006}$ & 0.995$_{-0.008}^{+0.008}$ & 2122 (2122) & 2.0$_{-0.0}^{+0.0}$ & \vlg & L0 \\
PSO~J237.1$-$23 & 1.056$_{-0.019}^{+0.019}$ & 1.045$_{-0.014}^{+0.014}$ & 1.026$_{-0.009}^{+0.009}$ & 0.999$_{-0.007}^{+0.007}$ & 2n22 (2n22) & 2.0$_{-0.0}^{+0.0}$ & \vlg & M7 \\
PSO~J239.7$-$23 & 1.032$_{-0.035}^{+0.034}$ & 1.101$_{-0.025}^{+0.025}$ & 1.065$_{-0.019}^{+0.019}$ & 0.974$_{-0.026}^{+0.025}$ & 2n12 (2n12) & 2.0$_{-1.0}^{+0.0}$ & \vlg & L0 \\
\enddata
\tablecomments{This table includes the discoveries for which our spectrum has
  high enough S/N to extract reliable measurements of the AL13 gravity indices,
  corroborated by visual inspection.  Our spectra for PSO~J228.6$-$29 and
  PSO~J229.2$-$26 and the 2MASS~J0437+2331 spectrum from \citet{Bowler:2014dk}
  did not yield reliable gravity scores, but do show visual indications of low
  gravity (Sections \ref{results.gravity} and~\ref{taurus.2m0437}).}
\tablenotetext{a}{Scores in parentheses were determined using the original AL13
  classification scheme, in which objects with index values corresponding to
  \intg\ but within $1\sigma$ of the \fldg\ value are classified with a score of
  ``?''.}
\tablenotetext{b}{The overall gravity score and the 68\% confidence limits were
  calculated as described in \citet{Aller:2016kg}.}
\end{deluxetable}

\input{tbl7_pm}

\input{tbl8_taurus_pm}

\input{tbl9_usco_pm}

\end{document}

%% file: tbl1_ps1phot.tex
\begin{deluxetable}{lccclcccc}
\tablecolumns{9}
\tablewidth{0pc}
\rotate
\tabletypesize{\tiny}
\tablecaption{\PS\ and AllWISE Photometry \label{tbl.ps1.wise}}
\tablehead{   % column headings
  \colhead{Name} &
  \colhead{\ips} &
  \colhead{\zps} &
  \colhead{\yps} &
  \colhead{AllWISE Name} &
  \colhead{$W1$} &
  \colhead{$W2$} &
  \colhead{$W3$} &
  \colhead{$W4$} \\
  \colhead{} &
  \colhead{(mag)} &
  \colhead{(mag)} &
  \colhead{(mag)} &
  \colhead{} &
  \colhead{(mag)} &
  \colhead{(mag)} &
  \colhead{(mag)} &
  \colhead{(mag)}
}
\startdata

\cutinhead{Taurus Discoveries}
PSO~J060.3200+25.9644 & $21.61\pm0.04$ & $20.06\pm0.02$ & $19.02\pm0.02$ & J040116.80+255752.0 & $14.99\pm0.04$ & $14.32\pm0.05$ & $>\!\!11.98$ & $8.58\pm0.39$ \\
PSO~J077.1033+24.3809 & $21.62\pm0.10$ & $20.18\pm0.05$ & $19.21\pm0.06$ & J050824.81+242251.3 & $15.15\pm0.05$ & $14.49\pm0.07$ & $>\!\!11.78$ & $>\!\!8.61$ \\

\cutinhead{Taurus (previously known)}
2MASS~J04373705+2331080\tablenotemark{a} & \nodata & $20.75\pm0.03$ & $19.67\pm0.10$ & J043737.06+233107.3 & $14.45\pm0.03$ & $13.90\pm0.05$ & $>\!\!11.68$ & $>\!\!9.10$ \\

\cutinhead{Scorpius-Centaurus Discoveries}
PSO~J228.6773$-$29.7088 & $21.41\pm0.04$ & $19.87\pm0.05$ & $18.74\pm0.03$ & J151442.59$-$294231.9 & $14.82\pm0.03$ & $14.22\pm0.05$ & $>\!\!11.83$ & $>\!\!8.90$ \\
PSO~J229.2353$-$26.6737 & $21.36\pm0.05$ & $20.14\pm0.04$ & $18.97\pm0.02$ & J151656.49$-$264025.3 & $14.73\pm0.04$ & $14.37\pm0.06$ & $>\!\!12.34$ & $>\!\!8.98$ \\
PSO~J231.7899$-$26.4494 & $20.73\pm0.04$ & $19.10\pm0.02$ & $18.06\pm0.02$ & J152709.57$-$262658.0 & $14.18\pm0.03$ & $13.79\pm0.04$ & $>\!\!12.55$ & $>\!\!8.99$ \\
PSO~J231.8942$-$29.0600 & $20.40\pm0.06$ & $18.81\pm0.02$ & $17.71\pm0.01$ & J152734.62$-$290335.8 & $13.88\pm0.03$ & $13.43\pm0.03$ & $>\!\!11.79$ & $>\!\!8.35$ \\
PSO~J237.1470$-$23.1489 & $18.69\pm0.01$ & $17.41\pm0.01$ & $16.61\pm0.01$ & J154835.30$-$230855.5 & $12.92\pm0.03$ & $12.40\pm0.03$ & $10.93\pm0.15$ & $8.75\pm0.45$ \\
\enddata
\tablecomments{\PS\ photometry is mean PSF magnitudes from \PS~Data Release 1.}
\tablenotetext{a}{Discovered by \citet{Luhman:2009cn}. Also known as
  PSO~J069.4044+23.5186.}
\end{deluxetable}

%% file: tbl7_pm.tex
\begin{deluxetable}{lccccccccc}
\tablecaption{Proper Motions, Luminosities, and Masses\label{tbl.kin.mass}}
\tablecolumns{10}
\tabletypesize{\scriptsize}
\tablewidth{0pt}
\tablehead{
  \colhead{Name} &
  \colhead{SpT} &
  \colhead{\mua\tablenotemark{a}} & 
  \colhead{\mud\tablenotemark{a}} &
  \colhead{\rchi} &
  \colhead{$N_{\rm ep}$} &
  \colhead{$\Delta t$} &
  \colhead{} &
  \colhead{log($L_{\rm bol}$/\lsun)} &
  \colhead{Mass\tablenotemark{b}}\\
  \colhead{} &
  \colhead{} &
  \colhead{(\my)} &
  \colhead{(\my)} &
  \colhead{} &
  \colhead{} &
  \colhead{(yr)} &
  \colhead{} &
  \colhead{(dex)} &
  \colhead{(\mjup)}
}
\startdata

\cutinhead{Taurus (age~$=1.5\pm0.5$~Myr, distance~$=145\pm15$~pc)}
PSO~J060.3+25 & L1 & $19.0\pm8.2$ & $-38.1\pm8.2$ & 0.8 & 43 & 17.2 &  & $-3.32^{+0.09}_{-0.10}$ & $6.0^{+0.9}_{-0.8}$ \\
 & & $14.3\pm3.1$\tablenotemark{c} & $-26.4\pm3.2$\tablenotemark{c} & & 29\tablenotemark{c} & 12.9\tablenotemark{c} \\ [6pt]
PSO~J077.1+24 & L2 & $14.1\pm12.5$ & $-27.1\pm12.1$ & 0.6 & 27 & 16.0 &  & $-3.34^{+0.09}_{-0.10}$ & $5.9^{+0.9}_{-0.8}$ \\ [3pt]
2MASS~J0437+2331 & L1 & $15.9\pm13.5$ & $-54.8\pm13.4$ & 0.5 & 16 & 15.0 &  & $-3.17^{+0.09}_{-0.10}$ & $7.1^{+1.1}_{-1.0}$ \\

\cutinhead{Upper Centaurus-Lupus (age~$=16\pm1$~Myr, distance~$=140\pm15$~pc)}
PSO~J228.6$-$29 & L1 & $-18.4\pm11.5$ & $-25.6\pm11.0$ & 1.3 & 18 & 15.4 &  & $-3.29^{+0.10}_{-0.10}$ & $15.1^{+0.6}_{-0.6}$ \\ [3pt]
PSO~J229.2$-$26 & L0 & $-30.8\pm10.9$ & $-38.4\pm10.3$ & 0.9 & 29 & 15.3 &  & $-3.18^{+0.09}_{-0.11}$ & $15.7^{+1.6}_{-1.5}$ \\ [3pt]
PSO~J231.7$-$26 & L0 & $-25.8\pm5.5$ & $-28.1\pm5.4$ & 1.0 & 35 & 16.2 &  & $-2.93^{+0.10}_{-0.10}$ & $20^{+4}_{-4}$ \\ [3pt]
PSO~J231.8$-$29 & L0 & $-21.1\pm2.8$ & $-36.3\pm2.3$ & 0.5 & 36 & 16.5 &  & $-2.82^{+0.10}_{-0.11}$ & $25^{+8}_{-6}$ \\
\enddata
\tablenotetext{a}{Proper motions from \PS\ Processing Version~3.2, calculated
  using PS1, 2MASS, and \textit{Gaia}~DR1 astrometry
  (Section~\ref{kinematics}).}
\tablenotetext{b}{Masses estimated using $L_{\rm bol}$ and the Lyon/DUSTY
  evolutionary models \citep{Chabrier:2000hq} as described in Sections
  \ref{taurus} and~\ref{scocen}.}
\tablenotetext{c}{From \citet{Bouy:2015gl}.  We adopt this proper motion for our
  analysis in Section~\ref{taurus}.}
\end{deluxetable}

%% file: tbl8_taurus_pm.tex
\begin{deluxetable}{lccccccccc}
\tablecaption{Proper Motions of Taurus Members \label{tbl.taurus.allpm}}
\tablecolumns{10}
\tabletypesize{\tiny}
\tablewidth{0pt}
\tablehead{
  \colhead{} &
  \multicolumn{3}{c}{Photometry} &
  \colhead{} &
  \multicolumn{5}{c}{Proper Motion} \\
  \cline{2-4} \cline{6-10}
  \colhead{Name} &
  \colhead{\ips} &
  \colhead{\zps} &
  \colhead{\yps} &
  \colhead{} &
  \colhead{\mua} & 
  \colhead{\mud} &
  \colhead{\rchi} &
  \colhead{$N_{\rm ep}$} &
  \colhead{$\Delta t$} \\
  \colhead{} &
  \colhead{(mag)} &
  \colhead{(mag)} &
  \colhead{(mag)} &
  \colhead{} &
  \colhead{(\my)} &
  \colhead{(\my)} &
  \colhead{} &
  \colhead{} &
  \colhead{(yr)}
}
\startdata

\cutinhead{Reliable fits: $0.3<\rchi<40$}
PSO J060.3200+25.9644 & $21.61\pm0.04$ & $20.06\pm0.02$ & $19.02\pm0.02$ & & $19.0\pm8.2$ & $-38.1\pm8.2$ & 0.8 & 43 & 17.2 \\
2MASS J04034997+2620382 & $16.27\pm0.01$ & $15.28\pm0.01$ & $14.73\pm0.01$ & & $15.6\pm6.3$ & $-16.0\pm5.3$ & 14 & 62 & 15.1 \\
2MASS J04064443+2540182 & $15.67\pm0.01$ & $14.68\pm0.01$ & $14.14\pm0.01$ & & $12.4\pm1.3$ & $-15.5\pm1.1$ & 2.0 & 87 & 15.2 \\
2MASS J04080782+2807280 & $15.19\pm0.01$ & $14.34\pm0.01$ & $13.91\pm0.01$ & & $2.6\pm1.5$ & $-7.7\pm1.5$ & 3.5 & 91 & 17.2 \\
2MASS J04102834+2051507 & $16.90\pm0.01$ & $15.87\pm0.01$ & $15.28\pm0.01$ & & $-2.3\pm2.6$ & $-7.3\pm1.0$ & 1.7 & 69 & 17.2 \\
2MASS J04105425+2501266 & $20.00\pm0.03$ & $19.74\pm0.07$ & $18.78\pm0.04$ & & $9.2\pm4.0$ & $-15.7\pm4.0$ & 0.7 & 34 & 17.2 \\
2MASS J04135328+2811233 & $20.37\pm0.03$ & $18.33\pm0.01$ & $16.93\pm0.01$ & & $13.3\pm9.9$ & $-39.8\pm12.3$ & 22 & 45 & 17.2 \\
2MASS J04135737+2918193 & $14.81\pm0.08$ & $14.79\pm0.06$ & $14.66\pm0.01$ & & $5.9\pm1.3$ & $-26.4\pm1.5$ & 1.8 & 64 & 17.2 \\
\enddata

\tablecomments{The objects in this table are taken from the catalog of
  \citet{Esplin:2014he}, except for PSO~J060.3200+25.9644 and
  PSO~J077.1033+24.3809 which are new discoveries presented in this paper.  We
  adopt a photometric precision floor of 0.01~mag for the PS1 photometry,
  following the analysis of \citet{Schlafly:2012da}.  The errors reported in the
  PS1 database are formal errors that do not include systematics, and are often
  smaller.}
\tablenotetext{a}{This object does not have a 2MASS detection.}
\end{deluxetable}

%% file: tbl9_usco_pm.tex
\begin{deluxetable}{lccccccccc}
\tablecaption{Proper Motions of Upper Scorpius Members \label{tbl.uppersco.allpm}}
\tablecolumns{10}
\tabletypesize{\tiny}
\tablewidth{0pt}
\tablehead{
  \colhead{} &
  \multicolumn{3}{c}{Photometry} &
  \colhead{} &
  \multicolumn{5}{c}{Proper Motion} \\
  \cline{2-4} \cline{6-10}
  \colhead{Name} &
  \colhead{\ips} &
  \colhead{\zps} &
  \colhead{\yps} &
  \colhead{} &
  \colhead{\mua} & 
  \colhead{\mud} &
  \colhead{\rchi} &
  \colhead{$N_{\rm ep}$} &
  \colhead{$\Delta t$} \\
  \colhead{} &
  \colhead{(mag)} &
  \colhead{(mag)} &
  \colhead{(mag)} &
  \colhead{} &
  \colhead{(\my)} &
  \colhead{(\my)} &
  \colhead{} &
  \colhead{} &
  \colhead{(yr)}
}
\startdata

\cutinhead{Reliable fits: $0.3<\rchi<40$}
2MASS J15350863$-$2532397 & $17.98\pm0.01$ & $16.86\pm0.01$ & $16.28\pm0.01$ & & $-10.2\pm6.8$ & $-13.5\pm4.8$ & 4.7 & 47 & 16.5 \\
2MASS J15355111$-$2021008 & $18.73\pm0.01$ & $17.32\pm0.01$ & $16.50\pm0.01$ & & $-11.2\pm1.6$ & $-17.6\pm2.2$ & 1.2 & 38 & 16.6 \\
2MASS J15411302$-$2308161 & $18.14\pm0.01$ & $17.02\pm0.01$ & $16.40\pm0.01$ & & $-12.1\pm2.2$ & $-22.8\pm1.9$ & 1.1 & 48 & 15.3 \\
2MASS J15420830$-$2621138 & $16.83\pm0.01$ & $15.85\pm0.01$ & $15.32\pm0.01$ & & $-10.9\pm3.3$ & $-16.8\pm6.6$ & 3.7 & 53 & 15.7 \\
2MASS J15423609$-$2108428 & $18.39\pm0.01$ & $17.02\pm0.01$ & $16.22\pm0.01$ & & $-11.1\pm6.3$ & $-13.5\pm5.0$ & 6.3 & 37 & 16.6 \\
2MASS J15433947$-$2535549 & $20.47\pm0.03$ & $18.85\pm0.02$ & $17.83\pm0.02$ & & $-15.9\pm4.5$ & $-16.6\pm4.4$ & 0.7 & 41 & 15.3 \\
2MASS J15442275$-$2136092 & $18.29\pm0.01$ & $17.23\pm0.01$ & $16.60\pm0.01$ & & $-13.2\pm2.2$ & $-28.2\pm2.1$ & 0.9 & 52 & 16.6 \\
2MASS J15453662$-$2510493 & $17.85\pm0.01$ & $16.70\pm0.01$ & $16.05\pm0.01$ & & $-10.9\pm1.8$ & $-20.2\pm2.1$ & 2.0 & 55 & 15.7 \\
\enddata
\tablecomments{The objects in this table are taken from the catalogs of
  \citet{Luhman:2012hj}, \citet{Dawson:2014hl}, and \citet{Rizzuto:2015bs},
  except for PSO~J237.1470$-$23.1489 and PSO~J239.7015$-$23.2665 which are new
  discoveries presented in this paper.  We adopt a photometric precision floor
  of 0.01~mag for the PS1 photometry, following the analysis of
  \citet{Schlafly:2012da}.  The errors reported in the PS1 database are formal
  errors that do not include systematics, and are often smaller.}
\end{deluxetable}